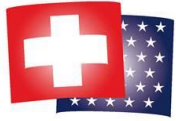

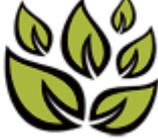

Project Submitted in Partial Fulfillment

of the Requirements of the

# Master of Business Administration

of the

University of Cumbria

The Disruptive Potential of FinTechs in the German

Consumer Finance Sector - A Blue Ocean Scenario?

Name:              Christian Wischnewski

Student ID:        [redacted for publication]

Submitted:         May 9th, 2017

# Abstract

Antony Jenkins, the former CEO of Barclays Bank Plc. stated in November 2015 "over the next 10 years, we will see a number of very significant disruptions in financial services (...)", hinting at the impact young and agile startup companies in the banking sector could have on existing players in the industry. Using the Blue Ocean strategy as an underlying strategic element, this dissertation analyses whether this statement holds true for the rather more conservative banking sector in Germany and the overall risk-averse mindset of the German population by using both quantitative and qualitative elements to assess the current market share of FinTech companies in the Federal Republic, as well as grasp a potential outlook on the future development. A literature review of the strategic framework, the banking sector in Germany and the FinTech sector is carried out accordingly. Subsequently, a formal verification as to whether the banking sector is a 'Red Ocean' and if the FinTech industry is a 'Blue Ocean' is carried out using case studies from both sectors. A quantitative analysis of banking customers in Germany and their use of FinTech companies is conducted by way of an online survey, with selected participants being interviewed thereafter to gain additional insights. Data evaluation is made using pivotal analysis and cross tabulation of survey results and interview findings, along with extrapolating indicators to reflect the full size of the German banking sector and transactional volumes per segment are provided and examined for signs of elevated FinTech use in the market. Despite several limitations from where ideas for future research are derived, the outcomes provide an overview of existing trends for the use of FinTechs in Germany. The main finding is that with the notable exception of payment solutions, Germans do not have a high affinity towards FinTechs, rendering them a byproduct of the financial service industry, with limited market share and low potential.

# Table of Contents

# Index of Figures

# Index of Tables

# Glossary / Acronyms

BASEL: European framework for banks and financial institutions providing minimum ratios / requirements for operations; defined by a consortium surrounding the Bank for International Settlements (BIS). Currently in revision three, also known as BASEL III.

BOS: Blue Ocean Strategy

BVR: Bundesverband der Volks- und Raiffeisenbanken (Federal Association of Communal and Cooperative Banks)

B2B: Business to Business; products & services available for enterprises only

B2C: Business to Consumer; products & services available for private individuals

CFS: Consumer Finance Sector

DSGV: Deutsche Sparkassen Giroverband (Association of German Savings Banks)

ECB: European Central Bank

EU: European Union

FinTech: Financial Technology, also used as a standalone term for 'Financial Technology Company'; an innovative company active in the financial sector, usually with a single specialization (unique product/service)

INSEAD: Institut Européen d'Administration des Affaires; a leading French business school

IPO: Initial Public Offering; the process of being listed at a Stock Exchange

KPI: Key Performance Indicator

KYC: Know Your Customer [principal], the legal requirement imposed on banks by regulatory frameworks such as BASEL to positively identify their customers to avoid / limit fraud, money laundering and other illegal activities

MiFID: Markets in Financial Instruments Directive, an EU legislation aimed to increase transparency in commission-based sales of financial products, requiring financial institutions to disclose commissions and kick-backs from product sales

P2P / p2p: Peer-to-Peer, a direct connection between two individuals without middleman. In the context of this study, the term is used for lending and money transfers from user to user without a financial institution.

USP: Unique Selling Proposition / Unique Selling Point; a feature unique to a certain product / service / company / brand

*"Over the next 10 years we will see a number of very significant disruptions in financial services. Let's call them Uber moments."*

*Antony Jenkins, Former CEO, Barclays Bank Plc.*

# 1.    Introduction and Background

The German banking sector comprises almost 6.8 trillion EUR in assets, and is – together with the French – the largest in the European Union. Judging from the market capitalization, it accounts for a total of 28% of the assets held by banks in the entire EU (Mersch 2016). However, the domestic market has been relatively immobile and inflexible over the past decades, and is largely characterized by fierce competition amongst the existing players in the market for a customer base that is rather shrinking than growing in size, as the German population growth is statistically negative since 1995; and only the recent influx of migrants, of which many are displaced from their respective home country, have contributed to an overall increase of the population in the years 2014 (+0,5%) and 2015 (+1,2%). As of the end of 2015, a total of 8,7 million out of the total 82.2 million people in Germany are foreigners (Statistisches Bundesamt 2015)[1].

Subsequently – and further fueled by the zero-interest policy of the European Central Bank, which is in place since March 2016 – profit margins in the consumer finance sector have been under pressure; forcing banks to engage in price fights on the one hand, while desperately trying to reduce operational and other expenses to maintain a positive profit margin. As of late, some banks have even begun imposing negative interest rates on savings accounts; while (re-) introducing several fees on previously free account types, or raising existing fees to generate additional income (Berlemann et al, 2014; Bos & Kool, 2006; Burghof et al., 2015; European Central Bank, 2015).

Consumers on the other hand are increasingly unhappy with the low return on savings and the changed fee structures, and are thusly more likely to move to any of the cheaper competitors. So called "online banks" or "branchless banks" as well as finance-related startup companies (so called "FinTechs") on both spectrums of the traditional consumer banking sector (lending and deposit taking as well as related services such as international transfers) are also increasingly targeting the German market, while their business models – utilizing the banking sector without being banks on their own, and as such avoiding cost-intensive licenses – allows for even greater margins and/or lower costs (Dorbian, 2016; Kang et al, 2016; McAuley, 2015).

---

[1] Federal Statistics Office, see http://www.destatis.de/en/

While FinTechs have gained in popularity only in recent years, branchless banks had their debut in the German market in 1999 with the emergence of the Netbank AG, which was also the first pure online bank in Europe (Netbank AG 2016); while other competitors – amongst them the now largest so-called "direct bank" ING DiBa – followed shortly after. In the years 2000 until 2015 an approximated 18.2 million people are now using online banks as opposed to traditional brick and mortar banks, comprising over 20% of the total population (Luber 2015).

Consecutively, the German consumer finance landscape can be divided into three segments: Traditional banks (including savings banks and cooperatives), online banks, and FinTechs, with various degrees of complexity in terms of product design, specialization, market shares and profitability / return. Taking the "Blue Ocean Strategy" coined by W. Chan Kim and Renée Mauborgne into consideration, which defines "Red Oceans" as established markets with a large number of actors competing for a fixed number of potential customers and "Blue Oceans" as newly created and emerging markets with no competitions, it appears at first glance as if the traditional banking sector would be a "Red Ocean", and the FinTech market largely a "Blue Ocean" (Kim & Mauborgne 2004; Sheehan & Vaidyanathan 2009).

The branchless banking market however seems to be in a transitional phase from a "Blue" to a "Red" ocean as more and more contestants enter the market, and even some traditional banks starting their own web-only subsidiaries to at least retain the customer base otherwise willing to change banking relationships – reduced income is still preferable to lost income. Now the Blue Ocean Strategy makes reference to the fact that it is not sufficient to create new markets to remain in a leading position but rather that continuous effort needs to be put into place to utilize the full potential – effectively recreating blue oceans one after another:

*'Are there lasting "excellent" or "visionary" companies that continuously outperform the market and repeatedly create blue oceans?' (Kim & Mauborgne 2004)*

This thesis aims to contribute to answer this question in relation to the German consumer finance sector, as will be further outlined in the following section.

## 1.1. Aims and Objectives

In view of the above background, the aim of this dissertation is threefold:

(1) Establish whether the German consumer finance market meets the textbook definition of the "Blue Ocean Strategy" for Red and Blue Oceans respectively, given the tools provided by Kim & Mauborgne;

(2) Find quantifiable indicators using a detailed survey amongst consumers (bank customers) in Germany, as to which part of the actively banked population are considered to be "pioneers", "migrators" and "settlers". Those terms are used by Kim & Mauborgne to describe the behavior of companies adapting to a new market environment, whereas this dissertation aims at replicating similar behavioral pattern amongst users of the FinTechs. When related to technological innovation, the terms "innovators", "early adopters", "early majority", "late majority" and "laggards" are also commonly used (Rogers 1995);

(3) Based on the outcome of (1) and (2), the research further aims to assess which part of the population is willing and/or likely to move from their current banks on towards the respective 'newer' form of banking. Identified indicators will be used to calculate the potential impact on banks' profitability as well as to provide an outlook on the potential market share a typical blue-ocean FinTech might initially acquire. This will help to answer the question of whether the emergence of new players in the banking sector truly has the potential to change the consumer finance behavior with a lasting impact, or whether the innovations presented are more likely to attract novelty customers but will, in the long run, only introduce new functions that all market players will sooner or later introduce into their general product range.

Following the aims and objectives, the next chapter provides a literature review regarding the Blue Ocean Strategy and its application using the tools developed by its creators as well as its common criticisms and limitations. Furthermore, an overview of the German consumer-banking sector will be provided, and aligned with the three core segments to be analyzed in the scope of this dissertation, with a particular emphasis on the FinTech segment.

Chapter 3 describes and justifies the adopted research methodology for a quantitative analysis of consumer behavior. Emphasis is placed on statistical sampling and methods for data evaluation in conjunction with the objectives of the research. One other core element of this chapter will be the details and processes for survey design, including the detailing of the choice of certain survey questions, scaling methods, and how results will be correlated with existing statistical material from various sources.

Chapter 4 presents the outcomes and the accompanying narrative of the research from the gathered data, and how the identified parameters contribute to the research objectives, giving detailed overview of the individual facts gathered.

The last chapter will draw critical conclusions as to which indications for the potentially disruptive element of FinTechs on the German consumer finance market are supported by the results of the research, and how likely it is to establish fundamentally new "blue oceans" in an industry that has been around for millennia.

Finally, the chapter will consider areas for future research not covered within this thesis.

## 1.2. Foreseen Limitations

- Focus on quantitative research with few qualitative components due to large sample size requirements and lack of time / resources for a full qualitative assessment given the scope of this dissertation
- Distribution of quantitative survey via digital channels only
- Lack of peer-reviewed articles particularly related to the FinTech sector due to its respective young age may lead to overproportional use of industry research through industry consulting services such as KPMG or Accenture
- FinTechs as often unregulated and very young companies are very secretive / restrictive about publishing hard data (profit & loss statement, annual statements etc.), so many information will have to be acquired through third parties and/or press releases, which are likely to contain only selected, favorable news

# 2.    Literature Review

This literature review aims at examining the recently emerging FinTech market from the viewpoint of the Blue Ocean Strategy in the German consumer finance sector. The principal goal of the review is to provide an answer to the first three of the following objectives:

1.) Establish the background of the strategic framework to be applied, starting from a definition of 'Strategic Management', and detailing the use of the BOS including its application, limitations, and scholarly criticism. To also provide an overview of the research subjects, namely the consumer finance sector in Germany and the FinTech industry.
2.) Factually establish whether the 'common' banking sector meets the textbook definition of a 'Red Ocean' and the FinTech sector a 'Blue Ocean' respectively;
3.) Derive the contextual requirements for further research (sampling) of the FinTech market as a potential share of the German consumer finance market, using a quantitative approach (to be addressed in the methodology and analysis).

Following the descriptive introduction to the first two objectives, the third aspect aims to critically reflect how the strategic framework correlates with the markets under review using the tools provided by Kim & Mauborgne in their original literature from 2005 as well as the amendment published in 2015[2].

Objective four of the literature review has a conclusive function: Under the assumption that the FinTech market is indeed a 'Blue Ocean' as per the literature, the cornerstones for the following quantitative research shall be established and serve as a transition to the research methodology in chapter 3.

## 2.1   Sources

The literature search was carried out using various available sources including the databases made available through the University of Cumbria One-Search, the academic communities on researchgate.net and mendeley.com, articles accessible through Google Scholar, Academic Search (EBSCO Publishing), the Directory of Open Access Journals (DOAJ; administered by the Lund University, Sweden), as well as using paid access to the Harvard Business Review (HBR) provided through the employer of the author.

---

[2] *The author will try to use as many English speaking sources as possible, but might need to revert to German sources as per the nature of the underlying research topic.*

Physical books were accessed mainly through the Grimm Library of the Humboldt University in Berlin and their partnering libraries in Germany, providing access to literature through inter-library loans. Electronic books in pdf, epub or mobipocket format have been used for referencing together with entries in Google Books where no physical copy of the resource could be obtained in due course.

Key terms for the research were 'The Blue Ocean Strategy', 'Blue Ocean Case Studies', 'Blue Ocean Scenarios in Banking', 'Financial Technology', 'FinTech Startups', 'Emerging Finance Technologies', 'FinTech Market Potential', 'Consumer Finance Trends', 'Banking Developments' and other related terms.

## 2.2   The FinTech Pioneers Bootcamp in Berlin, Germany

Additional insights specifically into the FinTech industry were gained through attendance of the 'FinTech Pioneers Bootcamp[3]' in Berlin, Germany on the 16[th] & 17[th] of February, 2017. Reference to this event will be made as (FinTech Pioneers 2017); the agenda is included as Appendix 1. An exemplary slide from the presentation on 'Disruptive Investments' (session at 9:10 – see agenda in Appendix 1) showing the global GDP vs. global volume of stocks traded in 2010 is displayed in Appendix 2. Further material, including the session overview and photos taken during the various presentations are made available via Dropbox[4].

The Bootcamp's primary aim was to provide a platform for established FinTechs to showcase their products and give insights into the industry; for emerging early-stage startups in the sector to connect with potential investors, as well as to allow interested members of the general public to interact with founders and mangers of FinTechs.

Especially N26 (previously Number26), a German FinTech that later obtained a full-fledged banking license and now operates as N26 Bank GmbH was of particular interest for the scope of this paper[5]. N26 was represented by co-founder and managing director, Maximilian Tayenthal, who provided valuable insights into the acceptance of FinTechs through 'Early Adopters'; especially his statement "not the founder starts a movement, but the first follower" (Tayenthal, 2017) reverberates with the topic of this dissertation.

---

[3] For further details, see also https://pioneers.io/events/fintech/ (accessed February 16, 2017)

[4] https://www.dropbox.com/sh/xdp42v0ve2x42oh/AAD3avzu67BbgOVev9dk38Dba?dl=0

## 2.3 Part I – Strategic Management & The Blue Ocean Strategy

### 2.3.1 Strategic Management

When Harvard Professor Michael E. Porter first published his book 'Competitive Strategy: Techniques for Analyzing Industries and Competitors' in 1980, he was able to set a first standard for defining business objectives through strategies. Amongst other factors, he mentions that a company can have three distinctive goals: Cost leadership, differentiation (from competitors) or focus (on few key services) (Lewis 1981).

Using the term 'strategy' in a business context however has been around long before, and has been discussed by scholars at length (Andrews 1971) – but in the context of progressive globalization and the emergence of modern communication technologies such as the ARPANET in the late 1960s / early 1970s and later the internet from the early 1990s (Hauben 1998), having a defined strategic roadmap became progressively more important to achieve business goals; and various researchers amongst multiple disciplines have developed tools to assist businesses in defining their strategic roadmaps, allowing business leaders and managers to increase efficiency and focus of their decision making (Selman et al. 2008; Yip 2004).

### 2.3.2 The Blue Ocean Strategy

INSEAD professors W. Chan Kim and Renée Mauborgne however held the view, that the strategic management tools during that time were not efficient enough to show the exact drivers within a business, and thus limiting the scope to enhance the underlying strategy. Tools such as the Ansoff's Product/Growth Matrix (Dehkordi et al, 2012; Nicolas, 2011) or Porter's Generic Strategies – particularly the differentiation strategy (Kabukin 2014) were deemed to descriptive in nature.

---

[5] See also: https://www.n26.com/

As such, the professors developed their own strategy to fill the gap with the goal in mind to provide hands-on methods to improve product & service specific strategies. Their research was initially published in the Harvard Business Manager (Kim & Mauborgne 2015), and due to the great success of the publication subsequently lead to a standalone volume the following year, titled "The Blue Ocean Strategy" (2005); with revisions and alterations incorporated into the most recent 2015 release (Kim & Maubourgne 2005; Kim & Mauborgne 2004).

The term "Blue Ocean" itself hereby refers to unexplored markets / market segments previously untouched by competitors, whereas "Red Oceans" are categorized as markets with multiple players, ongoing price wars and oftentimes predatory competition (Kim & Maubourgne 2005; Kim & Mauborgne 2004; Dehkordi et al. 2012). Table 1 below, based on the 2004 publication, shows the key descriptive points of both "Oceans":

**Table 1: Red vs. Blue Oceans**

| Red Ocean | Blue Ocean |
|---|---|
| Compete in existing market space | Create uncontested market space |
| Beat the competition | Make competition irrelevant |
| Focus on existing customers | Focus on non-customers |
| Exploit existing demand | Create and capture new demand |
| Make a value-cost tradeoff (i.e. sell high value at a higher price and lower value at a lower price) | Break the value-cost tradeoff (offer best value to customers due to no cost constraints) |
| Align the strategy on differentiation or low cost to remain competitive | Allow to differentiate from competitors while being cost-cautions |
| **Defend current position** | **Create new proposition** |

*Source: Adapted from "The Blue Ocean Strategy" (Kim & Mauborgne, 2004)*

### 2.3.2.1    The Company View

At the core of the BOS lies the so-called 'business strategy canvas', being the central element to visualize all factors pertaining to the product / service under review. By plotting the factors influencing the use/need of the product on a graph together with one (or more) competitor products, it allows even inexperienced audiences to clearly see the differentiation. The x-axis is hereby used to indicate high and low points, whereas the product-specific factors are plotted on the y-axis. The resulting overview enables the business strategist to focus on individual elements representing the overall scope of the product/service for modifications; figure 1 gives an example from the US wine industry that a new company "Yellow Tail" wanted to enter (Kim & Maubourgne 2005; Kim & Mauborgne 2004).

After visualizing the factors in this manner, the authors propose in a second step to review each factor individually in a so-called "four-actions-framework" in an attempt to streamline / realign the product or service:

> **First action:** Reduce – Which factor(s) should be reduced below the industry standard?
>
> **Second action:** Eliminate – Which factor(s) should be disregarded altogether?
>
> **Third action:** Raise – Which factor(s) should be increased above the industry standard?
>
> **Fourth action:** Create – Which additional factor(s) should be created that do not currently exist?

**Figure 1: The Strategy Canvas of the US wine industry**

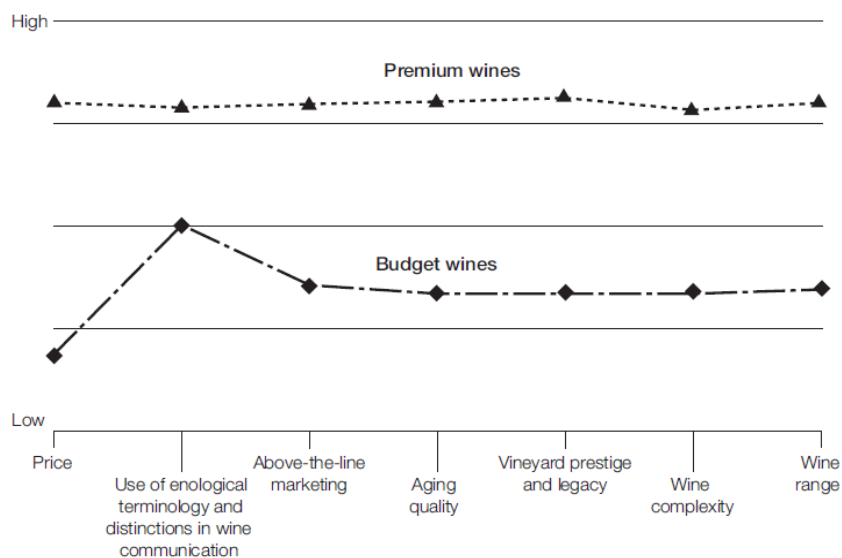

*Source: The Blue Ocean Strategy (Kim & Mauborgne, 2015)*

The rationale behind the four actions framework is simple:

Any factors identified not to provide significant benefit to the customer while at the same time incurring cost are to be eliminated in order to be able to provide the same or at least a similar value to the customer at a lower price; or to divert the cost savings into any measures increasing / generating tangible customer benefit. Along the same lines, elements providing some value to customers, but being of little importance or overutilized by competitors shall be reduced to a bare minimum – customers might expect the functionality to be present, but cost can easily be diverted (Nicolas 2011; Kabukin 2014).

Contrary to factors being reduced / eliminated, elements that can be raised to a provide a new positioning from other market players are those that are either not currently deemed of strategic importance but could become a major driver towards product diversification. Finally, adding previously nonexistent element to the product mix will play a key role in the differentiation from competing products and services. As such, those areas are likely to attract the highest investment / development cost, and at the same time bear the highest risk of failure (Burke et al. 2010; Kabukin 2014; Ayub et al. 2013).

Back to the example of the Yellow Tail wine, based on the seven initially identified indicators, the market entrant followed the four actions framework by completely eliminating wine-specific terminology to appeal to laymen, reduced marketing to the barest minimum, and disregarded any focus of the aging quality. At the same time – and in line with the focus on laymen – prestige, complexity and the product range were underweighted; while entirely new factors, namely the ease of drinking, ease of selection and a 'fun' component (expressed for example through the visual design of the bottles) were created.

Last but not least, Yellow Tail set the price at a lower mid-tier segment, out of competition with high-end products, but clearly above the budget varieties. The strategy proved to be highly effective, and enabled Yellow Tail to establish a large market share, bringing it into the Top 5 wine sellers of the USA (Statista 2015).

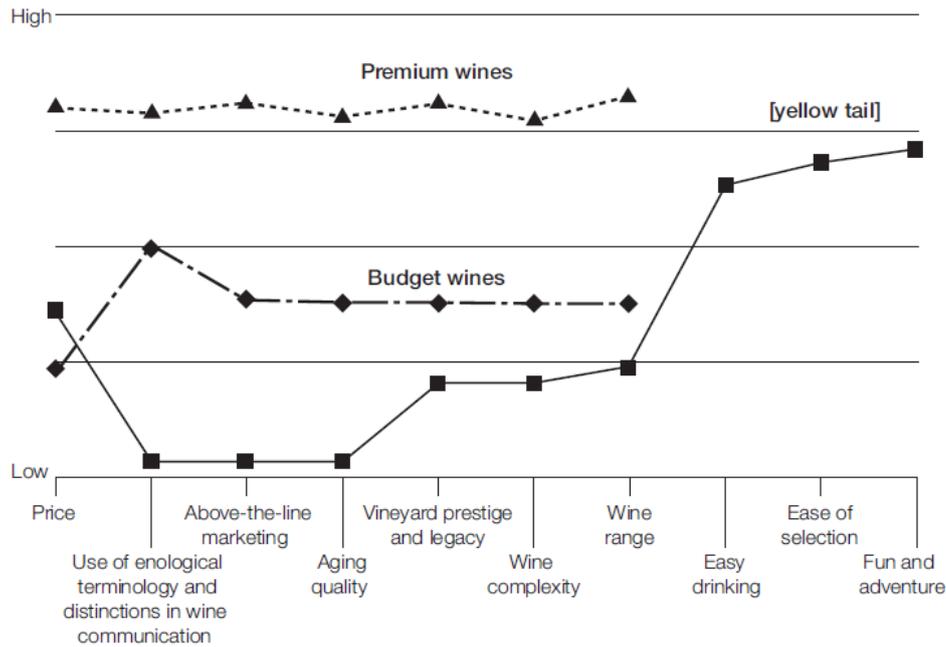

*Source: The Blue Ocean Strategy (Kim & Mauborgne, 2015)*

Implementation of the Blue Ocean Strategy however is detached from any particular industry, as a statistical analysis shows in the following case. Figure 3 displays an analysis conducted by Kim & Mauborgne (2015) amongst 108 business launches of already existing companies, which clearly underlines this trend – 86% of the launches were conducted in red ocean markets, added 62% to the revenue stream of the parent company, but only generated 39% in profits, whereas the opposite is true for blue oceans.

Figure 3: Profitability and growth of blue vs. red ocean companies

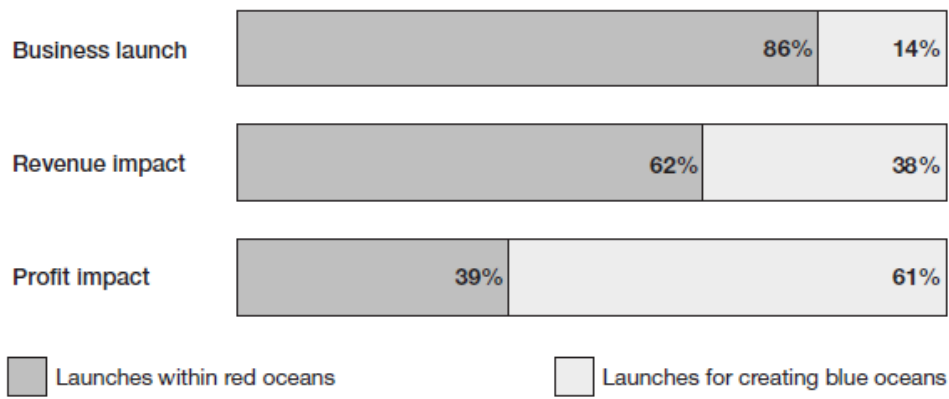

*Source: The Blue Ocean Strategy (Kim & Mauborgne, 2015)*

In addition, the company offering the new product as its solemn feature / USP will dedicate all available resources to improve it, thus further reducing cost and allowing for even larger margins (Mirrahimi 2013). Through modification and extension of the existing product, the company would effectively offer a new product or service that has not previously existed (so called "value innovation"). Customers interested in the proposed new value have little to no alternatives to choose from and hence, if satisfied with the product, have to pay the set price – as such it is of little surprise that the profitability of blue ocean companies is way higher, whereas the market entry is cheaper (Yang & Yang 2011).

Companies following the blue ocean strategy will however need to "continually monitor the needs of its target customers and competing value propositions" in order to remain ahead on the value curve, as new competitors will try to enter the market once they see avenues for revenue generation therein (Sheehan & Bruni-Bossio 2015). As such, companies / investors should also monitor their portfolio for products and services based on their current standing, as well as evaluate their future potential after applying the BOS in order to make strategic decisions and investments worthwhile.

Kim & Mauborgne further developed the so-called Pioneer-Migrator-Settler Map (or short, PMS-Map, see figure 4 below) which aims at visualizing the revenue contribution of various products (or from an investor's perspective, individual companies) to the overall income generation. At the same time it allows to indicate whether the portfolio is currently deeply set in a red ocean, or already well positioned for future growth (Kim & Mauborgne 2004; Kabukin 2014). In this concept, pioneers are products / companies with a large growth potential, and thereby offering the greatest probability of establishing a large market share devoid of competition; and thus fully in line with the BOS approach. Those pioneers are likely to become of crucial importance for the company, and should be monitored closely in order not to lose the future positioning.

Migrators on the other hand are currently generating revenue without having any distinctive USP or specific feature; but have not yet been fully engulfed in an overly competitive price war. With strategic investments and amendments, Migrators can be transformed into blue oceans. Ultimately, settlers are deeply set in red oceans, and face all the disadvantages of a saturated market (Kabukin 2014). While still potentially contributing to the revenue generation, profitability will be low, and investments to move settlers into even migrator stage are oftentimes enormous. For the overall course of the company it means having too many settlers and little migrators or even pioneers, a decline in future potential is highly likely; and strategic investments into new business areas are required to not lose the market positions to more agile competitors.

**Testing the growth potential of a portfolio of businesses**

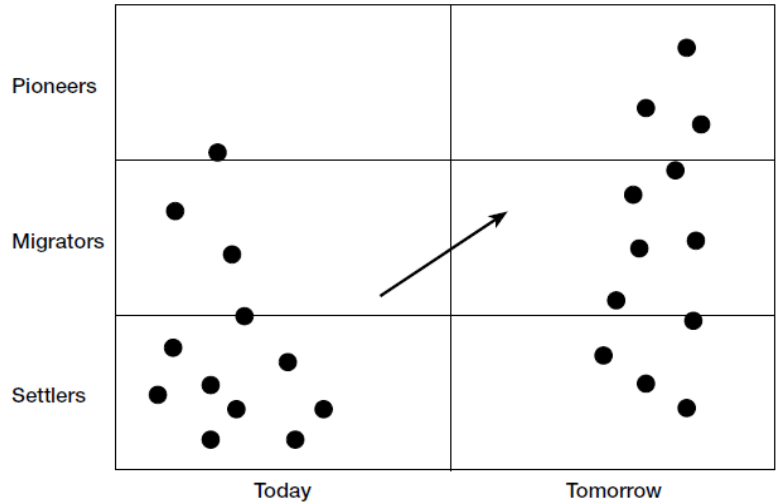

*Source: The Blue Ocean Strategy (Kim & Mauborgne, 2015)*

### 2.3.2.2    The Customer View

Following the company- and product / service centric viewpoint, Kim and Mauborgne further developed a concept to classify potential future customers – provocatively, they are referring to those as "non-customers", see figure 5 (Kim & Mauborgne 2004). The first category, "soon-to-be non-customers" are currently using an existing product, but are actively looking for added value, lower cost and other opportunities to diversify. Since they are actively aware of the services offered and engaged in monitoring the market development; concrete product development means can help retaining this group. At the same time, active engagement of the group can lead to valuable input for product diversification and blue ocean creation.

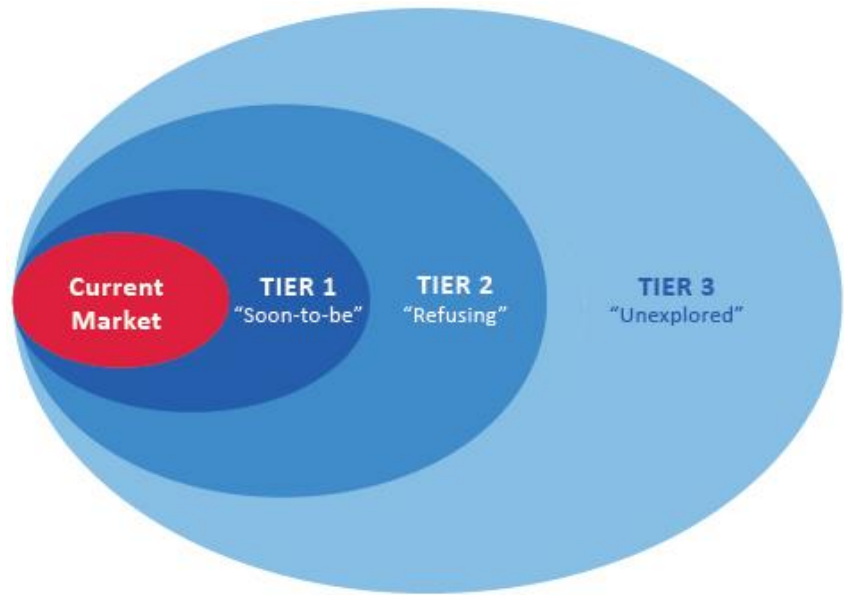

*Source: https://www.blueoceanstrategy.com/ (2014)*

Refusing non-customers have actively decided to not engage in the current market the products are being offered in, but are aware of its existence. Refusing non-customers can potentially be migrated to become active clients, if sufficient incentives through product modification / market diversification are made.

Unexplored non-customers finally are active in markets that are comparably far removed from the current market, and may need significant investments and diversification to engage.

## 2.4 Part II – Overview of the Research Objectives

Following the initial section of the review, the second main question to be addressed is how and which aspect(s) of the Blue Ocean Strategy is being applied in the Consumer Finance Sector in Germany, and what roles are being played by FinTechs? At a first glance, it appears as if traditional banks with their tens of thousands of employees and billions in assets represent red oceans and settlers (or migrators at best); while FinTechs appear to be agile, technology-backed pioneers aiming to create uncontested blue ocean markets. The following review of both concepts aims to provide a first overview.

### 2.4.1 The German Consumer Finance Market

Traditionally / historically, the German banking sector is divided into three pillars (Grill & Perczynski 2011; Lang et al. 2000):

Savings banks, which are majorly (if not wholly) owned by their respective municipalities, and aimed at providing savings & loans to the respective local population and industry; whereas 'local' is loosely defined as any person or enterprise within the municipality. As the need for larger clients to access advanced financing solutions and/or larger loan facilities arose, communal state banks within the respective national states have been established; those exclusively serve corporate clients within their respective state.

Cooperative banks, established as a private alternative to the municipally owned savings banks generally follow the same business principal, with the main difference that they are wholly owned by their respective members. A member is any person and/or enterprise that wishes to engage with the cooperative; membership is granted by purchasing shares (usually a single share is sufficient), which is a requirement prior to interacting with the bank. Cooperatives also have a relatively limited geographical focus, however usually enclosing a larger geographical area as they are overall fewer in number.

Commercial banks are any banks not fitting in the aforementioned categories, excluding special purpose institutions such as development banks or the central bank, which are not discussed as part of this dissertation. While commercial banks aim majorly at business clients operating in a supra-regional or international scope, they do serve private customers all over the country (and often beyond) as well. Online-only banks with no physical branches typically fall into the same category.

This structure has existed since the foundation of the Federal Republic of Germany and is analyzed in some more detail below (Lang et al. 2000). To visualize the overall development of financial institutes per category as well as the aggregated total, table 2 provides an overview of the respective progression.

**Table 2: Development of banks by type over time**

| Item | Type of bank | 1990 | 2016 | Development (abs.) | Development (rel.) |
|------|--------------|------|------|--------------------|--------------------|
| 1 | Savings banks | 769 | *396** | - 373 | - 48,50% |
| 2 | Cooperatives | 3.380 | 972 | - 2.408 | - 71,24% |
| 3 | Commercial banks | 338 | 288 | - 50 | - 14,79% |
| *Σ* | *All banks* | *4.487* | *1.656* | *- 2.831* | *- 63,09%* |

*Sources: Adapted from (1) DSGV Publication, 2017; (2) Statista Report, 2016; (3) Bethge, 2016*
*) January 2017 figure

With a total decline of over 63% or more than 2.800 banks since 1990, the overall trend in the German banking system is quite obvious (Berlemann et al. 2014; Straßberger & Sysoyeva 2016). Particularly cooperatives and savings banks have been affected due to their relative small size both in area coverage as well as in terms of market capitalization (measured in total assets under management / balance sheet total). With exception of few commercial banks that have been closed down due to business restructuring or bankruptcy of their parent companies, most notably during the financial crisis from 2007-2008 with subsequent ripple effects over the next years, banks disappeared from the markets mainly due to mergers / fusions (Berlemann et al. 2014; Lang et al. 2000). Since both cooperatives and savings banks are bound by their statutes to remain independent from influence outside of their respective networks, all of those mergers and fusions were carried out as horizontal integrations, i.e. within their existing ecosystems, thus creating supra-regional savings banks such as Sparkasse Köln-Bonn (merging the neighboring districts of Cologne and Bonn) or Volksbank Main-Taunus-Kreis (joining the two neighboring administrative districts in the greater Frankfurt area).

The overall necessity for consolidations in the financial sector can be narrowed down to two aspects – growing competition in the already congested markets, as well as a changing consumer behavior that banks are unable to address with uniform solutions other than cost reductions on their part; both of which are strong indicators of "red ocean" behavior in itself (Kim & Mauborgne 2015; Yang & Yang 2011).

Looking at the overall profitability situation from traditional consumer products (loans, current accounts, savings accounts and brokerage accounts) underlines this view further: Traditionally, Germans are avid savers and rather risk averse, keeping larger amounts of cash in savings accounts and term deposits while avoiding stock investments and other volatile investments (Mersch 2016; Bethge 2016; Benchimol 2014). Consequently, a major source of income for the banks used to be the so called term transformation –cheap deposits with maturities under one year and cost of borrowings (interest payable to clients) of 1.2-2.5% averagely (Statista - National Statistics Office 2016) were used to fund loans and mortgages with a maturity well in excess of 10 years and interest payable in the 7.57-10.54% range (FMH-Zinsberatung e.K. 2017; Deutsche Bundesbank 2010); see figure 6 below.

**Figure 6: Loan interest development in Germany up to the financial crisis 2008/09**

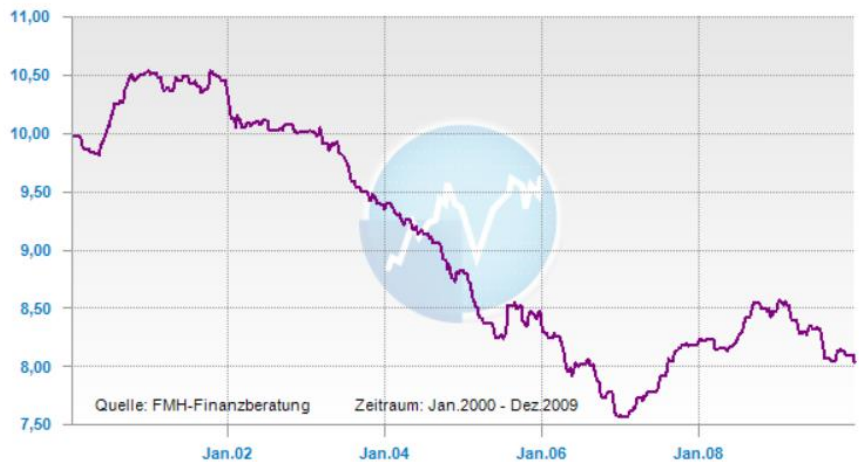

*Source: https://index.fmh.de/fmh-index/zinsentwicklung/detailversion/*

Since the deposit base never fluctuated much, there was little for the banks to do in terms of active liquidity management, and since loan defaults in Germany have traditionally been equally low (below 2.6% p.a. averagely over the past 15 years preceding the financial crisis (IMF 2011; FinTech Pioneers 2017), the banks were able to earn nicely on the interest spread between deposits to loans. As a direct result, interest income through term transformation before the financial crisis accounted for close to 70% of the income allocation of banks in Germany; while other fees and commissions together made up for the remaining ~30%.

However, following the financial crisis with the aforementioned low-interest period peaking in an imposed negative interest for overnight deposits at the European Central Bank triggered multiple events:

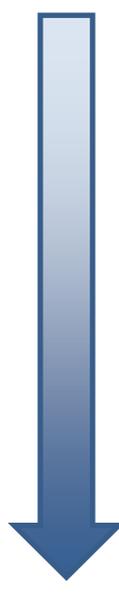

1.) Banks were penalized for keeping excess liquidity on central banks accounts;

2.) Banks could borrow money virtually for free from the central banks;

3.) Potential borrowers started requesting for cheaper interest rates on loans due to the 'free money';

4.) Banks were no longer interested in paying interest on customer deposits when they could refinance the loans for free and hence slashed interest on customer deposits;

5.) Revenue made from term transformation dropped drastically;

6.) Customers became increasingly unhappy due to the marginal returns on their deposits – taking the inflation rate of 0.9-2.3% into account, consumers were actively losing money by keeping it in the deposit accounts (Hau & Lai 2016; Arnold & Lemmen 2008);

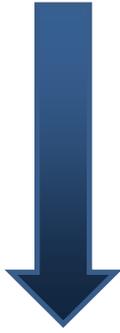

7.) Customers started looking for alternative solutions, and when finding those, often terminated their whole bank relationship – leaving the banks without the second leg of income (fees & commissions) as well, leaving yet another gap in their income generation.

At the same time, fiscal transparency initiatives such as MiFID (Markets in Financial Instruments Directive), a directive forcing banks to disclose all fees, commissions and kickbacks (incentives paid by distributors such as insurance companies or brokers) enabled consumers to easily compare prices and identify favorable alternatives to their house banks, creating yet another layer of cost sensitivity sparking a red-ocean price war amongst institutions (Kim & Mauborgne 2004; Karem El-Bastaweisy 2007):

And they are no longer alone in the market, as other players – FinTechs, young, agile, technology-backed companies – are increasingly trying to pushing into the gap left by the emancipating customers by offering cheap, scalable and international solutions (Kang et al. 2016; Freeman 1991; FinTech Pioneers 2017).

### 2.4.2 The FinTech Industry

Financial Technology Companies, or FinTechs in short, are a rather recent development in the financial service sector, coming up as one of the latest trends of the start-up scene.

Those companies aim to provide services from within the scope of traditional banks, while adding a new component / layer to it or fundamentally enhancing the existing scope. Looking at the business model of banks, FinTechs have four core functions to attach themselves to – payment solutions, savings, loans and investment / brokerage (Li et al. 2017; FinTech Pioneers 2017). Examples for companies offering the aforementioned enhancements in the German market are for example payment solutions such as PayPal or TransferWise, peer-to-peer lending providers such as Lendico or AuxMoney, or deposit pooling services such as Weltsparen.de (Mussler 2017; Li et al. 2017; Dorbian 2016). A common component of all FinTechs is that they do not operate under banking licenses themselves, but rather 'piggyback' on banks and/or credit card operators to utilize their services[6]; leading to reduced overhead cost for e.g. capital adequacy, KYC, risk etc. (Li et al. 2017; Freeman 1991)

The general model of FinTech companies hereby consists of picking one particular bank service that is unanimous amongst banks and most – if not all – jurisdictions and regulations, very consistent and simple in terms of product design, workflows and routines, as well as traditionally expensive, so that cost savings for potential customers can be realized. Consequently, those products and services can be near fully automated, limiting human interaction (=cost) to a bare minimum (Kang et al. 2016). It is therefore not surprising that traditional banks, where >50% of the workforce are working in retail sales branches and another 20% in backoffices managing the operations, are losing customers in big numbers to the competitive advantages of the FinTechs which have way slimmer setups – staff is highly concentrated in product development / programming and online marketing, and can sustain with limited support functions, thus reducing the cost base further (Bethge 2016; Mersch 2016).

In their 2016 "FinTech 100" report, the advisory & consulting company KPMG details the distribution of most influential FinTechs for the year 2015 as shown in figure 7 below; with lending and payment solution providers taking a combined 50% of the overall share worldwide (H2Ventures & KPMG 2016). Data & analytics as well as regtech companies are solutions offered on a B2B basis only and are hence excluded from the scope of this paper; the same applies for insurances as regulation in Germany allows banks merely to sell insurances based on commissions.

---

[6] One notable exception is PayPal Europe Sárl, which does operate as a licensed financial institution established in Luxembourg, with activities in the EU. Globally however they are a licensed payment provider and not a bank.

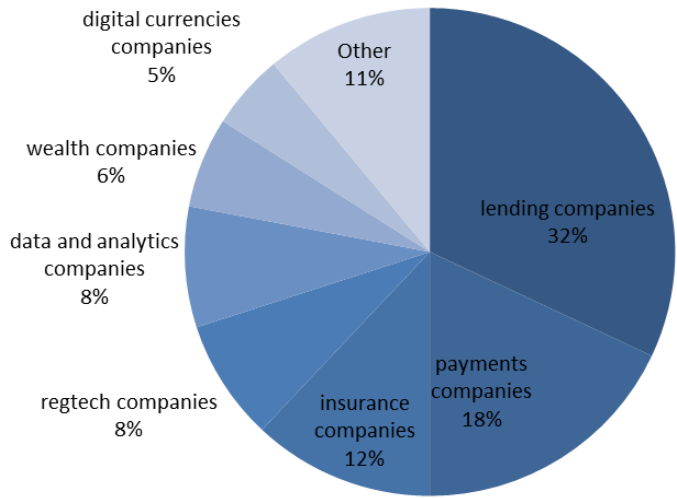

*Source: The 2015 FinTech 100 (KPMG, 2016)*

Another study conducted by research and consulting firm Accenture suggest that the way of customer interactions with their banks has changed dramatically with the emergence of smartphones and mobile channels in general, making people much more likely to consider digital alternatives to traditional brick and mortar banks. With a total of 15 monthly interactions customers have in Germany (see figure 8 below), only one is averagely conducted in the actual branch, following a worldwide trend (Gera et al. 2015; Accenture 2015; Freeman 1991). Mobile interactions are yet behind the international standard, whereas internet banking utilization is on par with the global average for developed economies. At the same time however the overall acceptance of mobile solutions for all types of tasks is increasing daily, leading to the assumption that it is only a matter of time until FinTechs reach the tipping point to utilize the increasing unhappiness of traditional banking customers and increase their market share.

**Figure 8: Number and channels of interactions with banks**

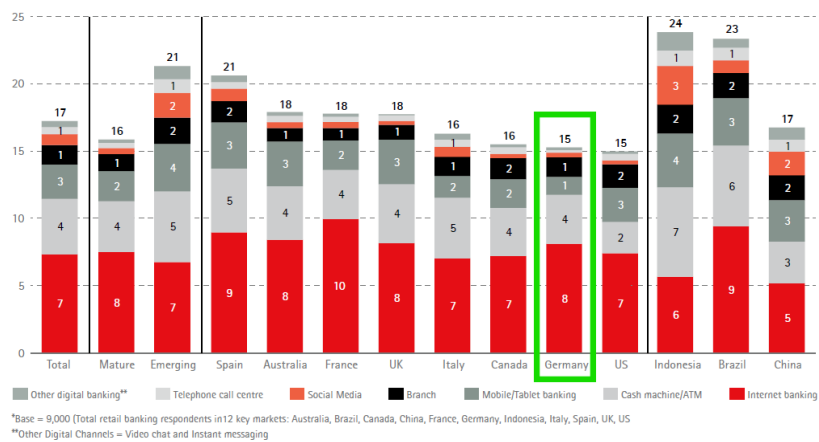

*Source: Retail Banking Pulse (Accenture, 2016)*

## 2.5    Case Studies – Mapping Theory and Practice

Having introduced the Blue Ocean Strategy, the traditional banking sector and FinTechs, this chapter aims to map the theoretic concept of the BOS with the banking sector to determine whether it represents a "red ocean" as already hinted in the previous chapter. Furthermore, extensions / adaptations of their business model through FinTechs are discussed with emphasis of typical "blue ocean" criteria.

Given that there are still ca. 1,650 banks in Germany (see 2.4.1), one representative of each group will be assessed to determine the current market average. For ease of referencing, the respective largest in terms of assets under management of each group (commercial, savings, communal banks) are chosen: Deutsche Bank (Bethge 2016), Hamburger Sparkasse (DSGV 2016) and Berliner Volksbank[7] (BVR 2017).

Reviewing their respective pricelists reveals the following services offered for private individuals:

Table 3: Product types for private consumers

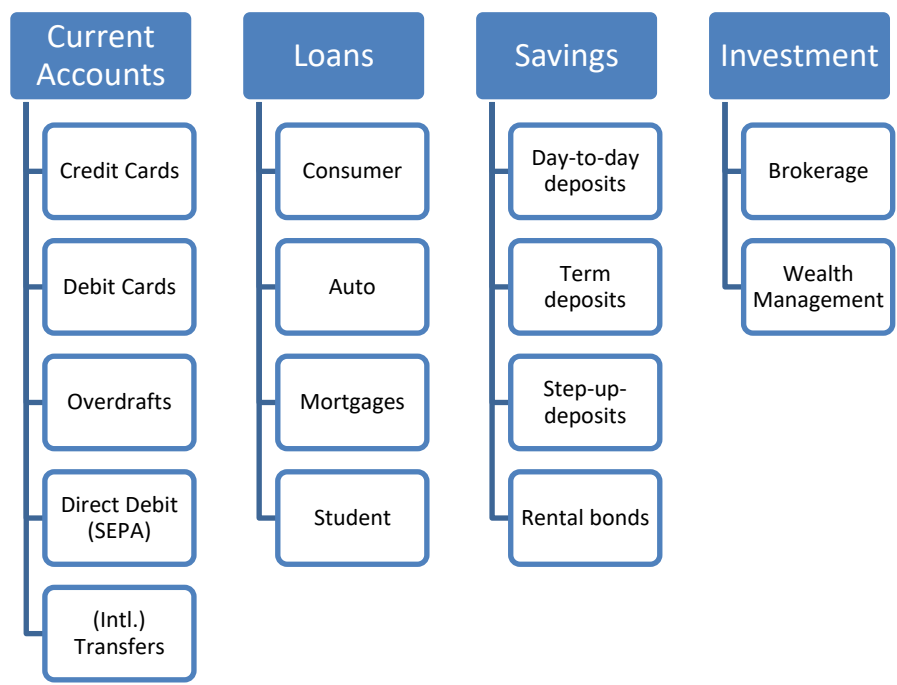

*Source: Pricelists of Deutsche Bank (2017), Hamburger Sparkasse (2017), Berliner Volksbank (2017)*

---

[7] Deutsche Ärzte und Apothekerbank (German Doctors and Pharmacists Bank) is the largest communal bank, but has been excluded from this review since services are rendered exclusively to healthcare professionals.

Mapping the target banks on the strategy canvas provided by Kim & Mauborgne shows the following picture:

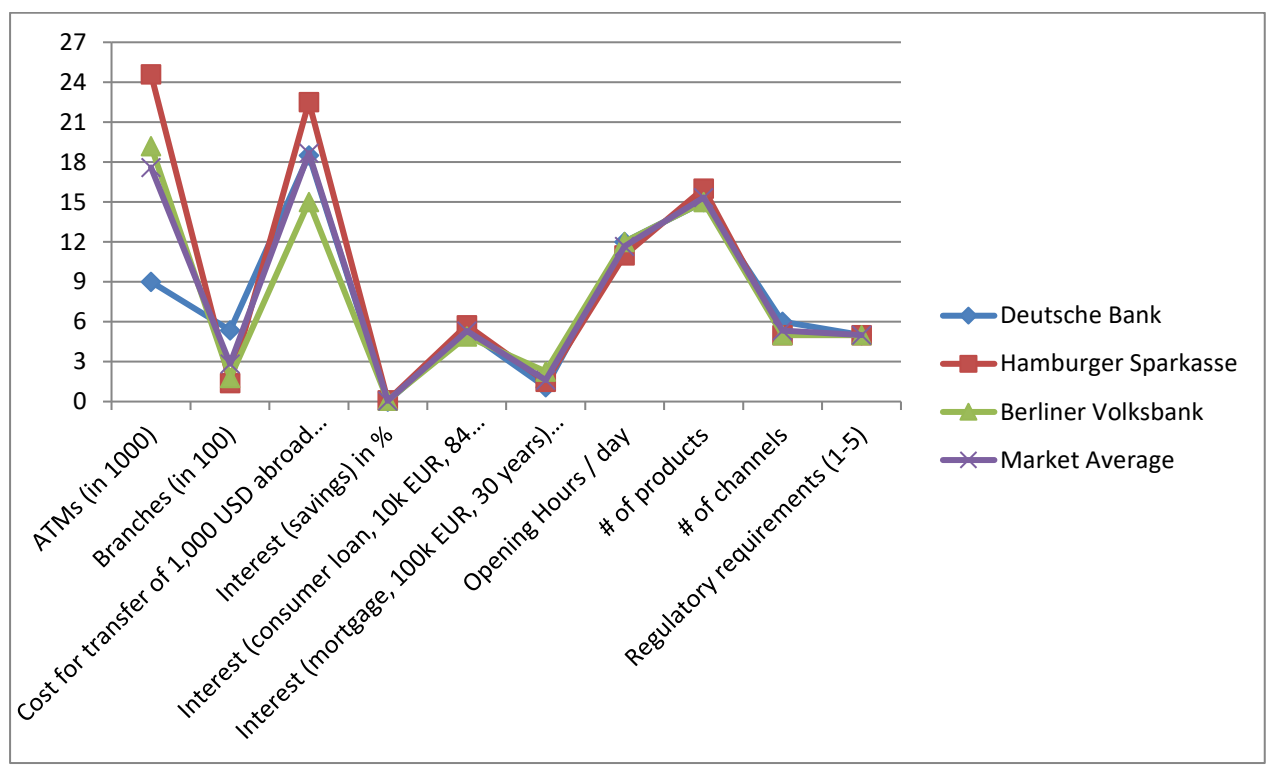

*Source: Based on term sheets & price lists of the respective banks*

The biggest variances are the number of ATMs and the number of branches respectively; where products are concerned, only international transfers stand out with a maximum variance of 7.50 EUR per transfer from the cheapest offer of 15 EUR (Berliner Volksbank) to the most expensive at 22.50 EUR (Deutsche Bank) – interest on savings, consumer loans and mortgages are nearly identical, as are the number of products, available channels and regulatory requirements. In fact, given how minimal the difference in interest rates is reflects both the market situation (low interest environment) as well as the pressure for margins to compete; both clear indicators for a hard-pressed market and thus textbook "red ocean" scenarios (Sheehan & Bruni-Bossio 2015; Kim & Mauborgne 2005; Kim & Mauborgne 2004).

The payment provider TransferWise should now serve as a proxy for the FinTech industry: Their business model is quite simple – using direct debit from current accounts (SEPA mandates) or credit card charges, TransferWise debits a client who wants to transfer money abroad, pools the amount in their own account together with all other customers from the originating country, and does the same in the receiving country. After a fixed time the transactions are netted, and the counterparts receiving the transaction amount are credited through the TransferWise collective account in the respective country; thus bypassing the international transfer altogether except for the surplus. The banks – previously the main benefactors of international transfers – are left with the marginal transaction charges on the cards or accounts, whereas the customers achieve the same goal (=transfer money abroad) in shorter time since ultimately only local transfers are executed (=benefit 1) at lower cost (=benefit 2). Furthermore, customers are now also able to request payments via email that can be replied to at the click of a button – a functionality no German bank has thus far offered.

In a Reduce-Eliminate-Raise-Create view, TransferWise's approach looks like this:

| Reduce | Eliminate |
|---|---|
| Channels | ATMs |
| Products | Branches |
| Cost | |
| Regulatory Requirements | |
| **Raise** | **Create** |
| Opening hours (=24/7) | Money request function |

And mapped onto a strategy canvas against the market average from figure 9, the resulting figure 10 below nicely visualizes the competitive advantage of the FinTech's business model:

Figure 10: Strategy Canvas of TransferWise

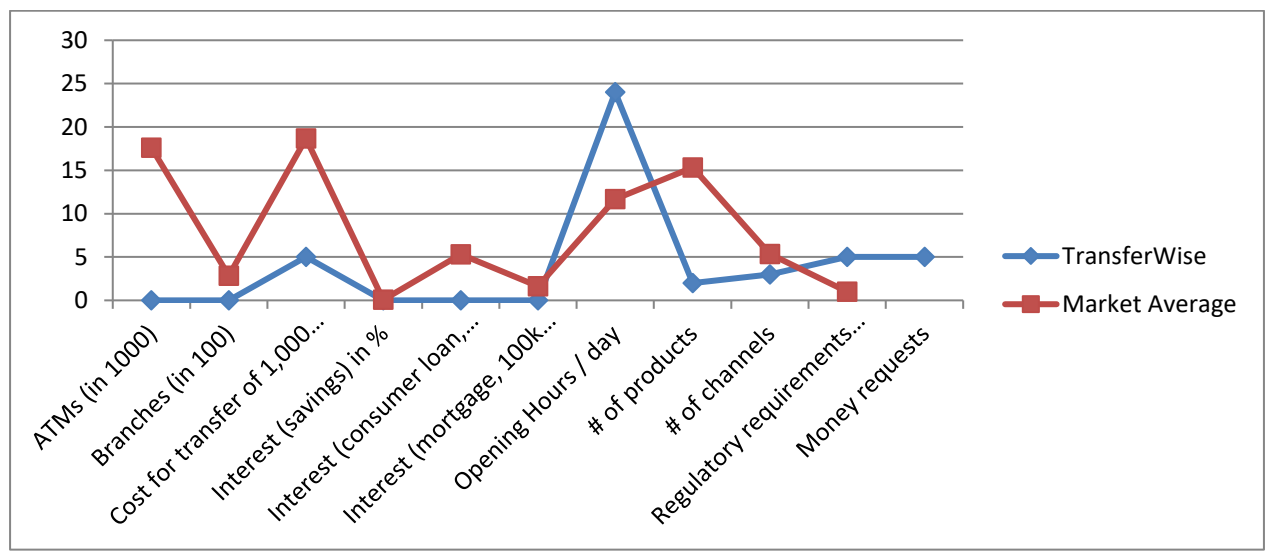

*Source: Based on price list of TransferWise (2017)*

With TransferWise being the first payment processor operating with the current business model, it was effectively established as de-facto market 'owner' and, in line with the blue ocean strategy, was able to occupy a previously uncontested market space (Kim & Mauborgne 2004; Kim & Maubourgne 2005; Dehkordi et al. 2012).

## 2.6    Summary of Review Findings

Having reviewed the existing literature and mapped the Blue Ocean theory onto the consumer finance sector in Germany by using the example of TransferWise, it has been shown that the traditional consumer finance sector is indeed characterized by an abundance of homogenous products with minimal variation in the interest- and fee structure; a very high level of competition as displayed by the number of financial institutions being forced into mergers or exiting altogether (see table 2 in chapter 2.4.1); as well as a constantly reducing profitability. Consequently, banks active in the consumer finance sector are caught up in a battle to reduce cost in order to stay competitive with one another, which ultimately slows innovation, which in turn drives customers away as they are becoming more and more demanding not only in terms of conditions, but also when it comes to the use of modern technologies (Bos & Kool 2006; Gera et al. 2015). All of those factors lead to the conclusion that the German consumer finance sector is indeed a verifiable case of a "red ocean" (Kim & Mauborgne 2015).

FinTechs on the other hand side aim to use their single-product focus and the agility of small teams and fast technology to leverage on the inertness of the banks in order to take a share of their market. Using innovative strategies aiming beyond the traditional outreach of banks, they are in a position to poach customers of financial institutions, leaving those with clients who are using the bank merely as a back-end through the naked current account, while other, more profitable transactions are carried out through the respective FinTech's ecosystem (Kang et al. 2016; Gera et al. 2015; Accenture 2015). All of those indicators speak for the fact that FinTechs can indeed be seen as "blue oceans" as per the underlying theory (Kim & Mauborgne 2015).

Nevertheless, as also outlined in the 'customer view' discussion in chapter 2.3.2.2, focusing on a clientele driven by the aim for low (transactional) cost as well as using a single-product strategy bears the risk of other market entrants copying the FinTech's business model by undercutting their price; creating yet another level of competition (see also chapter 2.3.2.1). Consequently, further research shall be conducted to identify how many customers are really willing to switch from banks to FinTechs; how many / what volume of transactions the banks are losing; if cost-awareness is really the driving force behind customers moving away from banks or if other factors play a deciding / contributing role; and which factors keep people away from FinTechs despite their knowledge about potential alternatives to the products offered by banks. This will be coupled with geographic and demographic information to identify categories of customers from least to most willing to adapt to new technologies (Pioneers, Migrators, Settlers). To allow for an indication whether the FinTechs have a chance to remain uncontested in their respective market – their "blue ocean" – the data will be collected using over 40 named solution providers.

# 3. Research Methodology

Research by itself is understood as the process of discovering new facts to increase knowledge and gain understanding of / develop new concepts and processes (Cooper & Schindler 2003; Business Dictionary 2017). By definition, research is carried out in a controlled environment in order to verify hypotheses with factually established data gathered by set standards (Baguley 2004; Johnson et al. 2007).

At the core of this dissertation lies the analysis of German residents' knowledge about FinTechs, their understanding of the product offers, the likelihood to make use thereof, as well as the potential market share the FinTechs can appropriate from traditional consumer banks. In conjunction with demographic information, a quantitative data assessment would provide input for a correlation analysis with profitability figures and market share discussions publicly disclosed by the banking industry through their governing bodies. A secondary layer of information from individual representatives of the customer groups – settlers, migrators and pioneers as discussed in chapters 1.1, 2.4 and 2.6 – will help to broaden the understanding of certain aspects related to technology adaptation. As discussed by Brewer et al (1999) and Ridenour & Newman (2008), there is no superior method of gathering information, it is rather the combination of factors leading to a holistic understanding of the research topic.

## 3.1. Research Questions & Hypotheses

By following the aforementioned mixed approach of quantitative with qualitative information, confirmation to the following hypotheses will be sought:

(1) The overall knowledge about FinTechs is high;

(2) The overall acceptance / use of FinTechs is – in line with the statement made in chapter 2.4.1. of Germans being rather risk-averse – still rather low;

(3) The primary driver for people using FinTechs are cost savings;

(4) People living in urban areas have a higher probability of using FinTechs than those in periurban or rural settings;

(5) There is a direct correlation between age and the likelihood to use FinTechs;

(6) People with a high satisfaction ratio over time are more likely to keep using FinTechs instead of banks;

(7) Despite the abundance of FinTechs in the market with seemingly unique business models, only a small fraction thereof gains significant users to be true competitors for banks

By way of confirming / refuting those hypotheses, the research questions to be answered are:

(1) What is the percentage share of the population in each of the three categories pioneers, migrators or settlers?

(2) Which volume of transactions / investments is potentially being processed by FinTechs now?

(3) How will the potential impact of the lost transactions reflect in the market share of banks?

(4) Are FinTechs likely to become / remain profitable using the acquired market share and remain blue oceans?

## 3.2. Research Theory & Background

The practical research part of the dissertation starts with the quantitative data collection and will conclude with a number of clarifying interviews with members of each proposed target group, as briefly indicated in the introduction to this chapter (3). The rationale behind this approach is that an accumulation of quantitative data will already provide strong indicators for the overall trend of the research, as 439 valid responses to the questionnaire from all over Germany (94 different cities, with peaks as expected in metropolitan areas of Berlin, Hamburg, Munich, Frankfurt, Stuttgart, Cologne-Bonn and Essen) have been gathered. Interviews conducted with a number of the participants of the survey therefore serve the function of deepening the understanding of the motives behind their choice of answers given. At the same time, the interview is used as a tool for data validation (Thomas 2003).

The overall response rate for the survey that has been running for just over a month was very high; with the highest number of replies in an individual day being March 12[th], with 68 individual replies. More details on these aspects will be provided below in the section on study design.

### 3.2.1    Data & Analytics

#### 3.2.1.1.    Primary & Secondary Data

The term 'primary data' refers to raw, unprocessed information, collected by the person conducting the research. This can be done using a variety of means, for example through field research using quantitative or qualitative gathering techniques such as interviews, surveys, questionnaires, focus group research or case studies as well as document review.

An essential component for primary data evaluation is to gather suitable information using a standardized approach that remains essentially identical to ensure unbiased information, high data integrity and quality for further processing in the analytical stage. Essentially, the quantitative methods are structured in a way that each participant is given the exact same conditions under which to answer the questions; meaning the same order, scales, wording, form etc. are used (Amick et al. 2015; McNeill & Chapman 2005).

Secondary data on the other hand are information collected by a third party. Those information are usually made available in an accumulated / aggregated form; and should ideally come from peer reviewed or otherwise audited sources to ensure a high degree of reliability (Amick et al. 2015; McNeill & Chapman 2005). Usually, secondary data has been collected and processed for research or statistical purposes. Typical sources for secondary data are scientific journal publications, government databases, or company data published following strict guidelines and/or regulatory oversight, such as annual financial statements for companies listed in stock exchange prime standards. Secondary data are commonly not verified by the user, but processed as-is; hence the importance of having a reliable source (ibid.).

### 3.2.1.2.  Quantitative & Qualitative Data

In addition to primary and secondary data sources, it is of high importance to distinguish between quantitative and qualitative data, their respective advantages and limitations and particular usefulness (Thomas 2003; McNeill & Chapman 2005).

Quantitative data, as the name implies, are accumulations of information numerical form, described as precisely as possible in order to mathematically / statistically process relationships between different data points. The idea behind quantitative data is to assess a problem by using a sample representing the total population large enough to account for variations within the subject under review, and allow for accurate conclusions of the behavior of the total population within a certain error margin (tolerance) (McNeill & Chapman 2005). Typical tools for the assessment of quantitative information are Microsoft Excel, the statistical programming language R or IBM's SPSS Statistical Software (Muijs 2005). Both Excel and SPSS will be used for the analysis of the data gathered for this thesis in chapter 4.

The below table provides an overview of the common advantages and disadvantages of a quantitative analysis:

Table 4: Advantages & Disadvantages of Quantitative Analysis

| Quantitative Methods | |
|---|---|
| **Advantages** | **Disadvantages** |
| Quantifiable results | Inflexible approach due to the high degree of standardization, reaction / clarification impossible |
| Allows calculation of statistical correlation through high level of objectivity and standardization | Difficult to assess information beyond numerical / statistical data due to limited number of responses provided. This can be eased through open questions (free text responses); which however limits the methods to freely compute data models, as free text requires individual interpretation. |
| Allows for representative information gathering and high data validity through evaluation of large sample size | Survey participants have only limited possibilities to provide suggestions / input on how to make the research subject more relevant / improve usefulness. |
| Comparably low cost / effort and time consumption to gather sufficient data | |

*Source: Adapted from Lienert & Raatz 1998; McNeill & Chapman 2005*

As opposed to quantitative methods, qualitative information gathering is characterized by allowing a broader focus and flexibility to adapt to variations of the subject under research. Using the example of interviews, the interviewer can react to certain responses from the interviewee in order to gain deeper insights and understanding for certain motivations and/or behavioral patterns. As such, qualitative data collection is characterized by an overarching topic and a number of core questions, which would then be used on each research subject individually to gather a maximum depth of responses. Given the comparatively low level of standardization in qualitative data gathering, this method yields best results when a relatively small but knowledgeable group is assessed, as larger audiences with limited knowledge can otherwise produce confusing / conflicting data of questionable quality and therefore usefulness (Thomas 2003).

The below table shows a summary of the common advantages found in qualitative analysis:

Table 5: Advantages & Disadvantages of Qualitative Analysis

| Qualitative Methods | |
|---|---|
| **Advantages** | **Disadvantages** |
| High flexibility due to adaptation of the method to the individual research subject allows for identification of entirely new angles / avenues for research and previously unobserved connections | Very time consuming |
| Apart from the overall topic, the focus is determined by the participants, making it more relevant for them | Review, analysis and assessment of data is rather complex |
| Since participants are not given a list of pre-selected answer possibilities, the chance of gathering unbiased information is much higher | The quality of data is dependent on the questioning techniques of the interviewer |
| Personal interaction with the participants allows for clarification should the topic be not entirely approachable | The replies to qualitative interviews are not universally applicable but only represent the opinion of the individual |
| Open questions allow for higher subjectivity of the responses | |

*Source: Adapted from Buber & Holzmüller 2009; McNeill & Chapman 2005*

### 3.2.2 Single or Mixed Approach

In order to maximize the efficiency / minimize the disadvantages of the two methods, it can be useful to combine both approaches of data collection, which is commonly referred to as a 'mixed approach'. Given that the scope of this dissertation is an extensive and comprehensive Germany-wide assessment of the market potential for FinTechs, those goals are predominantly achievable through a quantitative assessment; the fact that Germany currently has 77 cities with more than 100,000 inhabitants further prevents a qualitative assessment given the limited resources (cost, time, manpower) available for this dissertation (Bundesamt 2015). Consequently, a clear focus on a single, quantitative approach is made. Nevertheless, as part of the data validation, follow-up interviews with 'average' participants from each of the target customers (pioneers, migrators, settlers) are conducted to gain individual insights as reference points.

A broader qualitative assessment may also be considered as a potential follow-up research into the topic; this will be picked up in chapter 5 (opportunities for further research).

## 3.3. Sampling

### 3.3.1. Sample Selection

Given an estimated 50 million adults actively participating in the consumer finance sector in Germany, with a confidence level of 95% and an error margin in the 5-10% range a sample size between 96 and 384 participants would be sufficient to generate a satisfactory approximation to the market potential of the FinTechs within the country. This requirement has been overfulfilled by gathering a total of 519 answers, out of which 439 are relevant for the evaluation of the dissertation's scope (CRS Survey 2017).

Deutsche Bank, as already introduced in chapter 2.5, is the largest bank in the German market both in terms of asset size as well as number of customers. With over 27.4 million private clients in 2014 and therefore >50% of the relevant market, the bank and the development of the customer number over time will be used as a reference point for the market share development banks vs. FinTechs in chapter 4 (Statista 2014).

In order to gather data, a web-based survey was created using the open source platform "limesurvey" (Schmitz 2012). To bypass limitations of the free, centrally hosted version (number of replies, number of questions), a local copy has been installed on the author's webserver located at *http://survey.wischnew.ski*. Results have been collected until April 09, 2017 at 2 PM CEST.

Links to the survey have been published on a number of relevant forums and social media channels in German-centric groups such as LinkedIn, XING, reddit, facebook, the Robert-Kennedy-College discussion forum[8] as well as sent to private contacts of the author (friends, family, co-workers and acquaintances) via email, SMS, Skype, whatsapp, and other instant-messengers. Participants have been encouraged to distribute / forward the survey; making the exact origination of the results difficult – referral links have been tracked within limesurvey, yet links distributed directly (i.e. not via a certain forum or website but through email or a messenger) do not contain any trackable referrer.

Participants disqualified from the analysis fall into three different categories:

- Not residents in Germany (66 out of 519 total; 12.7%)
- Provided obviously wrong figures or 'fun' replies (14 out of 519; 2.7%)
- Did not fulfill minimum age requirement (1 out of 519; 0.2%)

Overall, therefore, 81 out of the 519 participants of the survey (15.6%) were excluded from the final analysis. Lacking or limited knowledge about FinTechs was not selected as a factor for removing participants, as those are beneficial to indicate which percentage of the population have actually encountered FinTechs in one way or another (penetration ratio) by implication.

### 3.3.2. Study design

#### 3.3.2.1. Ethical Considerations

In line with the Ethical Guidelines of the University of Cumbria, all participants of the survey and interviews were informed of the nature of their active participation in the survey, the desired outcomes, the survey methodology and their right to terminate the participation at any given time (Cooper & Schindler 2003). The author of the survey fully disclosed his contact information, the topic of the dissertation as well as the overseeing academic institution. All participants were assured of full confidentiality and anonymity of their identity as part of this work; unless they gave explicit permission to be contacted for follow-up interviews in a dedicated comment field at the end of the survey.

---

[8] The groups did not have a finance-related focus as not to falsify the overall market perspective

### 3.3.2.2. Questions

All survey questions have been posed in a manner as neutral as possible, and by way of communicating the survey to participants, people were encouraged to take part regardless of their background and prior knowledge on the topic to allow for a true representation of the average population sample. Given the market focus on Germany, the survey was made available in both English and German; the language choice of participants is also evaluated in chapter 4. As shown in Appendix 3, the survey contains a total number of 30 questions in four categories[9]:

- General Demographics (9 questions)
- Banking Specifics (max. 4 questions)
- FinTech Specifics (max. 14 questions)
- Self-Identification (3 questions)

As indicated above, the Banking and FinTech Sepcifics contain a *maximum* possible number of questions; this is a result of a program logic used to automatically hide irrelevant questions (i.e. participants were asked to indicate FinTech providers they are aware of; consequently they were not asked to give specific information regarding a company they do not even know exists. This technique was employed to reduce frustration and ensure a high number of completed surveys). With few exceptions, questions were asked in closed form using either a Likert Scale to grade / evaluate certain aspects under review; or given single / multiple choice answer possibilities (including free-text fields where applicable). Detailed questions to the number and volume of transactions under review (questions 22-24) were stated in a double-column free text format to indicate the number of transaction in column 1 and volume in column 2 respectively.

The self-identification is meant as a reference point for validation by giving the participants a retrospective option to grade themselves based on their input provided during the preceding survey. Given a limited choice of three options ('Early Adopter', 'Balanced' or 'Traditionalist'), a mapping to the user types 'Pioneer', 'Migrator' and 'Settler' is made possible (while the exact same wording was avoided to maintain a neutral tone). Furthermore, this review question allows to establish a baseline as to which number of FinTech interactions users actually consider to be low (settler) to high (pioneer).

---

[9] A copy of the survey, strictly for review purposes, has been made available under the following link and will be kept online until after the final grading of this thesis: http://survey.wischnew.ski/index.php/487379

Participants were given the option to leave comments in a free-text field at the end of the survey; and email addresses for potential follow-up interviews were collected on a strictly voluntary basis.

### 3.3.3. Interviews

As mentioned in section 3.2.2., 'average' members of each target group were interviewed to gather a deeper understanding of the individual use-case for FinTechs. Participants were defined as 'average' based on the aggregated results using Excel and SPSS.

Out of the 439 valid survey participations, 37 participants (~8.5%) indicated their willingness to participate in a further interview by leaving their contact email addresses. Out of those 37, a total of six individuals (~16.2%), two per target audience, were interviewed to provide the aforementioned qualitative layer (see section 3.2.1.2 and 3.2.2). The interview was designed based on a preliminary assessment of the data gathered from the survey in a staged approach, meaning that the interview questions were composed to complement as well as validate the outcome of the survey. Given that the survey has already accumulated large amounts of quantitative data, the interviews follow an informal / unstructured approach as described by McNeill & Chapman (2005) – while the overall topics of interest were predefined (using the four question categories used in the survey), the interviews themselves are meant to explore individual opinions of the interviewees based on circumstances and individual responses in greater detail. Transcripts of the interviews are available in Appendix 4.

Areas of interest were:

  Category 1:    General Demographics

- Confirm age, location, educational background and employment situation with entry made during the survey (validation)
- Field of study → potential relation to financial sector / FinTech (deepening)
- Current job → potential relation to financial sector / FinTech (deepening)
- Explore other areas of interest based on replies given

Category 2:    Banking Specific

- Confirm house bank, use of products, reasons (validation)
- Discuss particular examples for product use → where, since when, why with the particular bank / FinTech (deepening)
- Discuss examples for reasons why an established product with house bank was dropped in favor or another party (cost, image, brand, flexibility…) // discuss why particular product is still used with house bank regardless of cheaper, more flexible etc. alternatives in the market (deepening)
- Explore other areas of interest based on replies given

Category 3:    FinTech Specific

- Confirm knowledge about & use of FinTechs (Validation)
- Discuss 'first contact' with FinTech (particular case example; deepening)
- Discuss overall satisfaction / dissatisfaction (deepening)
- Discuss likelihood of using other / more FinTechs in the future based on current experience (deepening)
- Explore other areas of interest based on replies given

Category 4:    Self-Identification

- Discuss reasons for self-identification as early adopter / balanced / traditional (validation & deepening)
- Example from other areas / industries?
- Is there a difference between the area mentioned above and the financial service industry? What are the key differences / why are there no differences? (deepening)
- Explore other areas of interest based on replies given

### 3.3.4.   Mapping Hypotheses and Research Questions to Data

As indicated in the beginning of this chapter, the primary aim of the survey and the follow-up interview were to answer the research questions by way of finding sufficient evidence to satisfy the underlying hypotheses. Table 6 shows an overview of the key areas and discusses the particular usefulness of the questions asked during the survey and/or interview:

| Research Question 1: What is the percentage share of the population in each of the three categories pioneers, migrators or settlers? | |
|---|---|
| Supporting Hypothesis: | (1)   The overall knowledge about FinTechs |
| | (2)   The overall acceptance / use of FinTechs |
| | (3)   Satisfaction and length of use |
| | (4)   Direct correlation between age and likelihood to use FinTechs |
| Answer(s) from survey: | • Demographic information will provide the age required to assess (3) in conjunction with the other two hypotheses.<br>• Questions 14-24 have been specifically designed to provide as detailed as possible answers to (1) and (2) by listing the 40 most popular FinTechs in Germany, as well as allowing participants to enter any other FinTechs not part of the list.<br>• Questions 20 and 21 specifically ask for the satisfaction (using a Likert scale) and the length a particular service has been used.<br>• The survey logic would select all answers listed in Q14 (FinTechs the participant knows) to ask detailed follow-up questions regarding the use thereof (if any) with concrete number and volume of transactions to further indicate whether the acceptance / use is high, medium or low.<br>• The self-assessment question (28) will help to validate the categorization of the participant with the actual use of FinTechs as provided. |
| Answer(s) from interviews: | Two participants from each category (pioneer, migrator, settler) will be interviewed based on their self-assessment and categorization by calculated analysis; as such the interview has a validating function (categories 1, 3 and 4).<br>Furthermore, reasons for using / not using FinTechs (question categories 2 and 3) and the particular scenarios will underline the acceptance. |
| Research Question 2: Which volume of transactions / investments is potentially being processed by FinTechs now? | |
| Supporting Hypothesis: | (1)   The overall acceptance / use of FinTechs |
| | (2)   The primary driver for people using FinTechs are cost savings |
| Answer(s) from survey: | • Demographic information will provide the baseline to extrapolate the survey result to the total bankable population of Germany (from secondary source / country statistic). Comparison to the development of the banking sector data will indicate a total loss of turnover.<br>• Questions 11-13 have been designed to answer hypothesis (2) |
| | by asking for specific indicators for leaving house banks for third party solutions (FinTechs) |
| | • Questions 25 acts as a confirmation question for 11-13 by listing the answer option 'competitive pricing' as a key factor for trying out the services of a FinTech. |
| Answer(s) from interviews: | Discussing the reasons for using / not using FinTechs as part of question category 2 and 3 will provide further information regarding the key driver / resistance for the use of FinTechs, thus giving further validation to the hypothesis 'the primary driver are cost savings' |

**Research Question 3: How will the potential impact of the lost transactions reflect in the market share of banks?**

| | |
|---|---|
| Supporting Hypothesis: | (1) The overall acceptance / use of FinTechs |
| | (2) The primary driver for people using FinTechs are cost savings |
| | (3) People living in urban areas have a higher probability of using FinTechs than those in periurban or rural settings |
| Answer(s) from survey: | • The answers obtained as a result from research question 2 will provide a solid understanding of the volume of business banks are losing, particularly in conjunction with current market share information and population development trends for the next years |
| | • Questions 4-7 will allow to gain insights into the regional distribution of survey participants, and further allow to break the information down by age and gender (if relevant) |
| | • Identifying whether the age is an actual factor for the likelihood to use / not use FinTechs or if the geographical location is of higher importance will be answered by mapping the location (Q4) and age (Q7) with the outcome of Q14-24; thus showing if the average 'settler' is rather located in a rural setting and the average 'pioneer' in a |
| Answer(s) from interviews: | Two participants from each category (pioneer, migrator, settler) will be interviewed based on their self-assessment and categorization by calculated analysis, as such the interview has a validating function (categories 1 and 4). |
| | Furthermore, reasons for using / not using FinTechs (question categories 2 and 3) and the particular scenarios will underline the acceptance. |

**Research Question 4: Are FinTechs likely to become / remain profitable using the acquired market share, and remain blue oceans?**

| | |
|---|---|
| Supporting Hypothesis: | (1) The overall acceptance / use of FinTechs |
| | (2) The primary driver for people using FinTechs are cost savings |
| | (3) Despite the abundance of FinTechs in the market with |
| | seemingly unique business models, only a small fraction thereof gains significant users to be true competitors for banks |
| Answer(s) from survey: | • The answers obtained as a result from research question 2 will provide a solid understanding of the volume of business banks are losing; which in conjunction with secondary data (fee structure of typical FinTechs) will help to identify the long-term income structure<br><br>• Questions 14-18 will provide a solid understanding of the FinTechs in the various categories the participants are using. Seeing how those FinTechs rank in terms of acceptance / use (question 19) will help to understand if providers of similar services are universally exchangeable, or if certain providers are preferred; indicating that they have higher success rates of remaining on top of their segment, and thus occupying their 'Blue Ocean'<br><br>• Question 20 ('how long have you been using the FinTech') in combination with 21 (satisfaction with FinTech) will also provide further insights how dependent the user of a particular FinTech might be. Having a high satisfaction and using it for a long time speaks for itself; however using a FinTech for longer time while actually dissatisfied with it provides valuable information about how strong ties the FinTech has with the client (hard to replace / no alternative?). |
| Answer(s) from interviews: | The interview questions for this research question have a validating effect, and aim at further establishing to which extent FinTech users are loyal to their provider of choice, or if they have developed a high likelihood of moving on to another provider given anyone undercuts the price of their provider of choice. |

## 3.4.  Data Validation

High quality of data input is essential to achieve reliable and consistent output. Consequently, in order to eliminate misunderstandings related to the questions within the survey, two pilot group assessments with 12 and 15 participants respectively were conducted prior to the publication of the survey. Next to wording changes and minor changes to the survey logic, the feedback from the first pilot group also led to a design change of the survey itself; initially each question was asked on an individual page, which 8 out of the 12 testers (66.7%) said to reduce the willingness to complete the survey, not knowing how many more questions the category would entail.

The second pilot group test also included the German version of the text, all 12 testers of the first pilot group also participated in the second stage. Only minor wording changes and typographical errors have been identified in the second pilot.

Validation of data points is carried out in multiple stages:

Applicability:
As indicated earlier, participants who are not residents in Germany, do not operate a bank account in Germany and/or are under the age of 18 were excluded.

Completeness:
Only fully completed surveys are taken into consideration for the assessment. Participants that interrupted the survey for later completion but never returned were removed from the results.

Truthfulness:
The data was checked for logical inconsistencies based on multiple factors: Obviously fake data has been excluded from the data sheet. Examples for fake information were random text strings such as "asdasda" or slur-words in free-text fields. Participants that completed the survey while selecting most (if not all) FinTechs and providing input of questionable consistency, have also been removed. Examples for questionable consistency would be to enter "12" for transaction number and "34" as transaction volumes; even more so if the same numbers were found repeatedly.

Outliers:
Data points far below or far above the average of the respective category were excluded from the analysis.

Interviews:
As mentioned in section 3.3.3, the interviews were conducted with participants aimed not only at gathering a deeper understanding of the persons' particular relation to banks and FinTechs, but also to validate the information provided during the survey. Despite a small number of participants, this process of validation helps to ensure the overall integrity of the data gathered.

# 4. Results

Before the analysis could be conducted, the data had to be prepared and normalized in order to yield adequate results. The following steps have been performed:

Normalization of city names:

Participants were given the option to enter city names in either English or German. To avoid double-counting, the German entries were transformed into the English spelling (i.e. München → Munich, Düsseldorf → Dusseldorf etc.).

Normalization of nationality:

As with the city names, all nationalities were converted into the English form. Where participants gave the country name instead of the nationality (i.e. Germany instead of German), those were also converted.

Normalization of data input:

Instead of giving the time of use for FinTechs in months as per the question(s), some participants gave the year they started using the particular service instead. Those have been converted to months at 12 months/year plus 3 months for the current year (2017).

The analysis was further subdivided in correspondence with the four research questions as shown in chapter 3.1.; namely: What is the percentage share of the population in each of the three categories pioneers, migrators or settlers; which volume of transactions / investments is potentially being processed by FinTechs now; how will the potential impact of the lost transactions reflect in the market share of banks; and are FinTechs likely to become / remain profitable using the acquired market share and remain blue oceans. For the analysis, quantitative data plays a predominant role given the number of data points; qualitative information are used in a separate chapter to underline elements and provide deeper insights where applicable.

## 4.1. Data Analysis

As initially mentioned, a total of 519 people from 42 different nationalities participated in the study, with 439 participants being German citizens with banking relations to German banks. After applying several criteria to categorize and validate participants (more detail is given in section 4.2.), a total of 349 participants where ultimately found to be relevant for the analysis[10]. The male to female ratio of all participants was roughly 4:1, with 409 male and 96 female participants. Valid results have been contributed from all over Germany, with a total of 94 cities in scope of the survey, the 10 largest (by number of participations) are displayed in table 7 below (Kaiserslautern and Leipzig sharing 10th place). The high number of participants from Berlin can be attributed to a number of factors: Bordering on 4 million inhabitants, the city state of Berlin is 3x as big as the second largest city (Hamburg; also a city state), has a very active startup community that is eager to participate in related (tech-) surveys, and is also the residency of the author of this dissertation, leading to a large number of willing participants (friends & family).

With 253 completed questionnaires from March 11 – 18th, almost half of the results (48.75%) came within the first week of the survey publication; which was in line with expectations due to the 'novelty status' and channels used to distribute it.

**Table 7: Ranking of Cities by Number of Participants**

| Participants | City | Percentage |
|---:|---|---:|
| 160 | Berlin | 36,45% |
| 25 | Stuttgart | 5,69% |
| 23 | Munich | 5,24% |
| 17 | Hamburg | 3,87% |
| 16 | Cologne | 3,64% |
| 15 | Frankfurt | 3,42% |
| 14 | Essen | 3,19% |
| 12 | Bonn | 2,73% |
| 12 | Hannover | 2,73% |
| 10 | Kaiserslautern | 2,28% |
| 10 | Leipzig | 2,28% |

*Note: Only the top 10 cities are displayed here*

The language used to access the survey was almost equally distributed, with 244 participants opting for German against 275 English speakers; however the nationality distribution shows that out of the 275 actually 208 are German nationals or dual citizens; whereas 11 out of the 244 participants opting for the German language version were not German nationals.

---

[10] Note that some tables make reference to lower numbers of participants, as some made use of their right to withhold / not disclose information on some questions and were hence excluded from related calculations.

Overall, a total of 81,431 data points have been gathered for analysis over the collection period of one months and been analyzed to construct the following participant profiles.

## 4.2. Participant Profile Determination

To make use of all the individual data points collected and create suitable links and references, an individual scoring model was developed in order to first identify which participants belong into which of the three categories for evaluation (Pioneers, Migrators, Settlers). Out of the questions evaluated, the following rational was used to calculate scores, based on the supporting hypotheses outlined in section 3.3.4.:

**Table 8: Categorization of Participants**

| Category | Rationale |
|---|---|
| Knowledge about FinTechs and use thereof | The more individual FinTechs a participant knows and the higher the number of FinTechs is that s/he is using, the higher the likelihood of being attracted to new developments in the market. Participants who are already deeply engrained in the FinTech market are assumedly more assertive to trying out yet another new company, as the inhibition level has already been reduced by their exposure to other candidates. |
| | At the same time, FinTechs that are used by many participants carry a lower 'novelty factor' than those lesser known / lesser used. Table 25 in Appendix 6 shows the total number of participants being aware of the five different categories, and calculates a rank based on the weighted average. This rank was then used as a multiplier to weigh the knowledge of the individual categories; i.e. a participant knowing about three different payment solutions ranked 3, while a participant knowing two lending provider scored 6. |
| | Table 26 in Appendix 6 provides the same ranking for the use of FinTechs. Given that the actual use of FinTechs is an even stronger indicator for being a Pioneer, knowledge and use have been weighted at 1:5 to ensure overrepresentation of active users rather than people who have heard about FinTechs but are not interested in using them. |
| | Appendix 6 shows an excerpt of the scoring calculation based on the first 45 participants. |
| Loyalty and Satisfaction | The second parameter for the scoring exercise are the aggregated loyalty with a certain FinTech (measured in months used) and the |

| | satisfaction (on a scale of 1-10). The rationale behind this approach is that the longer a participant has been using FinTechs already, the higher is his/her affinity to the overall sector, whereas the satisfaction serves as an indicator for the likelihood of continuously using them. Both factors have been multiplied to achieve an aggregated total; and used as a second factor for the overall profile calculation.<br><br>Appendix 7 shows an excerpt of the scoring calculation based on the first 45 participants. Purely for improved readability the result has been divided by 10 in the 'subtotal' column; since the final ranking was conducted solemnly based on the relative number, the order of magnitude does not play a role here. |
|---|---|
| Volumes and Transaction Frequency | The final enumerator for the participant classification is the aggregated frequency and volumes used when transacting with FinTechs. By using these indicators it is ensured that participants who are using FinTechs more often or with substantial amounts have a higher overall utilization factor than those who merely conduct some minuscule transactions.<br><br>In order to also reflect the importance of lesser known FinTechs being utilized as a factor for being a pioneer as opposed to using 'mainstream' FinTechs, the same weight from table 10 has been applied; since deposit originators and lending platforms had the same rank, these two columns have been merged for the calculation as there is no difference in the resulting weight.<br><br>Appendix 8 displays the result of the scoring exercise for this category. |

### 4.2.1.    What is the percentage share of the population in each of the three categories pioneers, migrators or settlers?

After applying the calculations indicated in table 8 above, the participants got a rank for each of the three categories. The sum of all three ranks together was then used to indicate into which of the three different subgroups participants fall, providing the baseline to answer the first research question.

Appendix 10 shows the full list of ranks for all participants, sorted from largest to smallest grade. To validate the results, the self-assessment of the participants as displayed in Table 25 (Appendix 6) have been used; with the same number of participants in each calculated categorization. The column 'check' in Appendix 9 indicates whether the classification corresponds to the self-assessment or not. Overall, 43 out of 55 'Settlers' were correctly identified (78.2% accuracy); 255 out of 303 'Migrators' (84.2% accuracy) and 50 out of 81 'Pioneers' (61.7%); the overall accuracy thereby lies at 70.6% and confirms the validity of the calculation (Cooper & Schindler 2003). A second validation layer was added through the use of interviews; see chapter 4.2.

Table 9 below shows the key demographic criteria associated with the respective profiles.

**Table 9: Location, Gender and Age Correlation**

| Classification | Number | Share | Rural | Urban | Female | Male | 18-25 | 26-30 | 31-35 | 36-40 | 41-45 | 46-50 | >50 |
|---|---|---|---|---|---|---|---|---|---|---|---|---|---|
| Pioneer | 50 | 14,4% | 5 | 45 | 12 | 38 | 10 | 17 | 10 | 9 | 4 | | |
| Migrator | 255 | 73,3% | 29 | 227 | 48 | 208 | 34 | 40 | 56 | 47 | 41 | 19 | 19 |
| Settler | 43 | 12,4% | 6 | 37 | 9 | 34 | 2 | 2 | 8 | 11 | 10 | 5 | 5 |
| **Total** | **348** | **100,0%** | **40** | **309** | **69** | **280** | **46** | **59** | **74** | **67** | **55** | **24** | **24** |

*Note: One participant did not reveal their gender and was hence excluded from this table.*

As per the supporting hypothesis, a correlation between age groups and behavior is clearly visible, with younger participants in age groups 18 to 30 being statistically more likely to make use of FinTechs in their daily life; which is further supported by a Chronbach's Alpha of .976 showing that the age distribution is internally consistent amongst the sample (see also Appendix 6, Table 28). The distribution share per age group as shown in table 10 below illustrates this trend even more. A relationship between the three categories and the employment status or educational level could not be established.

**Table 10: Classification per Age Group**

| Classification | 18-25 | Share1 | 26-30 | Share2 | 31-35 | Share3 | 36-40 | Share4 | 41-45 | Share5 | 46-50 | Share6 | >50 | Share7 |
|---|---|---|---|---|---|---|---|---|---|---|---|---|---|---|
| Pioneer | 10 | 21,7% | 17 | 28,8% | 10 | 13,5% | 9 | 13,4% | 4 | 7,3% | | 0,0% | | 0% |
| Migrator | 34 | 73,9% | 40 | 67,8% | 56 | 75,7% | 47 | 70,1% | 41 | 74,5% | 19 | 79,2% | 19 | 79% |
| Settler | 2 | 4,3% | 2 | 3,4% | 8 | 10,8% | 11 | 16,4% | 10 | 18,2% | 5 | 20,8% | 5 | 21% |
| **Total** | **46** | **100%** | **59** | **100%** | **74** | **100%** | **67** | **100,0%** | **55** | **100%** | **24** | **100%** | **24** | **100%** |

*Note: The participant excluded in table 9 did however indicate their age and is included here.*

Using public data on the age distribution in Germany published by Federal Ministry for Education and Research (BMBF 2015), it is possible to extrapolate the percentual distribution of the overall population from the gathered data (table 11). Information on the share of people over 55 was too scarce to draw informed conclusions, with only two participants being in the 56-60 range and a single person over 60. Given the overall trend however it can be assumed with reasonable certainty that people above 55 are rather not falling into the pioneer domain; and are also not a typical target demographic for FinTechs.

| Public Data | | Relative | | | Absolute | | |
|---|---|---|---|---|---|---|---|
| Age Groups | Population | Pioneer | Migrator | Settler | Pioneer | Migrator | Settler |
| < 18 | 12.969.000 | #N/A | #N/A | #N/A | #N/A | #N/A | #N/A |
| 18-25 | 6.140.000 | 21,7% | 73,9% | 4,3% | 1.334.782,6 | 4.538.260,9 | 266.956,5 |
| 26-30 | 5.285.000 | 28,8% | 67,8% | 3,4% | 1.522.796,6 | 3.583.050,8 | 179.152,5 |
| 31-35 | 5.102.000 | 13,5% | 75,7% | 10,8% | 689.459,5 | 3.860.973,0 | 551.567,6 |
| 36-40 | 4.907.000 | 13,4% | 70,1% | 16,4% | 659.149,3 | 3.442.223,9 | 805.626,9 |
| 41-45 | 4.961.000 | 7,3% | 74,5% | 18,2% | 360.800,0 | 3.698.200,0 | 902.000,0 |
| 46-50 | 6.502.000 | 0,0% | 79,2% | 20,8% | - | 5.147.416,7 | 1.354.583,3 |
| 51-55 | 6.948.000 | 0,0% | 79,2% | 20,8% | - | 5.500.500,0 | 1.447.500,0 |
| 56-60 | 6.037.000 | #N/A | #N/A | #N/A | #N/A | #N/A | #N/A |
| 61-65 | 5.203.000 | #N/A | #N/A | #N/A | #N/A | #N/A | #N/A |
| > 65 | 17.292.000 | #N/A | #N/A | #N/A | #N/A | #N/A | #N/A |
| | 81.346.000 | | | | 4.566.988 | 29.770.625 | 5.507.387 |

*Source: Adapted from BMBF (2015), http://www.datenportal.bmbf.de/portal/de/Tabelle-0.15.xls and populated with data from own research*

## 4.2.2. Which volume of transactions / investments is potentially being processed by FinTechs now?

Having calculated the number of people within each of the categories per age group in connection with the average transaction amounts and investment volumes per product category allows for a quantification of the FinTech transaction shares. Table 29 in Appendix 6 provides a detailed overview of the individual breakdowns, whereas the aggregated averages per category are used for the further assessment here.

As seen in table 12, payment providers contribute the largest amount of transactional volumes, followed by deposit originators and FinTech banks; whereas loans and investment helpers currently play a minor role in the overall market.

Table 12: Average Transaction Volumes per Category

| Collected Averages | | | | | |
|---|---|---|---|---|---|
| Category | Payment Prov. | Investment H. | Banks | Loans | Deposits |
| Pioneer | 1.482 | 85,00 | 591,90 | 194,00 | 780,00 |
| Migrator | 289 | 3,52 | 60,27 | 8,20 | 47,85 |
| Settler | 2 | - | - | - | - |
| **Total** | **1.773** | **89** | **652** | **202** | **828** |

Mapping the average volumes per category to the overall number of people in the respective categories as assessed in 4.2.2. above provides an overview of the potential market share the FinTechs have acquired from the banks through their respective solutions.

**Table 13: Mapping Transaction Volumes to Population**

| Category | Population | Payment Prov. | Investment H. | Banks | Loans | Deposits |
|---|---|---|---|---|---|---|
| Pioneers | 4.566.988 | 6.768.732.814 | 388.193.974 | 2.703.200.157 | 885.995.659 | 3.562.250.587 |
| Migrators | 29.770.625 | 8.615.688.719 | 104.662.354 | 1.794.377.920 | 244.212.160 | 1.424.570.934 |
| Settlers | 5.507.387 | 8.325.120 | - | - | - | - |
| **Total** | **39.845.000** | **15.392.746.652** | **492.856.329** | **4.497.578.077** | **1.130.207.819** | **4.986.821.521** |

### 4.2.3. How will the potential impact of the lost transactions reflect in the market share of banks?

Overall, the volume of transactions lost by banks to FinTechs as per the above estimation amounts to roughly 26.5 billion EUR annually across the various segments, or about 665 EUR per person. To judge what effect this has on the individual shares in these markets however it is required to have a closer look at the overall volumes in the various segments as a whole.

*Payment Providers:*

The 2016 publication by the German Banker's Association as well as the aggregate figures of the German Central Bank (2016) are particularly useful in that regard, as they not only provide a breakdown of all transaction and investment volumes, but also lists of channels used for the various transactions (Deutsche Bundesbank 2016; Bethge 2016). According to the publication, the overall volume of cashless transactions constituted a total of 56.6 trillion EUR, out of which 92.5% (~52.36tn EUR) were conducted by means of bank transfers, and thus falling into the domain of payment providers. This volume however includes transfers from corporates, while only private households are relevant for this study. Another study from 2015 by the German Central Bank published representative information based on 2019 participants recording their expenses and modes of spending money, and found that out of 502,544.10 EUR spent, 26,404.67 EUR (5.25%) are spent via remittances on a weekly basis, and therefore susceptible to be taken over by FinTechs (Deutsche Bundesbank 2015). On a yearly basis this volume sums up to 26,132,293 EUR, or 12.943 EUR per participant of the study. Extrapolated to the total population aged 18 and over this leads to the assumption that ~46.5 bn EUR in remittances are used by private households; out of which FinTechs with their ~15.3 bn EUR volume have acquired roughly one third – related to the overall volume of transactions, this however accounts for only ~0.03%.

| Demographic data | |
|---|---|
| Total population | 81.346.000,00 |
| Total adults | 68.377.000,00 |
| Volume transfers | 56.600.000.000.000,00 |
| **Bundesbank study** | |
| Participants (number) | 2.019 |
| Volume of transactions (weekly) | 502.544,10 |
| out of which remittances | 26.404,67 |
| Share | 5,25% |
| Volume yearly | 26.132.293,20 |
| Volume per person | 12.943,19 |
| Total private households (adults) | 885.016.251.677,27 |
| Volume of remittances | 46.500.520.193,50 |
| Volume FinTech Payment Provider | 15.392.746.652,29 |
| *Share FinTech in Total Transfers* | *0,03%* |
| **Market Share FinTechs in Consumer Finance** | **33,10%** |

A 2013 survey conducted by the German e-commerce center for trade (ECC) attests PayPal a market share of 29.9% in the online trade sector (see figure 11), whereas a study conducted on behalf of the German Parliament in May 2016 lists PayPal with 19.6%, and SofortÜberweisung with 2.5% (Deutscher Bundestag 2016). While the variance is rather large; the studies at least confirm the order of magnitude for the accumulated data, given that (online) trade is the largest use case for payment providers in Germany (ibid.).

Figure 11: PayPal 2013 Market Share in e-Commerce

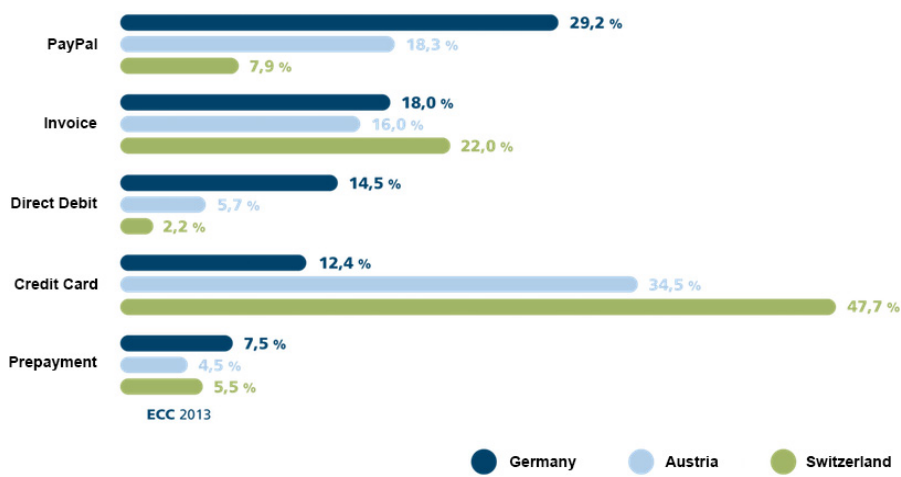

*Source: Adapted from http://www.ecckoeln.de/*

### Investment Helpers

As shown in table 13 above, investment helpers according to the assessed data account for roughly 0.5 bn EUR. A 2016 study conducted by Allianz Wealth Management and backed up with data from the German Central Bank shows that the total volume of liquid assets held by German consumers amounts to 5.49 trillion EUR in 2015 (Deutsche Bundesbank & Allianz SE 2016). Out of this volume, 2.6% are held in stocks, 4.0% in bonds and 5.0% in investment certificates (mainly investment funds), giving a combined total of 11.6%; other assets are held primarily in savings accounts, term deposits and savings certificates (combined 35.7%) or endowment policies (38.0%) (ibid.). Since only the former assets can be traded by using brokerage accounts where investment helpers come into play, their potential target market has a volume of ca. 636.8 bn EUR (table 15).

Table 15: Market Share of Investment Helpers

| Total assets (private households) | Relevant share (stocks, bonds, investment cert.) | Relevant market size | Volume of Investment Helpers | Share |
|---|---|---|---|---|
| 5.490.000.000.000 | 11,60% | 636.840.000.000 | 492.856.329 | 0,08% |

Unlike the volume of transfers covered by payment providers where FinTechs do have a significant impact on the consumer finance segment, the market share of investment helpers as per the analysis shows a marginal market share of only 0.08% covered to date.

### Lending Platforms

According to figures published by the German Central Bank and the German Bankers Association, the total outstanding volume of loans in Germany sums up to a total of ca. 1,1 trillion EUR (Deutsche Bundesbank 2016; Bethge 2016). The majority of these amounts were issued as long-term asset finance though, predominantly for real estate loans (ibid.), which are not useful for this analysis as the various lending platform have capped the maximum loan amount at 50k EUR (smava, auxmoney) and 30k EUR (Lendico). Consequently, only consumer loans have been assessed, a breakdown of which is provided by the German statistics office based on a detailed revision of the Central Bank data (Deutsche Bundesbank & Statista 2016). This review shows a total volume of outstanding consumer loans of 231.2 bn EUR for 2016, the German Credit Bureau indicates an average outstanding volume of 9,552 EUR per consumer loan (SCHUFA AG 2016).

Reviewing their websites (currently open p2p loans) shows that the average loan size is in a rather low range of 3-5k EUR (smava; with a single loan currently above 5k EUR). auxmoney currently has 394 p2p-loans open for funding with a range of 1,150 EUR to 62,800 EUR, although the bulk is below 10k EUR (smava 2017; auxmoney 2017). The survey conducted confirms those figures with an average of 4,700 EUR (smava) and 7,100 EUR (auxmoney) collected; but also indicates that the average loan typically requested by participants of this study is below the market average. Overall, with 1,13 bn EUR loans,

FinTechs according to the extrapolation of participants in this study account for 0.49% (see table 16).

**Table 16: Market Share of Lending Providers**

| Lending Provider | |
|---|---|
| Total Consumer Loans Outstading | 231.200.000.000 |
| FinTech volume coverage | 1.130.207.819 |
| FinTech market share | 0,49% |

### Deposit Originators

Using the same source as for the investment helpers and the total volume of deposits generated through the use of deposit originators (table 13) in conjunction with the total funds held in savings accounts (35.7% as indicated above) reflects in a total market size in the order of 2 trillion EUR, out of which ~0.25% have been acquired by deposit originators.

**Table 17: Market Share of Deposit Originators**

| Total assets (private households) | Relevant share (daily savings, term deposits) | Relevant market size | Volume of Deposit Originators | Share |
|---|---|---|---|---|
| 5.490.000.000.000 | 35,70% | 1.959.930.000.000 | 4.986.821.521 | 0,25% |

### Banks

FinTech Banks such as N26 and Fidor play a special role in this analysis, as they are not just piggybacking on bank accounts or cards to offer their services, but actually aim for the core product of the traditional banks itself – the current account. Owning the account enables banks to cross-sell solutions directly to their customers based on soft indicators (demographic, geographic) and hard indicators such as incoming salary, account turnover etc. (Ciarrapico & Cosci 2011; FinTech Pioneers 2017). As such it is not surprising that FinTech banks integrate other solutions from other FinTechs into their account – N26 for example is executing international transfers exclusively through TransferWise, offers investment portfolios through vaamo, and only offers loans and overdrafts on their own (N26 Bank 2017; FinTech Pioneers 2017). By doing so, they act as multipliers for third party solutions – as such it is not sufficient to look at the volume of transactions, but rather the numbers of people using them as 'bank of choice'.

As table 18 shows, online banks (which include N26, bunq and Fidor) but also those with a 'normal', not app- and integration driven online concept, already have the highest share in both the migrator and pioneer domain, while at the same time showing the lowest number of users in the settler-category, thus supporting the theory that participants in this category are rather tech-averse.

| Bank of Choice per Category | | |
|---|---|---|
| **Migrator** | **254** | **Share** |
| Commercial Bank (Deutsch Bank, Commerzbank…) | 50 | 19,7% |
| Communal bank (Volksbank, Raiffeisenbank…) | 59 | 23,2% |
| Online Bank (ING DiBa, DKB, N26…) | 81 | 31,9% |
| Public savings bank (Sparkasse) | 64 | 25,2% |
| **Pioneer** | **44** | |
| Commercial Bank (Deutsch Bank, Commerzbank…) | 10 | 22,7% |
| Communal bank (Volksbank, Raiffeisenbank…) | 10 | 22,7% |
| Online Bank (ING DiBa, DKB, N26…) | 15 | 34,1% |
| Public savings bank (Sparkasse) | 9 | 20,5% |
| **Settler** | **43** | |
| Commercial Bank (Deutsch Bank, Commerzbank…) | 12 | 27,9% |
| Communal bank (Volksbank, Raiffeisenbank…) | 9 | 20,9% |
| Online Bank (ING DiBa, DKB, N26…) | 8 | 18,6% |
| Public savings bank (Sparkasse) | 14 | 32,6% |
| **Total** | **341** | **100%** |

*Note: Eight participants did not indicate their house bank and were thus excluded from this table.*

Additionally, Table 19 shows the number of FinTech bank users amongst those participants that are already using a FinTech bank. Unsurprisingly, the relative share of the pioneers is highest amongst FinTech bank users, though the fact that amongst those users only 40% claim their house bank to actually be an online bank confirms what Maximilian Tayenthal, founder of N26 Bank mentioned during the FinTech Bootcamp in Berlin: "*We see a strong inflow of customers, but for many of those we are – as of yet, hopefully – only an interesting concept, and not yet the first bank. About one third of our customers are using N26 as their primary account, based on incoming salary payments.*" (FinTech Pioneers 2017).

**Table 19: FinTech Bank Users per Category**

| Share that uses FinTech Bank | | | | Migrators relative to total | Pioneers relative to total |
|---|---|---|---|---|---|
| | Settlers | Migrators | Pioneers | | |
| Communal Bank | 0 | 5 | 5 | 8,47% | 50,00% |
| Savings Bank | 0 | 7 | 1 | 10,94% | 11,11% |
| Commercial Bank | 0 | 8 | 6 | 16,00% | 60,00% |
| Online Bank | 0 | 5 | 6 | 6,17% | 40,00% |
| Total | - | 25 | 18 | | |
| Relative to sample | 0,00% | 9,84% | 40,91% | | |
| | | | Average | 10,40% | 40,28% |

According to their press release from March 2017, N26 Bank has acquired 300,000 customers in 17 countries since its inception in 2013;"*more than half outside of [their] initial core markets Germany and Austria*", making it "*a leading mobile bank in Europe*"[11] (Tayenthal & Stalf 2017).

Even though the bank remains vague about their client numbers per individual country, the given number allows for an assumption of the market share in Germany: Given that more than half of the customers (>150,000) are outside of Austria and Germany, and Austria with 8.8 million inhabitants has about 11% of Germany's population hints at a number of customers in the range of 110-130,000; out of which 1/3 are actively using N26 as their house bank, giving a range of 36,660 – 43,330 (averagely ca. 40,000).

Fidor Bank, the second large mobile bank in Germany, ended the fiscal year 2015 with 102,000 customers across all markets without indicating any breakdown (Fidor AG 2015). Assuming a similar distribution as N26, they would have had ca. 50,000 clients in Germany, with about 16,660 primary account holders. Other that N26 which promotes aggressive growth, Fidor has used a no-frills-marketing approach and is expecting a "moderate growth" without further indications (Fidor AG 2015). With lack of other information, a growth rate of 10% is assumed, increasing the number of active clients to ca. 18,260 by the end of 2016. Combined with the average active clients of N26 the total market share would thereby consist of ~58,260 primary account holders. Assuming each adult citizen has one primary bank account, the market share in this category would be ca. 0.09% (and ca. 0.26% for all active accounts).

---

[11] The quote has been literally translated to English, as the press release was only available in German.

| FinTech Bank - Market Share | |
|---|---|
| Total adults | 68.377.000 |
| FinTech Bank account hodlers | 175.000 |
| FinTech Bank primary a/c holders | 58.260 |
| **Market share account holders** | **0,26%** |
| **Market share primary a/c holders** | **0,09%** |

Neither the Dutch bunq not Finnish Holvi bank had any active users in Germany; category 'other' revealed that the German Wirecard Bank and British monese were known to participants, yet not used by any. Relative to each other, N26's market share would then be ~71.4% vs. ~28.6% for Fidor all things being equal (in terms of primary accounts).

## 4.2.4. Are FinTechs likely to become / remain profitable using the acquired market share and remain blue oceans?

FinTechs, as most other companies that have only recently been established, are very secretive about their actual cost and income structures, expansion plans and other KPIs, for two primary reasons: To keep investors curious, and to make assumptions for competitors as hard as possible. The latter is just as true for the analysis within this dissertation, and as such the focus here lies on the potential product margins to be realized, combined with the likelihood of people to abandon their FinTech again, based on the reasons given in table 21.

Table 21: Reason to Change Provider

| | Cost | | | Lower Interest | | | Faster Availability | | | Higher Revenue | | | 24/7 Available | | | Features | | |
|---|---|---|---|---|---|---|---|---|---|---|---|---|---|---|---|---|---|---|
| | No | Yes | % yes | No | Yes | % yes | No | Yes | % yes | No | Yes | % yes | No | Yes | % yes | No | Yes | % yes |
| Pioneer | 13 | 29 | 69,0% | 16 | 26 | 61,9% | 28 | 14 | 33,3% | 16 | 26 | 61,9% | 35 | 7 | 16,7% | 24 | 18 | 42,9% |
| Migrator | 61 | 174 | 74,0% | 148 | 87 | 37,0% | 187 | 48 | 20,4% | 110 | 125 | 53,2% | 186 | 49 | 20,9% | 145 | 90 | 38,3% |
| Settler | 11 | 29 | 72,5% | 25 | 15 | 37,5% | 32 | 8 | 20,0% | 21 | 19 | 47,5% | 29 | 11 | 27,5% | 29 | 11 | 27,5% |
| **Total** | **85** | **232** | **73,2%** | **189** | **128** | **40,4%** | **247** | **70** | **22,1%** | **147** | **170** | **53,6%** | **250** | **67** | **21,1%** | **198** | **119** | **37,5%** |

Starting from the aggregated market shares, it becomes obvious that other than in the category 'Payment Providers', FinTechs in the German Consumer Finance market have not been able to gain a significant foothold, while a deeper look into the market distribution within the individual categories indicates that each category sports one dominant player holding above 60% of the market, while the others are competing for the remaining 40% - many of which turned out to be entirely unused by participants of the questionnaire; despite the fact that some of them are at least known.

Table 22: Aggregated Overviews of Market Shares

| | Payment Providers | Investment Helpers | Lending Platforms | Deposit Originators | Banks |
|---|---|---|---|---|---|
| **Volume / *Number*** | 15.392.746.652 | **492.856.329** | 1.130.207.819 | **4.986.821.521** | *58.260* |
| **Market Share** | 33,10% | **0,08%** | 0,49% | **0,25%** | 0,09% |

Consequently, it can be said with reasonable certainty that the FinTech coming out on top of each group have indeed managed to occupy their 'Blue Ocean'; but the emergence of other players active in the same segment are a good indicator that competition has already moved into the markets. In fact, many FinTechs (and Startups in general) are not even striving to attain profitability, but are rather interested in creating a public 'proof of concept' and aim to be acquired by any other company – in this case most likely a traditional bank – that will then incorporate their new, agile functionality into their old, legacy business to open up new approaches for existing customers, and hence competing with the FinTechs directly (FinTech Pioneers 2017; Li et al. 2017; Suster 2011).

**Table 23: Breakdown of Market Shares per Category**

| Breakdown of Market Share | | | | | | | | | |
|---|---|---|---|---|---|---|---|---|---|
| **Payment Providers** | | **Investment Helpers** | | **Lending Platforms** | | **Deposit Originators** | | **Banks** | |
| Paypal | 61,9% | scalable.capital | 63,1% | auxmoney | 60,2% | ZINSPILOT | 12,3% | N26 | 71,4% |
| Transferwise | 17,8% | wikifolio | 32,0% | Smava | 39,8% | Weltsparen | 78,7% | Fidor | 28,6% |
| SofortÜberw. | 18,7% | vaamo | 4,9% | CrossLend | 0,0% | Savedo | 0,0% | Holvi | 0,0% |
| klarna | 1,7% | easyfolio | 0,0% | GIROMATCH | 0,0% | Other | 0,0% | Bunq | 0,0% |
| WorldFirst | 0,0% | quirion | 0,0% | Lendico | 0,0% | | | Other | 0,0% |
| CurrencyFair | 0,0% | whitebox | 0,0% | kapilendo | 0,0% | | | | |
| TorFX | 0,0% | fintego | 0,0% | Zopa | 0,0% | | | | |
| FrontierPay | 0,0% | LIQUID | 0,0% | Funding Circle | 0,0% | | | | |
| FC Exchange | 0,0% | cashboard | 0,0% | | | | | | |
| Azimo | 0,0% | growney | 0,0% | | | | | | |
| HiFX | 0,0% | VisualVest | 0,0% | | | | | | |
| Skrill | 0,0% | GINMON | 0,0% | | | | | | |
| Other | 0,0% | Other | 0,0% | | | | | | |

*Note (1): One participant mentioned "myPension" as a deposit originator; however upon further review they are offering a pension scheme under the umbrella of a life insurance. Since insurance companies are not in the scope of this dissertation, the entry has been excluded.*

*Note (2): One participant indicated to use Skrill but made use of the right to not disclose the transaction volume, hence the market share could not be assessed. Being the only user of Skrill in the study, it is however likely to be insignificant.*

Despite the initially mentioned difficulty to find relatable numbers for turnover and revenue, Appendix 5 provides a calculation of net income per product based on pricing / fee structures as published by the companies on their respective websites. These calculations confirm that the segment of payment solutions not only provides the largest turnover for FinTechs, but also the highest profits, regardless of the comparably small individual transaction amounts. Total income for klarna however could not be fully assessed as they do not provide their number of merchants (who pay 85 EUR per month in account maintenance fees; which can contribute a rather significant income on top of the transaction fees). Investment helpers are very unlikely to be profitable for the time being, and therefore potentially susceptible for new market entrants – in order to remain in a 'blue ocean', they would have to scale up significantly and get ahead of their emerging competitor; which at this time does not appear easy. Lending providers may attract gross profits in the 13-19 million EUR range, but given the overall low interest rate of loans at banks, are unlikely to increase their market share beyond the means of achieving self-sustainability.

Deposit originators and FinTech Banks again are very difficult to assess: While the former are completely free to use for their clients and only charge undisclosed commission to the final recipient of the deposits, the latter offer their account virtually for free and only earn money on cross-selling of other solutions and commission income from their customers' use of credit cards, loans and overdrafts. Given the comparably strong growths of the FinTech banks (see also 4.2.3), it would be possible for the competitors to indeed remain in 'Blue Ocean' scenarios; if they manage to keep adding new and relevant integrations of other technologies into their existing models.

## 4.3. Interviews

To gain a deeper understanding of FinTechs and their general use and acceptance, six peer group interviews have been conducted with two average participants for the target demographics settler, migrator and pioneer. The below table provides an overview of the classification of the participants, their attitude towards FinTechs and motivation to using / not using particular providers:

Table 24: Summary of Interviews

| Participant | Category | House Bank | Main FinTech Use | Main Concerns | Main Motivation |
|---|---|---|---|---|---|
| *AR* | Settler | Savings Bank | PayPal, SofortÜberweisung | Trust, security, control, transparency | Cost savings, Integrated into vendor platform |
| *MS* | Settler | Commercial Bank | None | Control, Transparency, Reputation | Return on investment |
| *PK* | Migrator | Online Bank | wikifolio, PayPal, SofortÜberweisung | Control | Cost savings, customer service quality, flexibility, features |
| *CF* | Migrator | Online Bank | N26, PayPal, transferwise | Reputation | Bypass transaction limit, cost savings, flexibility |
| *KL* | Pioneer | Online Bank | PayPal, N26, smava | None | Cost savings, features, efficiency, 24/7 availability |
| *MF* | Pioneer | Online Bank | Fidor, smava, PayPal, transferwise, wikifolio | Security, transparency, control | Cost savings, higher revenue, new feature, flexibility |

The interviews confirm that online banks are the predominant house bank of choice for both Migrators and Pioneers, whereas they are the least popular in the 'Settler' category. However, AR confirmed in the interview to be using a secondary account with an online bank for ease of access while abroad. All but one interviewee also make note of cost savings as one of the primary incentives to be using FinTechs; and the only person not to mention this reason does in fact not use any FinTechs – over the course of the interview he confirmed having heard about several in TV advertisements, but did until now not find a reason to use them; would however be open for investment opportunities due to the decline in deposit interests. The same reasoning was also provided by AR, who considers herself "too conservative for stock investments".

Apart from KL, who has no concerns about using FinTechs and "wants to try everything new", the majority of interviewees bring up on or more of the items transparency, control and security as primary concerns for using FinTechs. Reasons provided are e.g. that FinTechs which are not domiciled and regulated in Germany might be less approachable when any issues arise. On the other hand side, interviewees that have used FinTechs are generally convinced of their services and praise the additional functionality, reduced cost, higher efficiency and better accessibility. Only MF states that his FinTech Bank of choice still lacks in terms of customer service, and states that in the age of "machine learning and predictive analysis", their efficiency still has room for improvement.

## 4.4. Summary of Data Analysis and Interviews

As a summary of the findings from the structural analysis and interviews conducted it becomes apparent that participants of the survey have a varied knowledge and understanding of FinTechs, ranging from a basic understanding of their product / service offering without any particular personal use thereof, up to very enthusiastic advocacy of FinTechs as a structural enhancement if not replacement of traditional financial services.

Overall, the average user however is still quite skeptical of the promised innovations, as becomes evident in the evaluation of the market share for all but payment solutions – and even here, only PayPal, which has been around for nearly two decades and is as such far more 'mainstream' than any other FinTech managed to attract a significant number of users. Payment providers generally managed to benefit from the increasing internationalization of online shopping on a consumer level, where they add benefit in terms of guarantees and user support on top of the cost savings that have been mentioned as the primary reason across the whole sample to utilize any FinTech.

Yet despite the fact that over 73% of the participants indicated their primary incentive to leave their current bank to save transaction costs, over 53% to maximize their return and more than 40% to lower their interest expenses, only a fraction actually engaged with FinTechs outside of the payment domain with more than what one could call 'general curiosity'. And even across the payment solution providers, a handful of established companies divide the market amongst themselves – from the 14 providers mentioned, only 5 were used at all; which was further confirmed in the interviews as none of the selected candidates made any notion of using a FinTech outside of the commonly utilized companies.

The interviews ultimately reaffirmed the understanding voiced in chapter 2.4.1 that Germans are rather opposed to take risks, or only very calculated ones: All but one interviewee mentioned factors such as control, transparency, trust, reputation and security (or lack thereof) as a prime factor when engaging with FinTechs or new technology in general.

# 5. Conclusions & Critical Evaluation

*Outcome*

The goal of this dissertation was to find out whether the German consumer finance sector meets the textbook definition of the blue ocean strategy regarding the various market players as 'blue' and 'red' oceans respectively; find indicators to identify the number of 'settlers', 'migrators' and 'pioneers' amongst the population of the country, and ultimately try to find a way to quantify which percentage of the people would be likely to abandon their current financial institutions in favor of a FinTech company.

When it comes to blue ocean literature as part of the overall strategic management domain, an abundance of information and literature was available, with the original publication by Kim & Mauborgne and its recent republication being the most prominent books on the subject (Kim & Maubourgne 2005; Kim & Maubourgne 2015). Assessing the steady decline in traditional consumer banks as a result of competition-driven mergers and acquisitions due to reduced income and continuously increasing cost within chapters 2.4.1 and 2.5 confirmed that the traditional banking sector – across all segments – fulfills typical 'red ocean' criteria. Even the cost efficient online banks which are already making use of low-cost strategies without branches, without own ATM networks and with a very limited product offering (see chapter 2.5) are competing with each other, just on a different level. With traditional 'brick and mortar' banks operating in a polypole, online banks, due to their cost structure, are rather competing with each other amongst a limited number of market entrants in a growing oligopoly. Nevertheless, their margins and income are very restricted and the product offering varies only within tight constraints (product conditions), so still fulfilling the typical 'red ocean' classification.

FinTechs on the other hand have opened up the market by introducing new features and functionalities, while neglecting or deliberately ignoring core functions of banks, in line with the BOS' reduce-eliminate-raise-create-matrix derived from a business strategy canvas (see chapter 2.3.2, 2.4.2, 2.5 and 2.6). Focusing on those new and previously unknown features has indeed opened new markets – if limited in size – and as such established the respective company as a 'blue ocean' in line with the theory. Nevertheless, as the number of FinTechs now operating in a similar domain – yet with different features and peculiar characteristics – shows, have already begun competing for similar target customer groups; thus hinting at the fact that their respective 'blue oceans' might not be destined to last forever.

Subsequently, a survey was designed and distributed on multiple digital channels to assess the overall knowledge about FinTechs amongst German consumers, find out which products they are using and for what reasons and purposes, and ultimately to what extent in terms of transaction frequency and volume, with the aim of establishing the share of settlers, migrators and pioneers, where terms such as 'early adopters', 'balanced' and 'traditionalists' were also used in the survey to avoid potential bias amongst participants (see chapters 3.3.1 and 3.3.2). By assigning scores to the knowledge about- and use of FinTechs, the loyalty and satisfactions with their products, and the transaction frequency and volume, survey participants were grouped into the three aforementioned categories (see 4.2.1.), which were then cross-checked against their self-assessment, and all matching cases were then used for a thorough sampling of indicators. By applying the identified shares on the overall population of Germany using publically available data, all identified indicators have then been applied to the overall population of the country. In line with the second goal of this dissertation, the survey identified that the vast majority (73.3%) of German residents fall into the 'Migrator' category, thus being classified as adopting new technologies ones they have established themselves as a firm solution; whereas the share of pioneers and settlers was very similar with 14.4% (pioneers) and 12.4% (settlers) as shown in chapter 4.2.1.

Finally, applying all gathered indicators from the survey as well as follow-up interviews with 'average' candidates of each of the three categories to market data for transaction volumes (payments), investments, loans and savings (deposits) as well as use-cases for FinTech banks in chapters 4.2.2. – 4.2.4., the actual use of the respective FinTech solutions was quantified to determine how likely it would be for people to utilize FinTechs on a grand scale, and thus deduce the likelihood to abandon their currently used financial institutions in favor of a new competitor.

However, the data gathered reveals a somewhat unexpected and unspectacular outlook of the FinTech market: Despite the existing hype in the USA and Asia (Accenture 2015; FinTech Pioneers 2017; Kang et al. 2016), the ongoing investments into the industry during the last few years of over 100 billion USD from 2014-2016 and the constant murmurings of ongoing revolutions in the banking sector (FinTech Pioneers 2017; KPMG 2017; Reed 2012), Germans appear to be reluctant to follow this trend and rather pursue 'business as usual', with perhaps the exception of payment providers such as PayPal, which have been able to conquer a third of the transactions of private households in Germany. All other FinTechs display overall unspectacular numbers and volumes, indicating a very limited use amongst the population, and subsequently a very low likelihood to abandon their traditional banking channels. This phenomenon might however not be restricted to Germany alone – in February 2017, Warren Mead, partner of KPMG UK and co-head of their FinTech research unit commented that:

> "*There was a big rush of investment in fintech during 2014 and 2015 as investors globally bought into the idea of new and disruptive business models. Amid growing geopolitical and macroeconomic uncertainty, 2016 saw the investor sentiment tide turn, with investors seeming to want more proof that innovative solutions can be scaled and commercialized*" (KPMG 2017).

Amidst the perceived global enthusiasm, it appears after all as if the German consumer behavior and the outlook on profitability for FinTechs identified in chapter 4.2.4. reflects the paradigm change amongst investors: that a majority of the companies are not likely to become profitable and those will eventually disappear from the market or become acquisition targets for traditional banks to add agile services into their product portfolio; leading to blue oceans – and the big change in consumer behavior, the 'disruptive potential', dispersing into yet another – slightly differently composed – red ocean. FinTech banks, which have already begun integrating with other FinTechs might yet prove to be the biggest profiteers from this trend, as they would compete in the future red oceans with highly integrated and scalable, proven international solutions that are built to use only minimal human interaction and are thus set up to become the most cost efficient.

### Limitations & Challenges

This dissertation provides an outside view of the FinTech industry using publically available market information and insights gathered using quantitative and qualitative layers from (potential) users of their services and products. Given more time and opportunities, a third layer providing inside information from the FinTechs themselves would reduce the necessity for assumptions and deductions via third-party information, thus providing a higher level of validation to the calculations. Another limitation of the study was access to survey participants. Due to the market size under review, the survey was performed as a simple random sampling exercise as opposed to probability sampling, which might have provided a wider angle on the market under review. Probability sampling would have however required access to advanced research tools beyond the scope of a one-person project, where a random online sample was the only way to reach out to a significant number of people. The author tried to mitigate this to an extent by adding interviews as a qualitative layer, but the overall limitation of a single channel to approach participants remains a constraint. This also reflects in a third limitation, namely a more geographically diversified response that would lead to added benefits in terms of more concrete customer profiling per area (down to a district level). As it currently stands, more than 50 small towns / villages only have a single or maximum two respondents, which is not sufficient to make generalized statements about particular locations.

## *Outlook & Further Research*

Reviewing the limitations and challenges observed, several angles for further research present themselves. Starting with the outside-in perspective on FinTechs, dialogues and structured interviews with founders, leading employees and investors of the individual companies being reviewed would help to solidify the information gathered, and – if possible – allow a further assessment of the cost structure and therefore profitability of the service providers, thus creating a better understanding as to how future-proof the solutions are.

A deeper understanding of the motivation of FinTech users would also be worth pursuing by employing a true probability sampling technique across the whole federal republic of Germany. This would however require access to a qualified market research company with sufficient outreach and areal coverage; and would provide optimal results through structured interviews (i.e. via telephone) instead of self-selection through online questionnaires. This same approach will also be useful to unveil a more diverse perspective from rural and periurban areas, which are as of yet underrepresented.

# 7. Appendices

Appendix 1, Agenda of the 'FinTech Pioneers' Bootcamp in Berlin, Febraury 16th & 17th.

Appendix 2, Global Indicators of Stocks Traded vs. GDP, exemplary slide from FinTech Pioneers Fair

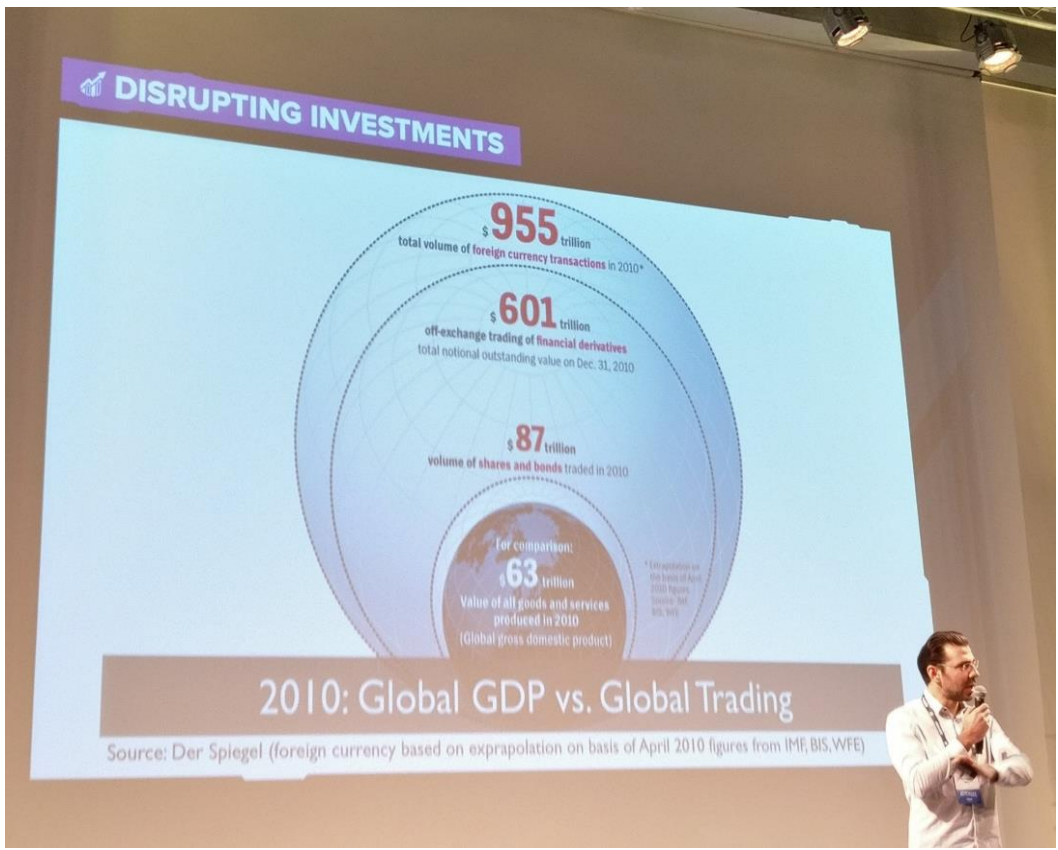

# Appendix 3 – FinTech Questionnaire

The survey will remain available for ca. 6 months following the submission of the dissertation under the following link for review purposes: http://survey.wischnew.ski/index.php/487379.

<div align="center">

**Question 0: Language Selection**

</div>

**[English] / [German]**

<div align="center">

**Questions 1 – 9: General demographics**

</div>

**Question 1:**
Are you at least 18 years old?
***[Yes] / [No]***

**Question 2:**
Do you currently possess a bank account in Germany?
***[Yes] / [No]***

**Question 3:**
Are you a resident in Germany?
***[Yes] / [No]***

**Question 4:**
In which city do you live? (if multiple, please enter main residency)
***[free text]***

**Question 5:**
What is your nationality?
***[free text]***

***Question 6:***
What is your gender?
***[male, female]***

**Question 7:**
Please select your year of birth:
***[dropdown]***

**Question 8:**
Please pick your current employment situation. If multiple apply (i.e. student and employed), please select the activity that accounts for >50% of your time. ***[single choice]***

- School
- University
- Employed (part-time)
- Employed (full-time)
- Self-employed
- Unemployed
- Parental leave
- Other, please specify:

**Question 9:**
What is your educational level? Please select highest applicable degree obtained. If you are currently studying, please select the degree you are currently pursuing. ***[single choice]***

- Secondary education
- Trade school
- Vocational school
- Bachelor's degree
- Master's degree
- PhD
- Habilitation

**Questions 10 – 13: Banking specifics**

**Question 10:**
Which of the following types of bank best describes your house bank? A house bank is the bank where your main current / checking account is located. Should you have multiple current accounts, please select the option where your primary source of income (such as salary, student allowance, and unemployment benefits) is paid into: *[single choice]*

- Public savings bank (Sparkasse)
- Communal bank (Volksbank, Raiffeisenbank…)
- Commercial Bank (Deutsch Bank, Commerzbank…)
- Online Bank (ING DiBa, DKB, N26…)

**Question 11:**
Please select all products you are using from a service provider other than your house bank. *[multiple choice]*

- None, I use all financial products through my house bank exclusively *[if selected, other options will be disabled]*
- Savings account (daily accessibility)
- Term deposit (fixed investment period)
- Brokerage account
- Consumer loan (personal consumption)
- Car loan
- Mortgage (housing loan)
- Credit card
- Money transfer services
- Other, please specify

**Question 12:**
Which of the following attributes attracted you to using a third-party provider rather than your house bank for the services / products stated above? *[multiple choice]* *[skipped if Q11 = none]*

- Cost / Revenue
- Speed
- Brand / image
- Flexibility
- Service quality
- Other, please specify

**Question 13:**
Which of the following factors do you feel can convince you to move your current banking products to another financial service provider? *[multiple choice]*

- Lower cost (per transaction)
- Lower interest (e.g. on loans, mortgages)
- Faster availability
- Higher return (e.g. better interest on savings account)
- 24/7 accessibility
- Advanced features not available from current bank
- Other, please specify
- None *[if selected, other options will be disabled]*

**Questions 14 – 27: FinTech specific**

**Introduction to FinTechs**
Next to traditional banks, other financial service providers have started offering products out of the banking domain, or are acting as intermediaries between you and other institutions and even private individuals. To be as cost-efficient as possible, those providers are offering their technology-based services via smartphone-apps, websites or other modern communication forms; hence the collective term "FinTech".

Some popular examples are the payment services PayPal and TransferWise, peer-to-peer loan originators Funding Circle and AuxMoney or savings accumulators Weltsparen and ZINSPILOT; as well as the German mobile banking solution Number26 (that recently transformed into a full bank and now operates under the name N26 Bank GmbH). Providers such as easyfolio or scalable.capital offer automated brokerage, also called "robo-trading" to maximize investment revenues.

Which of the following [*FinTechs from $category*] *are you aware of?* Leave blank if none. *[Individual checkboxes, unlimited choices; 5 individual questions for each of the categories; "FinTechs from $category" is replaced with the respective name of the category below]*

| Q 14 | Q 15 | Q 16 | Q 17 | Q 18 |
|---|---|---|---|---|
| **Payment services:** | **Loan platforms:** | **Deposit originators:** | **Investment helpers:** | **Banks:** |
| PayPal | Funding Circle | ZINSPILOT | easyfolio | N26 |
| TransferWise | auxmoney | Weltsparen | scalable.capital | Bunq |

| CurrencyFair | CrossLend | Savedo | wikifolio | Fidor |
|---|---|---|---|---|
| WorldRemit | Finnest | | quirion | Holvi |
| SofortÜberweisung | GIROMATCH | | whitebox | |
| WorldFirst | Lendico | | vaamo | |
| TorFX | Smava | | fintego | |
| FrontierPay | kapilendo | | LIQUID | |
| FC Exchange | Zopa | | cashboard | |
| Azimo | | | growney | |
| HiFX | | | VisualVest | |
| Skrill | | | GINMON | |
| klarna | | | | |
| *Other, <u>please specify</u>:* | *Other, <u>please sepcify</u>:* | *Other, <u>please sepcify</u>:* | *Other, <u>please sepcify</u>:* | *Other, <u>please sepcify</u>:* |

**Question 19:**
Which of the FinTechs you know about are you actively using?
*This question is skipped if Q14-18 are all left blank.*

*[list of selected items from Q 14-18 + "None, I know about them but do not use / have not used any".]*

**Question 20:**
How long have you been using the FinTech already?

*[list of selected items from Q 19; answer in months (free text, numbers only)].*
*This question is skipped if Q19 = none*

**Question 21:**
How satisfied are you with their service / product? **[1 = very dissatisfied, 10 = totally satisfied]**

*[list of selected items from Q 19 with scale from 0-10].*
*This question is skipped if Q19 = none*

**Question 22:**
How often are you using the Payment Providers, Investment Helpers and FinTech Bank?
What is your average transaction amount in Euro?

*[list of selected items from Q 14, 17, 18; two columns for free text (numbers only): column 1 = number of transactions per months, column 2 = average transaction volume].*
*This question is skipped if answers to Q14, 17, 18 as elements of Q19 = none*

**Question 23:**
Which total sum did you borrow over the last 12 months through the loan platform(s)?
Please give your answer in Euro.

*[list of selected items from Q 15; free text (numbers only)].*
*This question is skipped if answers to Q15 as elements of Q19 = none*

**Question 24:**
Which total sum did you deposit over the last 12 months through the deposit originator(s)?
Please give your answer in Euro.

*[list of selected items from Q 16; free text (numbers only)].*
*This question is skipped if answers to Q16 as elements of Q19 = none*

**Question 25:**
What was the initial reason that convinced you to use a FinTech provider? *[multiple choice]*
*This question is skipped if Q19 = none*

- Personal recommendations (friends, family, co-workers)
- Article in newspaper, TV, website, etc.
- Favorable online reviews
- Competitive pricing
- Offered as integrated solution of other service or website
- Transparency
- Product / service not available from house bank
- Personal curiosity

- Other, <u>please specify</u>

**Question 26:**
If you have never used a FinTech, what prevented you from doing so? *[multiple choice]*
*This question is skipped if <u>any FinTech is selected</u> in Q19*

- I was not aware of FinTechs and their services / products
- Simply no use / need for the functionality
- Product / service very new
- Nobody from personal environment (friends, family, co-workers) using it
- Negative online reviews
- Non-transparent product/service design
- Future development / acceptance unknown
- Product / service not (fully) regulated (e.g. through central bank or financial sector supervision)
- Security concerns (data protection, privacy, encryption…)
- General risk-awareness
- Technological limitations (no smartphone, no mobile internet etc.)
- Other, <u>please specify</u>:

**Question 27:**
If you are using a FinTech, did you still have any initial doubts? *[multiple choice]*
*This question is skipped if Q19 = none*

- No doubts whatsoever
- Product / service are very new
- Nobody from personal environment (friends, family, co-workers) using it
- Negative online reviews
- Non-transparent product/service design
- Future development / acceptance unknown
- Product / service not (fully) regulated (e.g. through central bank or financial sector supervision)
- Security concerns (data protection, privacy, encryption…)
- Other, <u>please specify</u>

**Questions 28 – 30: Self-Identification**

**Question 28:**
When it comes to general technological innovation, would you consider yourself to be: **[single choice]**

- An Early Adopter – 'I love technology and want to try everything new'
- Balanced – 'Once the technology has been around for a while and proven worthwhile, I might start using it'
- Traditionalist – 'I don't need new technologies as long as the current one is still doing its job, and will gladly skip the investment until absolutely necessary'

**Question 29:**
Please feel free to comment on any other aspect or issue relating to FinTechs and their presence in Germany.
**[free-text]**

**Question 30:**
If you would be available for follow-up interviews on this topic, either via phone, Skype or other personal contact (if feasible), please kindly state your email address:
**[free text]**

Many thanks for your participation!

| | Transactions per Person | | FinTech Volume | Market Share | Volume share | No. Transaction | Fee | Commission | Currency Conversion | Other | Commission & Fee Income | Other income | Total |
|---|---|---|---|---|---|---|---|---|---|---|---|---|---|
| | Month average | Year average | | | | | | | | | | | |
| PayPal | 3,24 | 38,87 | 15.392.746.652 | 61,9% | 9.527.478.612 | 245.123.974 | 0,35 EUR | 1,5-1,9% | 0,00-0,15% | | 252.524.267 | | 252.524.267 |
| Transferwise | 1,42 | 17,04 | | 17,8% | 2.733.705.790 | 160.428.744 | 0 | 0,50% | 0,02% | | 68.342.645 | | 68.342.645 |
| SofortÜberweisung | 2,23 | 26,76 | | 18,7% | 2.873.142.436 | 107.367.057 | 0,25 EUR | 0,90% | | 59,90 EUR Setup cost + 4,90 EUR monthly account fee | 52.700.046 | 381.150 | 53.081.196 |
| klarna | 1,9 | 22,80 | | 1,7% | 258.419.814 | 11.334.202 | | 1,5-3,0% | | 85 EUR monthly fee, if user wants to pay instalmentally, up to 20% p.a. loan interest | 255.020 | ? | 255.020 |
| | | | | | | | Fees | Commission | Third party commission | Performance fee | | | |
| scalable capital | irrelevant | irrelevant | 492.856.329 | 63,1% | 311.025.838 | | 0,75% | | 0,25% | 0 | 2.332.693,79 | | 2.332.694 |
| wikifolio | irrelevant | irrelevant | | 32,0% | 157.905.426 | | 0,95% | | 0-30% | | 1.500.101,54 | 789.527 | 2.289.629 |
| vaamo | irrelevant | irrelevant | 1.130.207.819 | 4,9% | 23.925.064 | | 0,79% | | 0,30% | 0 | 189.008,01 | | 189.008 |
| | | | | | | | Fees | Commission | Interest to Investor | Fee from interest retained | | | |
| smava | irrelevant | irrelevant | | 60,2% | 680.040.298 | | 0,00% | 2,5-3% | 3,0-7,9% | 2% | 18.701.108 | 32.641.934 | 19.353.947 |
| auxmoney | irrelevant | irrelevant | 4.996.821.521 | 39,8% | 450.167.521 | | 2,50 p.M. | 2,95% | 3,0-7,9% | 2% | 13.279.942 | 22.058.209 | 13.721.106 |
| weltsparen | irrelevant | irrelevant | | 12,3% | 613.379.047 | | | | 0 | 0 | | | - |
| ZINSPILOT | irrelevant | irrelevant | | 78,7% | 3.924.628.537 | | | | 0 | 0 | | | - |

Appendix 5 – Transcripts of Interviews

**Interview Protocol 001**

**Date:     April 19th, 2017**

**Interviewee:     Anna R.**

**Classification:     Traditionalist**

CW:     Dear Anna, thanks a lot for taking the time for this follow-up interview on my fintech study, really appreciate your support!

AR:     No problem, my pleasure.

CW:     Please let me confirm your demographic information first – you are born in 1987, live in Berlin, have a Master's degree and are employed full time?

AR:     Yes, that's correct. Though I actually pursue a second Master's degree at the same time in evening classes, but that takes less than 20% of my time.

CW:     Alright, thanks for clarifying that. May I ask what you are studying, and which Master you have completed already?

AR:     Sure. I have completed my first Master in Culture & Business and am currently pursuing a second degree in Sustainable Development.

CW:     Sounds great. Are you working in a related field then?

AR:     In a way – I am a HR strategy adviser for a finance holding, and we do have some operations in developing economies, which is why I wanted to deepen my understanding of the field.

CW:     Understand. Did working in a finance holding have any influence on taking part in my study to begin with?

AR:     No, not really. To be perfectly honest I was just curious and had some free time at work, that's why I participated. And since I did a number of interviews for my own degree, I understand how important it is to have some volunteers… Fintechs generally are not really my specialty as you might have noticed.

CW:     Well, yes, actually – according to my calculations you are a perfect example for what we call a 'traditionalist', which was also your self-assessment; that was one of the reasons to select you as a representative of this sample group.

CW:     Since you consider yourself a traditionalist when it comes to technology in general, does that also reflect on your preferences in banking and financial matters generally? Let's say for the selection of your house bank?

AR:     I guess you could say that, yes. In fact I'm not even sure you'd say I "selected" my house bank; I opened the account when I was 16 together with my parents at the local Sparkasse (savings bank) in the village where I was born, and I still have it to this day. I never bothered to change it; though I did open a second account with an online bank to have more options accessing it since I moved away.

CW:     But switching banks entirely was never an option? Not even to another savings bank here in Berlin?

AR:     Not really. I mean, of course I could, but I don't see the point. I'm still visiting my parents frequently, so if anything needs to be done I just wait, and if there is anything really urgent I could probably just resolve it by phone or send an email or so.

CW:     Why did you open the other account with the online bank then?

AR:     Basically because it's completely free, and they also offer a free credit card that I can use to withdraw money without any charges, anywhere in the world. I travel frequently, so that's a big plus - at my savings bank I pay fees even in other cities in Germany, and abroad it can get especially pricey, if it works at all.

CW:     So would you say it was a conscious decision to reduce cost that motivated you?

AR:     Yes, definitely.

CW:      And if you could save money by moving any of your other banking products from your house bank somewhere else, wouldn't that essentially be the same motive?

AR:     Yes, it would. And actually your survey did inspire me to look into some of the providers and see if there is anything that might be useful. Especially the investment helpers seem interesting; now that I get no interest on my savings account I will need to look for alternatives at some point.

CW:     I understand. So does that mean you are not currently investing through your house bank?

AR:     I do, but only in investment funds issued by them directly. Stocks are too volatile for my personal risk appetite.

CW:     Ok, fair enough. But generally speaking, if a fintech would offer a superior solution at a good price, you would consider it?

AR:     Well, at least I wouldn't disregard it completely; but I would make sure to read about it and get a second opinion. With a bank at least I know who is behind it, with the fintechs it's a bit obscure.

CW:     What do you mean with obscure?

AR:     I mean, take my Sparkasse or Deutsche Bank for example. You know where they are, you can find them anywhere, walk in and talk to people, they have been around since forever and such. If they screw up I can just walk into the next consumer protection office and they will resolve it for me, or take them to court. Those fintechs have only been around for a short time, are often not even headquartered here in Germany and I don't know anything about them except for what they write on the website… If anything goes wrong, how will I know if I ever see my money again?

CW:     Alright, I get your point. But if the fintechs find a way to mitigate your concerns, let's say by having a German office or some guarantees provided by reputable companies, you would consider them generally?

AR:     Yes, if that's the case and I can really save significant amounts, I would at least give it a try.

CW:     Awesome, thanks for clarifying that. Does that in turn mean you have never used a fintech?

AR:     No that's not entirely true; I have in fact used Sofortüberweisung to pay an invoice, but that was quite a while ago.

CW:     So what convinced you to use them?

AR:     Actually, they were the recommended way to pay in an online shops I used, and I wasn't even aware it was a different company. Only when I was forwarded to the login page of my bank's online banking terminal I actually became aware that this process is different than usual.

CW:     But that didn't concern you further?

AR:     It bothered me a bit, but since it was the only possibility to prepay my order for instant shipment, I decided to continue. And since I got an immediate confirmation from the online shop I figured it's not that big a deal after all. And since they are still charging my bank account at Sparkasse, I felt like if anything goes wrong I can just go there and have the charges reversed, so I'm always in control.

CW:     So with a deposit originator or investment platform where you would send them the money, it would be a concern?

AR:     Absolutely. Loans are one thing, if they give me money that I need to back, I am still in charge of the process. But if I lend my money to them, I have to make sure they are trustworthy and don't just run away.

CW:     So if I get that right - control, trust and security would be your keywords when it comes to financial decisions?

AR:     Yes, absolutely. I don't mind taking calculated risks, but I need full transparency to make up my mind.

CW:     Thanks a lot, that really helps me to get a clearer picture for my dissertation, many thanks for your time!

AR:     My pleasure, all the best and good luck!

---

**Interview Protocol 002**

**Date:      April 20th, 2017**

**Interviewee:       Matthias S.**

**Classification:      Traditionalist**

CW:     Dear Matthias, thanks a lot for taking the time for this follow-up interview on my fintech study, really appreciate your support!

MS:     Oh that's absolutely no problem, really.

CW:     Please let me confirm your demographic information first – you are born in 1971, live in Hamburg, have a Master's degree and are employed full time?

MS:     Yes, exactly.

CW:     Alright, thanks for your confirmation. May I ask in which field you are employed?

MS:     I work for the seaport administration, in cargo logistics - container routing, freight controlling, on- and offloading of vessels, those sort of things.

CW:     Wow, that sounds super interesting! I guess that must be a real high-tech job these days?

MS:     Well, yes and no. In terms of the software we are using, certainly - everything is logged in the SAP and monitored from the onset - we know which type of vessel loaded with which type of commodities is arriving on which particular day from which originating port, and so forth. But the people we are dealing with on the regular are oftentimes still "old sailor types" and they hate all the modern stuff.

CW:     Oh wow, I wasn't aware of that.

MS:     You need to have a talent for managing people, otherwise it is frustrating to no end...

CW:     Sure sounds like it! So, may I just be blunt and ask you how you developed an interest in participating in my fintech survey?

MS:     Oh yes, I saw that you posted it in the Lean Six Sigma discussion group on Xing that I'm following, and since fintechs are all over the news recently, I just wanted to check it out.

CW:     They sure are, and have been slowly emerging over the past 10 years if not more. Though I agree that the newsflow in the last 1-2 years has skyrocketed. Have you made any personal experience with them yet?

MS:     Well when I read your overview several of the names sounded familiar, but I wasn't 100% sure. Paypal and Smava definitely, they are showing a lot of advertisements on TV these days... But I haven't used any.

CW:     Any particular reason why you didn't, if you don't mind me asking?

MS:     No, of course not. Otherwise I wouldn't be talking to you, right?

CW:     Sounds logical.

MS:     So yes, well, actually I didn't use them because I didn't know them, and I don't trust those new companies too much. They need to prove that they are worth their while before I bother about it.

CW:     What does it take to prove it for you?

MS:     Good question. I think once they have gathered enough customers and have been around for some years without going bankrupt…

CW:     And how would you find out if they have managed to get sufficient customers?

MS:     Oh, I guess if enough of my friends are telling me about it, it would be a fair point to assume they are gaining traction… That and articles in Handelsblatt and Wirtschaftswoche [two popular business and trade newspapers in Germany], they seem to be reporting quite a bit on fintechs. The ones big enough to make headlines continuously can't be that bad.

CW:     Could I say that a positive reputation and good PR is it what attracts you then?

MS:     The positive reputation definitely.

CW:     And how long would you say a company needs to be around for you to consider them trustworthy enough to use them, since you brought it up?

MS:     That is a difficult question to answer. At least 5 years I would say, if not more… I really am a creature of habit, actually. Probably even up to 10 years, when they really become mainstream.

CW:     So what else would make you consider fintechs as opposed to traditional banks?

MS:     The conditions for sure. These days, banks are charging you more and more instead of less! Since 2 years already I don't get any interest on my savings account. Not even 0.1%, that's a joke. But new fees come up here and there… They are quick to invent those. And when I had to take a loan for a new car recently it was more expensive than I thought it would be, considering they get free money from the government!

CW:     But taking the loan from a fintech wasn't an option?

MS:     To be fair, I was shopping around for a better rate on check24.de [popular comparison website], but ultimately I just called my client advisor and asked him to see what he can do for me, and then I didn't want to bother any further.

CW:     What client adviser was that?

MS:     At the local branch of my bank.

CW:     Would you mind telling me which bank you are using?

MS:     Actually I would rather keep that to myself. One of the countrywide bigger ones.

CW:     Ok no problem, that will suffice for sure. So instead of the car loan, current and savings account, are you using any other products there?

MS:     Everything so far. I have a mortgage for almost 5 years now, and get most of my insurances there as well. It's simply the most convenient option.

CW:     But still you were saying that a fintech with good conditions might have a chance to move your business elsewhere?

MS:     Well, yes. It might, if they are really good.

CW:     Thanks a lot for your time, appreciate your support!

MS:     No problem. All the best for your paper!

**Interview Protocol 003**

**Date:**    **April 20th, 2017**

**Interviewee:**    **Patrick K.**

**Classification:**    **Balanced**

CW:    Dear Patrick, thanks a lot for taking the time for this follow-up interview on my fintech study, really appreciate your support!

PK:    Sure thing!

CW:    Please let me confirm your demographic information first – you are born in 1985, live in Berlin, have a Bachelor's degree and are self-employed?

PK:    Exactly.

CW:    Alright, thanks for the confirmation. So what business are you in?

PK:    I'm a filmmaker - with my agency [redacted] we are mainly doing short company profile videos, commercials, scenes for educational or documentary use, those kinds of things.

CW:    Wow, sure sounds interesting! So I guess you need to be always up to date with recent technological developments and stuff?

PK:    Yes, precisely. We work in a very fast-paced industry, and staying behind technological advancements can ultimately mean that we are losing customers to competitors simply because we are unable to fulfill their needs.

CW:    I see… And actually that makes me curious as to why you consider yourself to be "balanced" rather than an early adopter?

PK:    Well, that's the point. Not every new technology is going to stay, and as a business owner I need to invest in the future, not in a dead horse… Toshiba could afford to lose out on the Bluray vs. HD-DVD battle. I wouldn't survive that. So I observe the market and invest when the time is right.

CW:    That makes sense. And when it comes to personal financial decisions, you follow the same approach? Like with your house bank and other banking products?

PK:    Yes, I do. I have long since moved all my accounts to online banks, and it has been 5 years or more since I set foot into an actual bank.

CW:    What was the primary motivation behind it? And did you experience any downside to it?

PK:    Cost savings, plain and simple. I used to have an account with Deutsche Bank while I was studying, and it was completely free. The moment I graduated, they changed my account model to a paid one - without even informing me! I only noticed at the end of the quarter that they charged me almost 50 Euros for account maintenance fees, card fees for my mastercard, and even charged for every single remittance. It was ridiculous… I closed everything down instantly and moved it away.

CW:    That sure sounds like a good reason… Apart from the monetary aspects though, did you have any other motivation?

PK:    Well, the people at Deutsche Bank didn't really give me the feeling that I was welcome, more like an obstacle. They are clearly not interested in the smaller clients.

CW:    So bad customer service?

PK:    Yes, essentially. Maybe not outright bad, but very slow to say the least.

CW:    Got it. So, any downsides to using the online bank?

PK:     Not really, or nothing unexpected rather. Of course you don't have a personal contact, and the products are extremely standardized - custom solutions are pretty much impossible. Take it or leave it.

CW:     So a lack in flexibility?

PK:     Yes, in a way. But I know that it is required to get the free services, so I don't complain - instead when I need something that my bank can't offer, I go search it some other place. And that's where fintechs come in handy.

CW:     To add flexibility?

PK:     Yes, and make products available that my bank simply doesn't have.

CW:     Do you have an example for me?

PK:     Sure, let's take wikifolio for example. My bank has a brokerage account; but whatever I want to do, I have to do on my own. They don't have any stock recommendations, trading advice or whatever; so I follow someone on wikifolio in order to get a good performance on my investments. That's amazing, and it's happening in real time!

CW:     Sure sounds great. How long have you been using them?

PK:     Just over half a year, and I love it.

CW:     No downsides to it?

PK:     Not that I see it, no. So far, so good.

CW:     May I ask how much you actually trade with them? You indicated that you use them in your response to my questionnaire, but didn't provide any volumes.

PK:     Yes that's true, I didn't want to to be stored in any database online, I'm not too comfortable with that. But Usually I invest 250-350 EUR per months, depending on the income my company is generating. Sometimes when I get some large contracts that pay a lump sum I might to larger investments, but that doesn't happen too frequently, unfortunately.

CW:     Wikifolio has not even been around for 5 years. Doesn't that bother you?

PK:     Not much, no. After all I'm following an experienced trader, that's far more important to me. And if I decide to follow an investment strategy or not is still my own decision, after all.

CW:     What would you say if all investments would be fully automated? Is that something you could get behind as well?

PK:     Hmm, I'm not so sure. I still feel more comfortable when I'm in control about the final decision myself.

CW:     So what would happen to convince you otherwise?

PK:     [laughs] - I guess I would need to watch the market for a while and see that it works… Maybe I should try?

CW:     I didn't mean to imply you should, just want to identify motivators for you to switch to new technologies.

PK:     Yes I got that… Well as I said, for me it's about knowing when something is matured to a certain point. Doesn't have to be a final solution, but at least a solid foundation that has lost its childhood diseases. Once I can see that it works, and provides and actual value for me, I'm sold.

CW:     So, which other FinTechs are you using?

PK:     PayPal quite frequently, most online shops I'm buying from accept it, and it's just convenient for me. Other than that, I've been checking out N26 and lending circle, but haven't used them. Sofortüberweisung comes up once in a while.

CW:     Thanks a lot for the information, appreciate your time!

PK:    Anytime, take care!

---

**Interview Protocol 004**

**Date:    April 22nd, 2017**

**Interviewee:    Carsten F.**

**Classification:    Balanced**

CW:    Dear Carsten, thanks a lot for taking the time for this follow-up interview on my fintech study, really appreciate your support!

CF:    That's fine, glad to be of additional use.

CW:    Please let me confirm your demographic information first – you are born in 1994, live in Cologne and go to university for your Bachelor's degree?

CF:    Sounds about right; though I actually live in Frechen, which is a small village just next to Cologne. I study in Cologne though, and that's where I spend most of my time.

CW:    Alright, thanks for clearing that up! May I ask what you are studying?

CF:    Electrical engineering and mathematics, though I will likely drop maths. Getting too much doing both at the same time.

CW:    Oh sure, I can understand. Is it also the engineering part that got you interested in new technologies and such?

CF:    No, not that much actually. Technologies have always been a hobby, if anything they got me to pursue the studies, but not the other way around.

CW:    Ah, got you. And fintechs were a natural extension of your love for technology?

CF:    Well, I guess you can say that. To be perfectly honest I never before considered fintechs to be something special, they are just new companies coming up with some nice features, so naturally I go for it.

CW:    But still you consider your profile to be "balanced"?

CF:    Oh, ok. That. Yes, I chose balanced because you were asking about technology in general; and I don't always need to have the newest phone, don't need to use the newest instant massaging app or social media site… I will use them eventually if I see a reason for it, but I don't have to be the first one. Some of my friends are like that, buy the newest iPhone, a new Laptop every year, have a million different accounts on twitter, snapchat, facebook, mastodon, kik and whatever and annoy everybody. I don't really like that.

CW:    Fair point, duly noted. But you mentioned something interesting – fintechs for you are just "other companies". So a bank for you is just a company as well, nothing else?

CF:    Yes, absolutely. They are a company in charge of my money, and if I'm not happy with them, I go somewhere else. Just like that.

CW:    Alright, good to know! Can I ask which bank you are using then yourself?

CF:    I recently switched to N26, and love it so far.

CW:    Oh wow, an actual fintech – can I ask why you chose them?

CF:    Sure. I was with DKB before [author's note: Deutsche Kreditbank Berlin is another online-only bank], but they recently introduced a withdrawal limit – I can only take out a minimum of 50 EUR from the ATM, and that sucks. I don't want to carry so much cash around, and N26 has no such limits. And I love the app! I get instant notifications on my phone for any transactions, I can send money to my friends as simple as a text message and request it from them as well, etc. Basically I have everything I need right here in my phone.

CW:     Sure sounds convincing! So the flexibility was your primary reason to switch? Cost or any other factors didn't play a role?

CF:     I'm naturally cost sensitive, I only make about 500 EUR per month with my side job, so I think twice what I do. That's why I still live with my parents as well.

CW:     Alright, fair point. You still seem to be using some fintechs though, how come?

CF:     That's exactly the point. I try to maximize the use of my available money, so instead of buying stuff on Amazon I order it from China with Aliexpress or through ebay from the USA. So especially Paypal and Transferwise come in super handy.

CW:     I see. Do you think that might change after your studies, when money becomes less of a bottleneck?

CF:     I wouldn't see why. I like to keep up to date and try out new things, and when it helps me to save money on top, even better!

CW:     So are there any services or products you would not get from fintechs?

CF:     I'm not sure. Right now I wouldn't see any… Maybe if I were to buy a house or something it would help me to discuss with a real person, that kind of big investments. But then again, if a fintech offers exactly that and has a good reputation, I might not need it. I guess I will have to see what there is when the time comes, not right now.

CW:     But since you manage reputation already – that is something you factor in generally?

CF:     Oh yes, definitely. Independent reviews are one of my key decision making criteria; I learned that the hard way when buying things from abroad – when you have no direct access or control, it's all about reviews.

CW:     Thanks a lot for the additional input, really appreciate it!

CF:     Let me know if you need anything else. Bye!

---

**Interview Protocol 005**

**Date:     April 22nd, 2017**

**Interviewee:     Kathrin L.**

**Classification:     Early Adopter**

CW:     Dear Kathrin, thanks a lot for taking the time for this follow-up interview on my fintech study, really appreciate your support!

KL:     Nice to hear from you, no problem!

CW:     Please let me confirm your demographic information first – you are born in 1982, live in Frankfurt, have a Master's degree and are employed full time?

KL:     Correct.

CW:     Thanks for confirming that. May I ask where you work at?

KL:     I'm a business analyst working for DEKA Bank [author's comment: DEKA is an investment fund administrator and wholly owned by the German savings' bank association]

CW: Oh wow, that's interesting! And now it actually makes me curious as to why you are using an online bank instead of a savings bank when you are part of the association…

KL: Well, I guess I do both technically. I have my account with 1822direct, which is the online-daughter of the Frankfurter Sparkasse. Though many of my colleagues are using banks completely outside of our network, that's not really a big thing these days.

CW: Got it. Can you use the full product range of a Sparkasse through the online bank though?

KL: It's a bit limited, I have a savings account there, brokerage, credit cards and access to consumer loans; though at the time I don't have one.

CW: I see. You mentioned in the questionnaire though that you are using a mortgage for the time being; may I ask where you took that?

KL: That one is actually with Frankfurter Sparkasse together with LBS [author's note: LBS – Landes Bau Sparkasse – is a building society owned by the Sparkassen Association]. As an employee within the network I got preferred rates there.

CW: Oh alright, that's a nice feature for sure… I can see that in fact you mention cheap conditions as one of the key aspects that would motivate you to take your business elsewhere, is that right?

KL: Yes, that's true.

CW: Is there anything else that would make you reconsider your bank of choice? Or at least using a third party to take care of part of the functions?

KL: Well, certainly. What I'm especially keen about are brand new things that are revolutionizing the industry. That's where my analyst's heart beats faster.

CW: Do you have an example for me?

KL: The biggest one would be the blockchain – if properly used, they could make core banking functions like lending, borrowing and transactions irrelevant. Imagine that – a completely decentralized world, without any effort needed for clearing and settlements, in near-real time. That is amazing!

CW: So new features, efficiency and 24/7 accessibility?

KL: Yes, in a nutshell. I'm tired of being bound to traditional opening ours and such, banking doesn't have to be a bureaucratic affair in this day and age.

CW: I can see why you consider yourself an early adopter…

KL: Oh totally. And that's not just in banking, I am generally ready to explore just about anything, and I find technology inspirational. Especially if it can make my life easier.

CW: Have you ever used any finance-related products that you were initially totally hyped about but then realized they are not as great as you thought? Or any fintech that you started using that went out of business again?

KL: Yes, actually that happens quite frequently. I sign up for new services when I come across them, and try using them for a bit. If I decide they are not for me, I will simply disregard them shortly after. One such example was savedroid. They are here from Frankfurt, and I met one of the founders on a conference, so I got into the beta-program. Basically whenever you use your card to pay for something, they will take an additional percentage of the purchase out of your card and transfer it to your savings account. They promise to help building savings without me even knowing, but that's simply not the case. When I pay 20 EUR for something and 22 are charged to my card, I find that pretty annoying. And after 2 months all I got in the savings account was about 80 EUR, while I regularly put a couple hundred on a savings plan… So they didn't fulfill the objective in my opinion, and I stopped using it again.

CW: Interesting! But that doesn't stop you from trying something new?

KL:     Not at all! Experiencing new things are what keeps me going. It's a lifestyle, in a way.

CW:     So which one is your favorite fintech then?

KL:     From the list you provided it is definitely PayPal. That was a game changer for me. Instant payments worldwide, plus a guarantee & refund mechanism if the recipient of the money doesn't keep his end of the deal? That's genius. Couldn't be happier. But I have also started using N26 a while ago, just to see what the new players are up to, and the app is really amazing – wouldn't want to move my whole account just yet simply because I have so many payments linked to my 1822 account, but as a transaction account N26 is awesome for sure. I have also been using smava in the past as an investor, and funded some smaller individual loans, though I wasn't too happy with that one.

CW:     Why what happened with smava?

KL:     One of the loans ended up in default, and the communication from smava's end was rather bad. Utimately I lost almost 30% of my investment; I didn't make that much from investments in the other handful of loans, so I gave it up.

CW:     Certainly a valid reason to give up on them, I understand. May I ask where you inform yourself about novelties on the market?

KL:     One of the good things of living in Frankfurt is that you are really close to the market pulse [author's note: Frankfurt is German's financial hub]. There are conferences every other week, trade fairs, we get infomails and magazines daily, etc. There is so much innovation going on I can barely follow up with it.

CW:     Awesome, thanks a lot for all the information!

KL:     Anytime, glad to help!

---

**Interview Protocol 006**

**Date:**      April 22nd, 2017

**Interviewee:**      Markus F.

**Classification:**      Early Adopter

CW:     Dear Markus, thanks a lot for taking the time for this follow-up interview on my fintech study, really appreciate your support!

MF:     Sure thing, one of my favorite topics.

CW:     Please let me confirm your demographic information first – you are born in 1988, live in Berlin, have a Bachelor's degree and are self-employed?

MF:     Not exactly – I have not yet completed my studies, and am too busy working to tell you whether I will actually manage anytime soon. But whatever.

CW:     OK, thanks for clarifying that. And now I'm curious – you said that fintechs are one of your favorite topics? How come?

MF:     I started out as a software developer (which I still do) but am mainly operating as a software security tester, and earn a living by reviewing source code, perform penetration testing of IT systems etc. Basically my employers hire me to try breaking into their systems as a means to improve their security standards and identify potential vulnerabilities before any unauthorized person does and creates real damage – either by stealing customer data, business secrets or money, or through the ensuing reputational damage. And fintechs are a natural target; after all they 'live' entirely online, are available around the clock from virtually everywhere in the world, and they can only remain on top of the group by being one step faster than any of their competitors – as a result, not all tests are always performed to the highest standards. That creates a lot of opportunities for me, so I make sure to stay up to date with all current developments.

CW:     Wow, that really is a good reason to follow the market closely. But yet at the same time you are actively using a lot of the fintechs, despite the risks you mention and observe at the same time... Or are you using only companies that you have personally reviewed?

MF:     I can't reveal which companies I reviewed and which ones I didn't, that's a business secret. But even if I don't personally review all of them, there are certain standards to look out for that tell me if a company did their homework, such as a strong encryption, SSL certificates, dedicated server architecture and so on. If they are also located in a jurisdiction that pays attention to security standards, I am ready to trust them enough to start using them, if only with some transactions in the beginning to find out if they are suitable for me.

CW:     Ok, those are super interesting insights... So would you say that trust and security are the key things to look out for when deciding whether to work with a fintech or not?

MF:     Well, not exactly. In the first place I need to have a use case for them – they need to add a benefit to me personally or my company, be it an entirely new function, higher revenue, lower cost or greater flexibility. If that doesn't exist, there is no real need for me to use the platform other than testing them out generally in order to approach their IT department with a business proposal from my end to review their system for security flaws.

CW:     Got it. And may I now ask which bank you decided to trust with your business?

MF:     I chose Fidor; though they are far from perfect.

CW:     Why, what happened? And why did you choose them?

MF:     They offer completely free business accounts, and operate as a licensed and registered bank in Germany, so they have a deposit insurance in place and comply with our regulation; so that's a big plus. The downside is that they haven't been around for too long, and all of their processes are still in development. Reaching out to an actual human if something needs to be sorted out can take forever; that's something they will definitely have to improve.

CW:     Wouldn't you think it is a general tradeoff between low cost and high service quality?

MF:     To an extent, certainly. But with modern technology, machine learning, predictive analytics and all that, there is still a lot of room for improvement without creating more overhead.

CW:     Do you have any examples for that?

MF:     Sure – look at Smava for example. They are a p2p lending operator, people can use them to take loans funded through other users of the platform, while Smava themselves are providing the security layer – they are scoring the borrower automatically through their automated mechanisms paired with data from Schufa [author's note: Schufa is the largest German credit bureau] and demographic information, and assign a risk-based interest rate for the loan. There is no personal involvement of any employee required. Payout of the loan, repayments etc. are all totally automated.

CW:     Sounds like you have some experience with them?

MF:     In the past; I am not currently active there. Though I did recommend them to some friends recently, and they are quite satisfied.

CW:     What was your reason to use them?

MF:     I needed a small business loan to get another side company started that I have. Production based, with goods made in China. My house bank at the time wouldn't hear from it, so I had to look elsewhere – in the end I got a few quotes from banks and one from Smava that was 30% cheaper. Easy decision, and no regrets whatsoever.

CW:     Great, that's really good to hear. Would you say that this positive experience contributed to the continued advocacy for fintechs?

MF:     Sure, there are definitely spillover effects, though I keep a natural awareness up at all time. I embrace new technology, but as I said I am well aware of potential flaws and downsides – and one of my key paradigms is never to put all eggs in one basket, especially not in such a fast-spinning industry. Otherwise you might wake up one day and find yourself between a rock and a hard place, if one of the key enablers for your business is simply gone. For private use that may not be as drastically, but thinking from my business's perspective, I can't afford any downtime.

CW:     So you always make keep alternatives at hand?

MF:     Absolutely, for each core company I am using, I have at least one alternative lined up, starting from the bank account itself. I love technology, but I'd never let it outpace me…

CW:     That sounds like a quote I might use in my dissertation.

MF:     I'd be delighted!

CW:     According to your participation in my study, you are also using PayPal, transferwise, wikifolio and N26, is that correct?

MF:     Yes I do, N26 as my private bank account, don't want to have one at the same bank as my business account. Simply in case there would be any issues I'd have access to money elsewhere. Transferwise was my alternative to PayPal, though actually I'm using both at the same time now, depending on what I need to do. Wikifolio is an interested concept and I signed up recently, though haven't come around to making actual investments yet. No time.

CW:     Thanks a lot for your input and taking the time, highly appreciate it.

MF:     No problem!

Appendix 6 – FinTech Questionnaire – Summary Tables

**Table 25: Ranking of FinTech Awareness**

| | Totals | | | | |
|---|---|---|---|---|---|
| | **Payment** | **Lending** | **Deposit** | **Invest** | **Banks** |
| **# Aware** | 1205 | 536 | 144 | 339 | 408 |
| **# Provider** | 14 | 10 | 4 | 13 | 5 |
| **Weighted Ø** | 86,07143 | 53,6 | 36 | 26,07692 | 81,6 |
| **Rank** | 1 | 3 | 4 | 5 | 2 |
| | | | | | |

*Source: Based on Processed Data from Survey*

**Table 26: Ranking of FinTech Use**

| | Totals | | | | |
|---|---|---|---|---|---|
| | **Payment** | **Lending** | **Deposit** | **Invest** | **Banks** |
| **# Aware** | 545 | 20 | 8 | 36 | 54 |
| **# Provider** | 14 | 10 | 4 | 13 | 5 |
| **Weighted Ø** | 38,92857 | 2 | 2 | 2,769231 | 10,8 |
| **Rank** | 1 | 5 | 5 | 3 | 2 |
| | | | | | |

*Source: Based on Processed Data from Survey*

**Table 27: Self-Identification of Participants**

| Self-Identification | Count | Percentage |
|---|---|---|
| An Early Adopter – 'I love te | 81 | 18,5% |
| Balanced – 'Once the techn | 303 | 69,0% |
| Traditionalist – 'I don't need | 55 | 12,5% |
| **Grand Total** | **439** | **100%** |

*Note: These numbers were taken before matching participants with the calculated categories, hence they do not reflect the number of 349 used elsewhere in the study.*

**Table 28: Reliability of Age Groups**

Reliability Statistics

| Cronbach's Alpha | Cronbach's Alpha Based on Standardized Items | N of Items |
|---|---|---|
| ,976 | 1,000 | 7 |

**Table 29: Breakdown Tables of Volume per Category**

| | | Transaction Volumes - Payment Providers | | | | | |
|---|---|---|---|---|---|---|---|
| Category | Number | Paypal | Transferwise | SofortÜberw. | klarna | Sum | Average |
| Pioneer | 50 | 42.745 | 19.950 | 10.045 | 1.365 | 74.105 | 1.482,10 |
| Migrator | 256 | 49.020 | 6.380 | 17.563 | 1.124 | 74.087 | 289,40 |
| Settler | 43 | - | - | 65 | - | 65 | 1,51 |
| **Total** | **349** | **91.765** | **26.330** | **27.673** | **2.489** | **148.257** | **424,81** |

| | | Transaction Volumes - Investment Helpers | | | | |
|---|---|---|---|---|---|---|
| Category | Number | scalable.capital | wikifolio | vaamo | Sum | Average |
| Pioneer | 50 | 2.750,00 | 1.250,00 | 250,00 | 4.250,00 | 85,00 |
| Migrator | 256 | 500,00 | 400,00 | - | 900,00 | 3,52 |
| Settler | 43 | - | - | - | - | - |
| **Total** | **349** | **3.250** | **1.650** | **250** | **5.150,00** | **14,76** |

| | | Transaction Volumes - FinTech Banks | | | |
|---|---|---|---|---|---|
| Category | Number | N26 | Fidor | Sum | Average |
| Pioneer | 50 | 25.150,00 | 4.445,00 | 29.595,00 | 591,90 |
| Migrator | 256 | 10.455,00 | 4.975,00 | 15.430,00 | 60,27 |
| Settler | 43 | - | - | - | - |
| **Total** | **349** | **35.605** | **9.420** | **45.025** | **129,01** |

| | | Transaction Volumes - Loans | | | |
|---|---|---|---|---|---|
| Category | Number | Auxmoney | Smava | Sum | Average |
| Pioneer | 50 | 6.500,00 | 3.200,00 | 9.700,00 | 194,00 |
| Migrator | 256 | 600,00 | 1.500,00 | 2.100,00 | 8,20 |
| Settler | 43 | - | - | - | - |
| **Total** | **349** | **7.100** | **4.700** | **11.800** | **33,81** |

| | | Transaction Volumes - Deposits | | | |
|---|---|---|---|---|---|
| Category | Number | ZINSPILOT | Weltsparen | Other | Sum | Average |
| Pioneer | 50 | | 26.500 | 12.500 | 39.000,00 | 780,00 |
| Migrator | 256 | 4.750 | 7.500 | | 12.250,00 | 47,85 |
| Settler | 43 | | | | - | - |
| **Total** | **349** | **4.750** | **34.000** | **12.500** | **51.250,00** | **146,85** |

*Note*: This table contains the number of <u>all</u> valid respondents, including those who opted to not disclose the transaction / investment / loan amounts. For the calulations in chapter 4, those have been disregarded and only inputs with tangible volume indicators were used.

| Participant | Valid participant? | Knowledge | | | | | | | Use | | | | | | | Subtotal |
|---|---|---|---|---|---|---|---|---|---|---|---|---|---|---|---|---|
| Weight | | 1 | 3 | 4 | 5 | 2 | | | 1 | 5 | 5 | 3 | 2 | | | Average |
| | | Payment | Lending | Deposit | Invest | Bank | Total | Weighted | Payment | Lending | Deposit | Invest | Bank | Total | Weighted | 23,885845 |
| 1 | Not Resident | 3 | 0 | 0 | 1 | 0 | 4 | #N/A | 0 | 0 | 0 | 0 | 0 | 0 | #N/A | #NA |
| 2 | Valid | 3 | 0 | 0 | 0 | 1 | 4 | 5 | 0 | 0 | 0 | 0 | 1 | 1 | 1 | 15 |
| 3 | Valid | 6 | 2 | 0 | 0 | 4 | 12 | 20 | 2 | 0 | 0 | 0 | 0 | 2 | 2 | 30 |
| 4 | Not Resident | 1 | 0 | 0 | 0 | 0 | 1 | #N/A | 0 | 0 | 0 | 0 | 1 | 1 | #N/A | #NA |
| 5 | Valid | 4 | 2 | 2 | 0 | 1 | 7 | 12 | 2 | 0 | 0 | 0 | 0 | 2 | 3 | 32 |
| 6 | Valid | 2 | 2 | 0 | 1 | 0 | 5 | 13 | 2 | 0 | 0 | 0 | 0 | 2 | 2 | 23 |
| 7 | Valid | 4 | 3 | 2 | 2 | 2 | 13 | 35 | 3 | 1 | 1 | 0 | 1 | 6 | 15 | 110 |
| 8 | Bogus | 2 | 0 | 0 | 0 | 0 | 2 | #N/A | 0 | 0 | 0 | 0 | 1 | 1 | #N/A | #NA |
| 9 | Valid | 2 | 0 | 0 | 1 | 0 | 3 | 7 | 2 | 1 | 0 | 0 | 0 | 3 | 7 | 42 |
| 10 | Valid | 2 | 0 | 0 | 0 | 0 | 2 | 2 | 0 | 0 | 0 | 0 | 1 | 1 | 1 | 7 |
| 11 | Ace | 0 | 0 | 0 | 0 | 0 | 0 | #N/A | 0 | 0 | 0 | 0 | 0 | 0 | #N/A | #NA |
| 12 | Valid | 3 | 0 | 0 | 0 | 1 | 4 | 5 | 2 | 0 | 0 | 0 | 0 | 2 | 2 | 15 |
| 13 | Valid | 6 | 2 | 0 | 0 | 4 | 12 | 20 | 2 | 0 | 0 | 0 | 0 | 2 | 2 | 30 |
| 14 | Valid | 2 | 0 | 0 | 0 | 0 | 2 | 2 | 1 | 0 | 0 | 0 | 0 | 1 | 1 | 7 |
| 15 | Valid | 2 | 0 | 0 | 1 | 0 | 3 | 7 | 2 | 1 | 0 | 0 | 0 | 3 | 7 | 42 |
| 16 | Not Resident | 1 | 0 | 0 | 0 | 0 | 1 | #N/A | 0 | 0 | 0 | 0 | 1 | 1 | #N/A | #NA |
| 17 | Valid | 5 | 2 | 0 | 6 | 2 | 15 | 45 | 2 | 0 | 0 | 0 | 0 | 2 | 2 | 55 |
| 18 | Valid | 2 | 1 | 0 | 0 | 0 | 3 | 5 | 1 | 0 | 0 | 0 | 0 | 1 | 1 | 10 |
| 19 | Valid | 2 | 0 | 0 | 0 | 0 | 2 | 2 | 1 | 0 | 0 | 0 | 0 | 1 | 1 | 7 |
| 20 | Valid | 2 | 0 | 0 | 0 | 2 | 4 | 6 | 1 | 0 | 0 | 0 | 0 | 1 | 1 | 11 |
| 21 | Valid | 3 | 4 | 0 | 1 | 2 | 10 | 24 | 1 | 0 | 0 | 0 | 0 | 1 | 1 | 29 |
| 22 | Valid | 2 | 0 | 0 | 1 | 0 | 3 | 7 | 2 | 1 | 0 | 0 | 0 | 3 | 7 | 42 |
| 23 | Valid | 4 | 2 | 1 | 1 | 2 | 10 | 23 | 1 | 0 | 0 | 0 | 0 | 1 | 1 | 28 |
| 24 | Valid | 3 | 2 | 0 | 0 | 3 | 8 | 15 | 1 | 0 | 0 | 0 | 0 | 1 | 1 | 20 |
| 25 | Valid | 5 | 5 | 1 | 3 | 2 | 16 | 43 | 1 | 0 | 0 | 1 | 1 | 3 | 6 | 73 |
| 26 | Valid | 2 | 4 | 1 | 4 | 1 | 12 | 40 | 1 | 1 | 0 | 1 | 0 | 3 | 9 | 85 |
| 27 | Valid | 4 | 0 | 0 | 2 | 2 | 8 | 18 | 2 | 0 | 1 | 1 | 1 | 5 | 12 | 78 |
| 28 | Valid | 2 | 2 | 0 | 1 | 0 | 5 | 13 | 2 | 0 | 0 | 0 | 0 | 2 | 2 | 23 |
| 29 | Valid | 2 | 1 | 0 | 0 | 0 | 3 | 5 | 0 | 0 | 0 | 0 | 1 | 1 | 1 | 15 |
| 30 | Valid | 2 | 0 | 1 | 0 | 0 | 3 | 6 | 2 | 0 | 0 | 0 | 0 | 2 | 2 | 15 |
| 31 | Valid | 1 | 0 | 0 | 0 | 0 | 1 | 1 | 0 | 0 | 0 | 0 | 1 | 1 | 1 | 11 |
| 32 | Valid | 4 | 3 | 2 | 5 | 2 | 16 | 50 | 1 | 1 | 0 | 0 | 0 | 2 | 6 | 80 |
| 33 | Valid | 2 | 0 | 1 | 0 | 0 | 3 | 6 | 1 | 0 | 0 | 0 | 0 | 1 | 1 | 20 |
| 34 | Valid | 2 | 4 | 0 | 0 | 0 | 6 | 14 | 2 | 0 | 0 | 0 | 0 | 2 | 2 | 24 |
| 35 | Not Resident | 2 | 0 | 0 | 0 | 0 | 2 | #N/A | 0 | 0 | 0 | 0 | 1 | 1 | #N/A | #NA |
| 36 | Valid | 3 | 2 | 1 | 3 | 0 | 9 | 28 | 1 | 0 | 0 | 0 | 0 | 1 | 1 | 32 |
| 37 | Valid | 2 | 1 | 2 | 2 | 2 | 9 | 27 | 1 | 0 | 0 | 0 | 0 | 1 | 1 | 32 |
| 38 | Valid | 2 | 0 | 0 | 0 | 1 | 3 | 4 | 2 | 0 | 0 | 0 | 1 | 3 | 4 | 24 |
| 39 | Valid | 1 | 2 | 2 | 9 | 1 | 15 | 62 | 0 | 0 | 0 | 1 | 0 | 1 | 3 | 77 |
| 40 | Valid | 3 | 0 | 0 | 0 | 2 | 5 | 7 | 1 | 0 | 0 | 0 | 0 | 1 | 1 | 12 |
| 41 | Not Resident | 1 | 0 | 0 | 0 | 1 | 2 | #N/A | 0 | 0 | 0 | 0 | 1 | 1 | #N/A | #NA |
| 42 | Valid | 3 | 0 | 0 | 0 | 0 | 3 | 3 | 3 | 0 | 0 | 0 | 0 | 3 | 3 | 20 |
| 43 | Valid | 2 | 4 | 1 | 4 | 1 | 12 | 40 | 1 | 1 | 0 | 1 | 0 | 3 | 9 | 85 |
| 44 | Bogus | 13 | 9 | 3 | 12 | 4 | 41 | #N/A | 1 | 0 | 0 | 1 | 1 | 3 | #N/A | #NA |
| 45 | Valid | 3 | 2 | 0 | 0 | 2 | 7 | 13 | 0 | 0 | 0 | 1 | 1 | 2 | | 23 |

| Participant | Valid participant? | Loyalty | | | | | | Satisfaction | | | | | | Subtotal |
|---|---|---|---|---|---|---|---|---|---|---|---|---|---|---|
| Weight | | | | | | | | | | | | | | Average |
| | | Payment | Lending | Deposit | Invest | Bank | Average >0 | Payment | Lending | Deposit | Invest | Bank | Average >0 | 28,67087 |
| 1 | Not Resident | 0,0 | 0 | 0 | 0 | 0 | #N/A | 0 | 0 | 0 | 0 | 0 | #N/A | #NA |
| 2 | Valid | 0,0 | 0 | 0 | 0 | 8 | 8,0 | 0 | 0 | 0 | 0 | 9 | 9,0 | 7,2 |
| 3 | Valid | 54,5 | 0 | 0 | 0 | 0 | 54,5 | 5 | 0 | 0 | 0 | 0 | 5,0 | 27,25 |
| 4 | Not Resident | 0,0 | 0 | 0 | 0 | 0 | #N/A | 0 | 0 | 0 | 0 | 0 | #N/A | #NA |
| 5 | Valid | 15,0 | 0 | 0 | 0 | 24 | 19,5 | 4,5 | 0 | 0 | 0 | 7 | 5,8 | 11,2125 |
| 6 | Valid | 55,0 | 0 | 0 | 0 | 0 | 55,0 | 5 | 0 | 0 | 0 | 0 | 5,0 | 27,5 |
| 7 | Valid | 12,0 | 6 | 6 | 0 | 18 | 10,5 | 8,6666667 | 8 | 10 | 0 | 7 | 8,4 | 8,8375 |
| 8 | Bogus | 0,0 | 0 | 0 | 0 | 0 | #N/A | 0 | 0 | 0 | 0 | 0 | #N/A | #NA |
| 9 | Valid | 21,0 | 36 | 0 | 0 | 0 | 28,5 | 9,5 | 10 | 0 | 0 | 0 | 9,8 | 27,7875 |
| 10 | Valid | 8,0 | 0 | 0 | 0 | 0 | 8,0 | 8 | 0 | 0 | 0 | 0 | 8,0 | 6,4 |
| 11 | Ace | 0,0 | 0 | 0 | 0 | 0 | #N/A | 0 | 0 | 0 | 0 | 0 | #N/A | #NA |
| 12 | Valid | 2,0 | 0 | 0 | 0 | 0 | 2,0 | 7 | 0 | 0 | 0 | 0 | 7,0 | 1,4 |
| 13 | Valid | 98,5 | 0 | 0 | 0 | 0 | 98,5 | 4,5 | 0 | 0 | 0 | 0 | 4,5 | 44,325 |
| 14 | Valid | 120,0 | 0 | 0 | 0 | 0 | 120,0 | 9 | 0 | 0 | 0 | 0 | 9,0 | 108 |
| 15 | Valid | 48,0 | 24 | 0 | 0 | 0 | 36,0 | 5,5 | 9 | 0 | 0 | 0 | 7,3 | 26,1 |
| 16 | Not Resident | 0,0 | 0 | 0 | 0 | 0 | #N/A | 0 | 0 | 0 | 0 | 0 | #N/A | #NA |
| 17 | Valid | 71,0 | 0 | 0 | 0 | 0 | 71,0 | 7 | 0 | 0 | 0 | 0 | 7,0 | 49,7 |
| 18 | Valid | 60,0 | 0 | 0 | 0 | 0 | 60,0 | 9 | 0 | 0 | 0 | 0 | 9,0 | 54 |
| 19 | Valid | 24,0 | 0 | 0 | 0 | 0 | 24,0 | 9 | 0 | 0 | 0 | 0 | 9,0 | 21,6 |
| 20 | Valid | 72,0 | 0 | 0 | 0 | 0 | 72,0 | 10 | 0 | 0 | 0 | 0 | 10,0 | 72 |
| 21 | Valid | 60,0 | 0 | 0 | 0 | 0 | 60,0 | 1 | 0 | 0 | 0 | 0 | 1,0 | 6 |
| 22 | Valid | 36,0 | 40 | 0 | 0 | 0 | 38,0 | 9,5 | 10 | 0 | 0 | 0 | 9,8 | 37,05 |
| 23 | Valid | 36,0 | 0 | 0 | 0 | 0 | 36,0 | 8 | 0 | 0 | 0 | 0 | 8,0 | 28,8 |
| 24 | Valid | 16,0 | 0 | 0 | 0 | 0 | 16,0 | 8 | 0 | 0 | 0 | 0 | 8,0 | 12,8 |
| 25 | Valid | 120,0 | 0 | 0 | 8 | 10 | 46,0 | 8 | 0 | 0 | 10 | 8 | 8,7 | 39,866667 |
| 26 | Valid | 66,0 | 12 | 0 | 20 | 0 | 32,7 | 1 | 1 | 0 | 8 | 0 | 3,3 | 10,888889 |
| 27 | Valid | 24,0 | 6 | 12 | 24 | | 16,5 | 6 | 0 | 1 | 1 | 3 | 2,8 | 4,5375 |
| 28 | Valid | 37,0 | 0 | 0 | 0 | 0 | 37,0 | 10 | 0 | 0 | 0 | 0 | 10,0 | 37 |
| 29 | Valid | 0,0 | 0 | 0 | 0 | 0 | 0 | 0 | 0 | 0 | 0 | 0 | 0 | 0 |
| 30 | Valid | 18,0 | 0 | 0 | 0 | 0 | 18,0 | 9 | 0 | 0 | 0 | 0 | 9,0 | 16,2 |
| 31 | Valid | 0,0 | 0 | 0 | 0 | 0 | 0 | 0 | 0 | 0 | 0 | 0 | 0 | 0 |
| 32 | Valid | 120,0 | 84 | 0 | 0 | 0 | 102,0 | 9 | 6 | 0 | 0 | 0 | 7,5 | 76,5 |
| 33 | Valid | 60,0 | 0 | 0 | 0 | 0 | 60,0 | 3 | 0 | 0 | 0 | 0 | 3,0 | 18 |
| 34 | Valid | 180,0 | 0 | 0 | 0 | 0 | 180,0 | 8 | 0 | 0 | 0 | 0 | 8,0 | 144 |
| 35 | Not Resident | 0,0 | 0 | 0 | 0 | 0 | #N/A | 0 | 0 | 0 | 0 | 0 | #N/A | #NA |
| 36 | Valid | 120,0 | 0 | 0 | 0 | 0 | 120,0 | 2 | 0 | 0 | 0 | 0 | 2,0 | 24 |
| 37 | Valid | 72,0 | 0 | 0 | 0 | 0 | 72,0 | 1 | 0 | 0 | 0 | 0 | 1,0 | 7,2 |
| 38 | Valid | 240,0 | 0 | 0 | 0 | 12 | 126,0 | 5,5 | 0 | 0 | 0 | 5 | 5,3 | 66,15 |
| 39 | Valid | 0,0 | 0 | 0 | 5 | 0 | 5,0 | 0 | 0 | 0 | 5 | 0 | 5,0 | 2,5 |
| 40 | Valid | 156,0 | 0 | 0 | 0 | 0 | 156,0 | 9 | 0 | 0 | 0 | 0 | 9,0 | 140,4 |
| 41 | Not Resident | 0,0 | 0 | 0 | 0 | 6 | #N/A | 0 | 0 | 0 | 0 | 8 | #N/A | #NA |
| 42 | Valid | 20,7 | 0 | 0 | 0 | 0 | 20,7 | 9 | 0 | 0 | 0 | 0 | 9,0 | 18,6 |
| 43 | Valid | 60,0 | 18 | 0 | 24 | 0 | 34,0 | 8 | 8 | 0 | 0 | 0 | 8,0 | 27,2 |
| 44 | Bogus | 240,0 | 0 | 0 | 222 | 2 | #N/A | 10 | 0 | 0 | 1 | 1 | #N/A | #NA |
| 45 | Valid | 0,0 | 0 | 0 | 0 | 5 | 5,0 | 0 | 0 | 0 | 0 | 8 | 8,0 | 4 |

# Appendix 9 – FinTech Questionnaire – Scoring Excerpt – Transaction Volumes & Frequency

| | | Payments | | | | Bank Movements | | | | Investments | | | | Loan / Deposit | | | | Subtotal |
|---|---|---|---|---|---|---|---|---|---|---|---|---|---|---|---|---|---|---|
| **Weight** | | **5** | | | | **4** | | | | **3** | | | | **1** | | | | **158,66713** |
| Participant | Valid participant? | P. Freq. | P. Amnt. | P. Vol. | Score | B. Freq. | B. Amnt. | Bank | Score | I. Freq. | I. Amnt. | Invest | Score | Lending | Deposit | Vol | Score | Average |
| 1 | Not Resident | 0 | 0 | #N/A | #N/A | 0 | 0 | #N/A | #N/A | 0 | 0 | 0 | #N/A | 0 | 0 | 0 | #N/A | #N/A |
| 2 | Valid | 0 | 0 | 0 | 0,0 | 15 | 50 | 750 | 187,5 | 0 | 0 | 0 | #N/A | 0 | 0 | 0 | 0,0 | 46,875 |
| 3 | Valid | 8 | 85 | 680 | 136,0 | 0 | 0 | 0 | 0,0 | 0 | 0 | 0 | 0,0 | 0 | 0 | 0 | 0,0 | 34 |
| 4 | Not Resident | 0 | 0 | #N/A | #N/A | 0 | 0 | #N/A | #N/A | 0 | 0 | 0 | #N/A | 0 | 0 | 0 | #N/A | #N/A |
| 5 | Valid | 11 | 110 | 1210 | 242,0 | 5 | 60 | 300 | 75,0 | 0 | 0 | 0 | 0,0 | 0 | 0 | 0 | 0,0 | 79,25 |
| 6 | Valid | 7 | 400 | 2800 | 560,0 | 0 | 0 | 0 | 0,0 | 0 | 0 | 0 | 0,0 | 0 | 0 | 0 | 0,0 | 140 |
| 7 | Valid | 9 | 755 | 6795 | 1359,0 | 3 | 30 | 90 | 22,5 | 0 | 0 | 0 | 0,0 | 600 | 7500 | 8100 | 8100,0 | 2370,375 |
| 8 | Bogus | 0 | 0 | #N/A | #N/A | 0 | 0 | #N/A | #N/A | 0 | 0 | 0 | #N/A | 0 | 0 | 0 | #N/A | #N/A |
| 9 | Valid | 2 | 170 | 340 | 68,0 | 0 | 0 | 0 | 0,0 | 0 | 0 | 0 | 0,0 | 5000 | 0 | 5000 | 5000,0 | 1267 |
| 10 | Valid | 1 | 80 | 80 | 16,0 | 0 | 0 | 0 | 0,0 | 0 | 0 | 0 | 0,0 | 0 | 0 | 0 | 0,0 | 4 |
| 11 | Bogus | 0 | 0 | #N/A | #N/A | 0 | 0 | #N/A | #N/A | 0 | 0 | 0 | #N/A | 0 | 0 | 0 | #N/A | #N/A |
| 12 | Valid | 2 | 600 | 1200 | 240,0 | 0 | 0 | 0 | 0,0 | 0 | 0 | 0 | 0,0 | 0 | 0 | 0 | 0,0 | 60 |
| 13 | Valid | 7 | 300 | 2100 | 420,0 | 0 | 0 | 0 | 0,0 | 0 | 0 | 0 | 0,0 | 0 | 0 | 0 | 0,0 | 105 |
| 14 | Valid | 8 | 50 | 400 | 80,0 | 0 | 0 | 0 | 0,0 | 0 | 0 | 0 | 0,0 | 0 | 0 | 0 | 0,0 | 20 |
| 15 | Valid | 3 | 240 | 720 | 144,0 | 0 | 0 | 0 | 0,0 | 0 | 0 | 0 | 0,0 | 1200 | 0 | 1200 | 1200,0 | 336 |
| 16 | Not Resident | 0 | 0 | #N/A | #N/A | 0 | 0 | #N/A | #N/A | 0 | 0 | 0 | #N/A | 0 | 0 | 0 | #N/A | #N/A |
| 17 | Valid | 6 | 400 | 2400 | 480,0 | 0 | 0 | 0 | 0,0 | 0 | 0 | 0 | 0,0 | 0 | 0 | 0 | 0,0 | 120 |
| 18 | Valid | 15 | 20 | 300 | 60,0 | 0 | 0 | 0 | 0,0 | 0 | 0 | 0 | 0,0 | 0 | 0 | 0 | 0,0 | 15 |
| 19 | Valid | 10 | 50 | 500 | 100,0 | 0 | 0 | 0 | 0,0 | 0 | 0 | 0 | 0,0 | 0 | 0 | 0 | 0,0 | 25 |
| 20 | Valid | 0 | 15 | 0 | 0,0 | 0 | 0 | 0 | 0,0 | 0 | 0 | 0 | 0,0 | 0 | 0 | 0 | 0,0 | 0 |
| 21 | Valid | 1 | 50 | 50 | 10,0 | 0 | 0 | 0 | 0,0 | 0 | 0 | 0 | 0,0 | 0 | 0 | 0 | 0,0 | 2,5 |
| 22 | Valid | 5 | 310 | 1550 | 310,0 | 0 | 0 | 0 | 0,0 | 0 | 0 | 0 | 0,0 | 2500 | 0 | 2500 | 2500,0 | 702,5 |
| 23 | Valid | 5 | 500 | 2500 | 500,0 | 0 | 0 | 0 | 0,0 | 0 | 0 | 0 | 0,0 | 0 | 0 | 0 | 0,0 | 125 |
| 24 | Valid | 2 | 20 | 40 | 8,0 | 0 | 0 | 0 | 0,0 | 0 | 0 | 0 | 0,0 | 0 | 0 | 0 | 0,0 | 2 |
| 25 | Valid | 2 | 25 | 50 | 10,0 | 2 | 20 | 40 | 10,0 | 1 | 500 | 500 | 166,7 | 0 | 0 | 0 | 0,0 | 46,666667 |
| 26 | Valid | 2 | 30 | 60 | 12,0 | 0 | 0 | 0 | 0,0 | 0 | 2000 | 0 | 0,0 | 3000 | 0 | 3000 | 3000,0 | 753 |
| 27 | Valid | 10 | 140 | 1400 | 280,0 | 15 | 30 | 450 | 112,5 | 2 | 250 | 500 | 166,7 | 0 | 1250 | 1250 | 1250,0 | 452,29167 |
| 28 | Valid | 1,2 | 600 | 720 | 144,0 | 0 | 0 | 0 | 0,0 | 0 | 0 | 0 | 0,0 | 0 | 0 | 0 | 0,0 | 36 |
| 29 | Valid | 0 | 0 | 0 | 0,0 | 0 | 0 | 0 | 0,0 | 0 | 0 | 0 | 0,0 | 0 | 0 | 0 | 0,0 | 0 |
| 30 | Valid | 0 | 250 | 0 | 0,0 | 0 | 0 | 0 | 0,0 | 0 | 0 | 0 | 0,0 | 0 | 0 | 0 | 0,0 | 0 |
| 31 | Valid | 0 | 0 | 0 | 0,0 | 0 | 0 | 0 | 0,0 | 0 | 0 | 0 | 0,0 | 0 | 0 | 0 | 0,0 | 0 |
| 32 | Valid | 3 | 40 | 120 | 24,0 | 0 | 0 | 0 | 0,0 | 0 | 0 | 0 | 0,0 | 0 | 0 | 0 | 0,0 | 6 |
| 33 | Valid | 3 | 30 | 90 | 18,0 | 0 | 0 | 0 | 0,0 | 0 | 0 | 0 | 0,0 | 0 | 0 | 0 | 0,0 | 4,5 |
| 34 | Valid | 3 | 50 | 150 | 30,0 | 0 | 0 | 0 | 0,0 | 0 | 0 | 0 | 0,0 | 0 | 0 | 0 | 0,0 | 7,5 |
| 35 | Not Resident | 0 | 0 | #N/A | #N/A | 0 | 0 | #N/A | #N/A | 0 | 0 | 0 | #N/A | 0 | 0 | 0 | #N/A | #N/A |
| 36 | Valid | 7 | 20 | 140 | 28,0 | 0 | 0 | 0 | 0,0 | 0 | 0 | 0 | 0,0 | 0 | 0 | 0 | 0,0 | 7 |
| 37 | Valid | 5 | 30 | 150 | 30,0 | 0 | 0 | 0 | 0,0 | 0 | 0 | 0 | 0,0 | 0 | 0 | 0 | 0,0 | 7,5 |
| 38 | Valid | 4 | 400 | 1600 | 320,0 | 10 | 2000 | 20000 | 5000,0 | 0 | 0 | 0 | 0,0 | 0 | 0 | 0 | 0,0 | 1330 |
| 39 | Valid | 0 | 0 | 0 | 0,0 | 0 | 0 | 0 | 0,0 | 0 | 0 | 0 | 0,0 | 0 | 0 | 0 | 0,0 | 0 |
| 40 | Valid | 2 | 40 | 80 | 16,0 | 0 | 0 | 0 | 0,0 | 0 | 0 | 0 | 0,0 | 0 | 0 | 0 | 0,0 | 4 |
| 41 | Not Resident | 0 | 0 | #N/A | #N/A | 4 | 10 | #N/A | #N/A | 0 | 0 | 0 | #N/A | 0 | 0 | 0 | #N/A | #N/A |

# Appendix 10 – FinTech Questionnaire – Grading

| Participant | K&U | Ranking_KU | L&S | Ranking_LS | Vol | Ranking_Vol | Final Score | Classification | Validation | Check | PMS | Email |
|---|---|---|---|---|---|---|---|---|---|---|---|---|
| 38 | 24 | 20 | 66,15 | 162 | 1330 | 184 | 366 | Early Adopter | An Early Adopter – 'I low | Yes | Pioneer | 0 |
| 67 | 41 | 33 | 68,4 | 165 | 304 | 159 | 357 | Early Adopter | An Early Adopter – 'I low | Yes | Pioneer | 0 |
| 145 | 20 | 17 | 72 | 166 | 605 | 173 | 356 | Early Adopter | An Early Adopter – 'I low | Yes | Pioneer | 0 |
| 169 | 51 | 40 | 31,73333 | 114 | 3066 | 192 | 346 | Early Adopter | An Early Adopter – 'I low | Yes | Pioneer | 0 |
| 454 | 21 | 17 | 60 | 158 | 450 | 165 | 340 | Early Adopter | An Early Adopter – 'I low | Yes | Pioneer | 0 |
| 22 | 42 | 34 | 37,05 | 127 | 702,5 | 177 | 338 | Early Adopter | An Early Adopter – 'I low | Yes | Pioneer | 0 |
| 460 | 46 | 38 | 29,13333 | 102 | 2805,75 | 191 | 331 | Balanced | Balanced – 'Once the tec | No | Migrator | 0 |
| 316 | 43 | 35 | 45 | 142 | 192,5 | 153 | 330 | Early Adopter | An Early Adopter – 'I low | Yes | Pioneer | 0 |
| 301 | 25 | 21 | 31,66667 | 113 | 2436 | 189 | 323 | Early Adopter | An Early Adopter – 'I low | Yes | Pioneer | 0 |
| 17 | 55 | 41 | 49,7 | 147 | 120 | 132 | 320 | Balanced | Balanced – 'Once the tec | No | Migrator | 0 |
| 110 | 7 | 5 | 329,4 | 180 | 125 | 134 | 319 | Early Adopter | An Early Adopter – 'I low | Yes | Pioneer | 0 |
| 9 | 42 | 34 | 27,7875 | 99 | 1267 | 182 | 315 | Early Adopter | An Early Adopter – 'I low | Yes | Pioneer | 0 |
| 52 | 47 | 39 | 26,63333 | 92 | 1274,5 | 183 | 314 | Early Adopter | An Early Adopter – 'I low | Yes | Pioneer | 0 |
| 120 | 24 | 20 | 43,5 | 139 | 175,125 | 150 | 309 | Early Adopter | An Early Adopter – 'I low | Yes | Pioneer | 0 |
| 96 | 69 | 42 | 31,2 | 110 | 196,313 | 154 | 306 | Early Adopter | An Early Adopter – 'I low | Yes | Pioneer | 0 |
| 103 | 69 | 42 | 19,73611 | 70 | 3691,88 | 194 | 306 | Early Adopter | An Early Adopter – 'I low | Yes | Pioneer | 0 |
| 75 | 41 | 33 | 54 | 152 | 92,75 | 120 | 305 | Balanced | Balanced – 'Once the tec | No | Migrator | 0 |
| 61 | 47 | 38 | 18,2 | 66 | 6346 | 197 | 301 | Early Adopter | An Early Adopter – 'I low | Yes | Pioneer | 0 |
| 328 | 26 | 23 | 40,33333 | 135 | 144,5 | 142 | 300 | Early Adopter | An Early Adopter – 'I low | Yes | Pioneer | simon.a.ker |
| 299 | 14 | 12 | 58,5 | 157 | 115,5 | 130 | 299 | Early Adopter | An Early Adopter – 'I low | Yes | Pioneer | 0 |
| 300 | 51 | 39 | 21,66667 | 79 | 1102,5 | 181 | 299 | Early Adopter | Balanced – 'Once the tec | No | Migrator | 0 |
| 178 | 26 | 22 | 29,33333 | 104 | 585 | 171 | 297 | Early Adopter | An Early Adopter – 'I low | Yes | Pioneer | 0 |
| 289 | 47 | 38 | 16,5 | 63 | 4016 | 195 | 296 | Early Adopter | An Early Adopter – 'I low | Yes | Pioneer | 0 |
| 13 | 30 | 27 | 44,325 | 140 | 105 | 128 | 295 | Early Adopter | An Early Adopter – 'I low | Yes | Pioneer | iblue@gma |
| 86 | 69 | 42 | 34,88333 | 122 | 116,625 | 131 | 295 | Early Adopter | An Early Adopter – 'I low | Yes | Pioneer | 0 |
| 118 | 110 | 50 | 15,67917 | 57 | 1964,75 | 187 | 293 | Early Adopter | An Early Adopter – 'I low | Yes | Pioneer | 0 |
| 54 | 47 | 38 | 16,46667 | 62 | 3087,75 | 193 | 293 | Balanced | Balanced – 'Once the tec | No | Migrator | 0 |
| 309 | 12 | 9 | 50,4 | 148 | 129,5 | 136 | 293 | Early Adopter | An Early Adopter – 'I low | Yes | Pioneer | 0 |
| 168 | 25 | 21 | 40 | 134 | 131,25 | 137 | 292 | Early Adopter | An Early Adopter – 'I low | Yes | Pioneer | 0 |
| 43 | 85 | 48 | 27,2 | 94 | 168,75 | 149 | 291 | Early Adopter | Balanced – 'Once the tec | No | Migrator | 0 |
| 167 | 25 | 22 | 48,53333 | 145 | 96,25 | 124 | 291 | Early Adopter | Balanced – 'Once the tec | No | Migrator | 0 |
| 121 | 24 | 20 | 32,625 | 115 | 222,125 | 155 | 290 | Early Adopter | An Early Adopter – 'I low | Yes | Pioneer | kathrin.t.luc |
| 155 | 55 | 40 | 26,4 | 91 | 224,083 | 156 | 287 | Early Adopter | An Early Adopter – 'I low | Yes | Pioneer | 0 |
| 15 | 42 | 33 | 26,1 | 90 | 336 | 162 | 285 | Early Adopter | Balanced – 'Once the tec | No | Migrator | 0 |
| 182 | 26 | 23 | 51 | 149 | 82,5 | 113 | 285 | Early Adopter | An Early Adopter – 'I low | Yes | Pioneer | cosmin.olte |
| 131 | 78 | 46 | 14,26563 | 51 | 1374,58 | 185 | 282 | Early Adopter | An Early Adopter – 'I low | Yes | Pioneer | 0 |
| 256 | 13 | 11 | 48 | 144 | 100 | 127 | 282 | Early Adopter | An Early Adopter – 'I low | Yes | Pioneer | 0 |
| 279 | 27 | 23 | 54,6 | 154 | 65,25 | 105 | 282 | Early Adopter | An Early Adopter – 'I low | Yes | Pioneer | 0 |
| 321 | 25 | 21 | 34,66667 | 121 | 140 | 140 | 282 | Balanced | Balanced – 'Once the tec | No | Migrator | 0 |
| 108 | 110 | 50 | 10 | 34 | 4227,94 | 196 | 280 | Early Adopter | An Early Adopter – 'I low | Yes | Pioneer | 0 |
| 186 | 14 | 12 | 76 | 168 | 60 | 100 | 280 | Early Adopter | Balanced – 'Once the tec | No | Migrator | fernolimits@ |
| 179 | 26 | 22 | 22,53333 | 82 | 649 | 174 | 278 | Early Adopter | An Early Adopter – 'I low | Yes | Pioneer | 0 |
| 244 | 20 | 17 | 26 | 89 | 586 | 172 | 278 | Early Adopter | Balanced – 'Once the tec | No | Migrator | 0 |
| 222 | 78 | 46 | 12,79688 | 45 | 1940,38 | 186 | 277 | Early Adopter | An Early Adopter – 'I low | Yes | Pioneer | 0 |
| 240 | 18 | 16 | 62,4 | 159 | 60,75 | 101 | 276 | Early Adopter | Balanced – 'Once the tec | No | Migrator | 0 |
| 147 | 93 | 49 | 21,8625 | 81 | 153,667 | 145 | 275 | Early Adopter | An Early Adopter – 'I low | Yes | Pioneer | 0 |
| 171 | 29 | 26 | 39 | 130 | 92,5 | 119 | 273 | Early Adopter | An Early Adopter – 'I low | Yes | Pioneer | fh@futix.eu |
| 74 | 27 | 24 | 67,5 | 164 | 45 | 85 | 273 | Early Adopter | An Early Adopter – 'I low | Yes | Pioneer | 0 |
| 97 | 18 | 16 | 96 | 172 | 45 | 85 | 273 | Balanced | Balanced – 'Once the tec | No | Migrator | 0 |
| 267 | 20 | 17 | 25,2 | 88 | 497,75 | 168 | 273 | Early Adopter | An Early Adopter – 'I low | Yes | Pioneer | 0 |
| 69 | 85 | 48 | 15 | 55 | 506 | 169 | 272 | Early Adopter | An Early Adopter – 'I low | Yes | Pioneer | 0 |
| 122 | 110 | 50 | 9,75 | 32 | 2621,38 | 190 | 272 | Early Adopter | An Early Adopter – 'I low | Yes | Pioneer | 0 |
| 213 | 18 | 16 | 41,6 | 136 | 90 | 117 | 269 | Early Adopter | Balanced – 'Once the tec | No | Migrator | 0 |
| 7 | 110 | 50 | 8,8375 | 29 | 2370,38 | 188 | 267 | Early Adopter | An Early Adopter – 'I low | Yes | Pioneer | 0 |
| 57 | 10 | 8 | 96 | 172 | 45 | 85 | 265 | Early Adopter | Balanced – 'Once the tec | No | Migrator | 0 |
| 26 | 28 | 48 | 10,88889 | 38 | 753 | 175 | 264 | Early Adopter | Balanced – 'Once the tec | No | Migrator | 0 |
| 419 | 35 | 30 | 60 | 158 | 38 | 75 | 263 | Early Adopter | Balanced – 'Once the tec | No | Migrator | 0 |
| 25 | 73 | 44 | 39,86667 | 132 | 46,6667 | 86 | 262 | Early Adopter | An Early Adopter – 'I low | Yes | Pioneer | 0 |
| 217 | 18 | 16 | 65,55 | 160 | 45 | 85 | 261 | Early Adopter | An Early Adopter – 'I low | Yes | Pioneer | 0 |
| 241 | 20 | 17 | 23,4 | 84 | 319 | 160 | 261 | Early Adopter | An Early Adopter – 'I low | Yes | Pioneer | 0 |
| 23 | 28 | 25 | 28,8 | 101 | 125 | 134 | 260 | Early Adopter | An Early Adopter – 'I low | Yes | Pioneer | 0 |
| 70 | 62 | 42 | 19,8 | 71 | 156,5 | 147 | 260 | Early Adopter | An Early Adopter – 'I low | Yes | Pioneer | 0 |
| 237 | 24 | 20 | 54 | 152 | 47,5 | 88 | 260 | Early Adopter | An Early Adopter – 'I low | Yes | Pioneer | 0 |
| 146 | 93 | 49 | 16,2 | 61 | 156,667 | 148 | 258 | Early Adopter | An Early Adopter – 'I low | Yes | Pioneer | 0 |
| 413 | 33 | 29 | 66 | 161 | 33,25 | 68 | 258 | Early Adopter | Balanced – 'Once the tec | No | Migrator | 0 |
| 6 | 23 | 20 | 27,5 | 97 | 140 | 140 | 257 | Early Adopter | An Early Adopter – 'I low | Yes | Pioneer | 0 |
| 172 | 26 | 22 | 21,3 | 77 | 296,25 | 158 | 257 | Early Adopter | An Early Adopter – 'I low | Yes | Pioneer | 0 |
| 243 | 20 | 17 | 18 | 65 | 651 | 175 | 256 | Early Adopter | An Early Adopter – 'I low | Yes | Pioneer | 0 |
| 412 | 25 | 22 | 72 | 166 | 33,25 | 68 | 256 | Early Adopter | Balanced – 'Once the tec | No | Migrator | 0 |
| 275 | 20 | 20 | 36 | 125 | 72 | 110 | 255 | Early Adopter | Balanced – 'Once the tec | No | Migrator | 0 |
| 42 | 20 | 17 | 18,6 | 67 | 578,25 | 170 | 254 | Early Adopter | An Early Adopter – 'I low | Yes | Pioneer | fintechmail@ |
| 200 | 8 | 6 | 80 | 170 | 40 | 78 | 254 | Early Adopter | An Early Adopter – 'I low | Yes | Pioneer | 0 |
| 236 | 24 | 20 | 51 | 149 | 44 | 84 | 253 | Early Adopter | An Early Adopter – 'I low | Yes | Pioneer | 0 |
| 416 | 29 | 26 | 60 | 158 | 34 | 69 | 253 | Early Adopter | Balanced – 'Once the tec | No | Migrator | 0 |
| 441 | 14 | 12 | 48 | 144 | 52,5 | 96 | 252 | Early Adopter | Balanced – 'Once the tec | No | Migrator | 0 |
| 387 | 28 | 25 | 75,6 | 167 | 25 | 59 | 251 | Early Adopter | An Early Adopter – 'I low | Yes | Pioneer | 0 |
| 442 | 22 | 19 | 97,2 | 173 | 25 | 59 | 251 | Early Adopter | An Early Adopter – 'I low | Yes | Pioneer | 0 |
| 99 | 23 | 19 | 21,6 | 78 | 187,5 | 152 | 249 | Early Adopter | An Early Adopter – 'I low | Yes | Pioneer | 0 |
| 158 | 72 | 42 | 12 | 43 | 444,167 | 164 | 249 | Early Adopter | An Early Adopter – 'I low | Yes | Pioneer | 0 |
| 283 | 18 | 16 | 28,8 | 101 | 120 | 132 | 249 | Early Adopter | An Early Adopter – 'I low | Yes | Pioneer | 0 |
| 298 | 14 | 12 | 35,4 | 123 | 85,75 | 114 | 249 | Early Adopter | An Early Adopter – 'I low | Yes | Pioneer | 0 |
| 68 | 78 | 46 | 6,075 | 22 | 959,833 | 180 | 248 | Balanced | Balanced – 'Once the tec | Yes | Migrator | 0 |
| 277 | 27 | 23 | 37,8 | 128 | 57 | 97 | 248 | Balanced | Balanced – 'Once the tec | Yes | Migrator | 0 |
| 452 | 21 | 17 | 60 | 158 | 36 | 72 | 247 | Balanced | Balanced – 'Once the tec | Yes | Migrator | 0 |
| 115 | 24 | 20 | 25,2 | 88 | 135,375 | 138 | 246 | Balanced | Balanced – 'Once the tec | Yes | Migrator | 0 |
| 285 | 18 | 16 | 33,6 | 119 | 78 | 111 | 246 | Balanced | Balanced – 'Once the tec | Yes | Migrator | 0 |
| 76 | 80 | 46 | 11,875 | 42 | 261,25 | 157 | 245 | Balanced | An Early Adopter – 'I low | Yes | Pioneer | 0 |
| 142 | 38 | 32 | 24,375 | 87 | 98,5 | 126 | 245 | Balanced | An Early Adopter – 'I low | No | Pioneer | 0 |
| 136 | 38 | 32 | 18,75 | 68 | 148 | 144 | 244 | Balanced | Balanced – 'Once the tec | Yes | Migrator | 0 |
| 219 | 18 | 16 | 51 | 149 | 41,25 | 79 | 244 | Balanced | Balanced – 'Once the tec | Yes | Migrator | 0 |
| 64 | 85 | 47 | 5,533333 | 20 | 687,5 | 176 | 243 | Balanced | An Early Adopter – 'I love | No | Pioneer | 0 |
| 84 | 73 | 43 | 19,6 | 69 | 109,417 | 129 | 241 | Balanced | An Early Adopter – 'I love | No | Pioneer | 0 |
| 380 | 20 | 17 | 108 | 174 | 18,75 | 50 | 241 | Balanced | Balanced – 'Once the tec | Yes | Migrator | 0 |
| 430 | 9 | 7 | 84 | 171 | 27 | 63 | 241 | Balanced | Balanced – 'Once the tec | Yes | Migrator | 0 |
| 32 | 80 | 47 | 76,5 | 169 | 6 | 24 | 240 | Balanced | Balanced – 'Once the tec | Yes | Migrator | 0 |
| 144 | 93 | 48 | 14,45 | 53 | 136,25 | 139 | 240 | Balanced | Balanced – 'Once the tec | Yes | Migrator | 0 |
| 450 | 33 | 29 | 60 | 158 | 20 | 53 | 240 | Balanced | Balanced – 'Once the tec | Yes | Migrator | 0 |
| 164 | 26 | 23 | 35,6 | 124 | 50 | 92 | 239 | Balanced | Balanced – 'Once the tec | Yes | Migrator | 0 |
| 315 | 34 | 30 | 54,15 | 153 | 22,5 | 56 | 239 | Balanced | An Early Adopter – 'I love | No | Pioneer | 0 |
| 93 | 73 | 43 | 20,53333 | 74 | 93,2917 | 121 | 238 | Balanced | Balanced – 'Once the tec | Yes | Migrator | 0 |
| 139 | 72 | 43 | 16,15 | 60 | 127,5 | 135 | 238 | Balanced | Balanced – 'Once the tec | Yes | Migrator | 0 |
| 332 | 35 | 31 | 108 | 174 | 10 | 33 | 238 | Balanced | Balanced – 'Once the tec | Yes | Migrator | 0 |
| 106 | 42 | 34 | 52,59375 | 150 | 20 | 53 | 237 | Balanced | Balanced – 'Once the tec | Yes | Migrator | 0 |
| 261 | 17 | 14 | 37,8 | 128 | 52,25 | 95 | 237 | Balanced | Balanced – 'Once the tec | Yes | Migrator | 0 |
| 429 | 11 | 9 | 96 | 172 | 22,5 | 56 | 237 | Balanced | Balanced – 'Once the tec | Yes | Migrator | 0 |
| 407 | 28 | 25 | 48 | 144 | 30 | 67 | 236 | Balanced | An Early Adopter – 'I love | No | Pioneer | 0 |
| 434 | 12 | 9 | 48 | 144 | 42,75 | 83 | 236 | Balanced | An Early Adopter – 'I love | No | Pioneer | 0 |
| 113 | 44 | 36 | 30,8 | 109 | 49,25 | 90 | 235 | Balanced | Balanced – 'Once the tec | Yes | Migrator | 0 |
| 312 | 7 | 5 | 84 | 171 | 25 | 59 | 235 | Balanced | Balanced – 'Once the tec | Yes | Migrator | 0 |
| 435 | 8 | 6 | 67,2 | 163 | 29,25 | 68 | 235 | Balanced | Balanced – 'Once the tec | Yes | Migrator | 0 |
| 221 | 30 | 27 | 30,25 | 107 | 59,5 | 99 | 233 | Balanced | Balanced – 'Once the tec | Yes | Migrator | 0 |
| 14 | 7 | 5 | 108 | 174 | 20 | 53 | 232 | Balanced | Balanced – 'Once the tec | Yes | Migrator | irmina.abra |
| 254 | 14 | 12 | 50,4 | 148 | 36 | 72 | 232 | Balanced | Balanced – 'Once the tec | Yes | Migrator | 0 |
| 27 | 78 | 46 | 4,5375 | 18 | 452,292 | 166 | 230 | Balanced | Balanced – 'Once the tec | Yes | Migrator | 0 |
| 447 | 7 | 5 | 60 | 158 | 30 | 67 | 230 | Balanced | Balanced – 'Once the tec | Yes | Migrator | 0 |
| 269 | 7 | 5 | 84 | 171 | 20 | 53 | 229 | Balanced | Balanced – 'Once the tec | Yes | Migrator | 0 |
| 274 | 23 | 20 | 28,8 | 101 | 68,75 | 108 | 229 | Balanced | Balanced – 'Once the tec | Yes | Migrator | 0 |
| 104 | 32 | 28 | 20,9875 | 75 | 97,375 | 125 | 228 | Balanced | Balanced – 'Once the tec | Yes | Migrator | 0 |
| 154 | 72 | 43 | 22,95 | 83 | 61,6667 | 102 | 228 | Balanced | Balanced – 'Once the tec | Yes | Migrator | 0 |
| 259 | 17 | 14 | 31,35 | 111 | 62,5 | 103 | 228 | Balanced | Balanced – 'Once the tec | Yes | Migrator | 0 |
| 268 | 20 | 16 | 9,777778 | 33 | 912 | 179 | 228 | Balanced | Balanced – 'Once the tec | Yes | Migrator | 0 |
| 34 | 24 | 21 | 144 | 178 | 7,5 | 28 | 227 | Balanced | Balanced – 'Once the tec | Yes | Migrator | 0 |
| 62 | 18 | 16 | 49,05 | 146 | 28,75 | 65 | 227 | Balanced | An Early Adopter – 'I love | No | Pioneer | 0 |
| 226 | 20 | 17 | 27 | 93 | 90 | 117 | 227 | Balanced | Balanced – 'Once the tec | Yes | Migrator | 0 |
| 112 | 44 | 36 | 33,33333 | 118 | 34,75 | 71 | 225 | Balanced | Balanced – 'Once the tec | Yes | Migrator | 0 |
| 235 | 18 | 16 | 108 | 174 | 12 | 35 | 225 | Balanced | Balanced – 'Once the tec | Yes | Migrator | 0 |
| 247 | 23 | 20 | 28,8 | 101 | 65 | 104 | 225 | Balanced | Balanced – 'Once the tec | Yes | Migrator | 0 |
| 273 | 23 | 20 | 39,9 | 133 | 35 | 72 | 225 | Balanced | Balanced – 'Once the tec | Yes | Migrator | 0 |
| 319 | 22 | 19 | 20,25 | 73 | 123,75 | 133 | 225 | Balanced | Balanced – 'Once the tec | Yes | Migrator | 0 |
| 66 | 41 | 33 | 15,3 | 56 | 127,5 | 135 | 224 | Balanced | Balanced – 'Once the tec | Yes | Migrator | 0 |
| 231 | 24 | 20 | 39,6 | 131 | 35 | 72 | 223 | Balanced | Balanced – 'Once the tec | Yes | Migrator | 0 |
| 239 | 18 | 16 | 55,2 | 155 | 19,25 | 51 | 222 | Balanced | Balanced – 'Once the tec | Yes | Migrator | 0 |
| 282 | 47 | 39 | 62,4 | 159 | 5,775 | 23 | 221 | Balanced | Balanced – 'Once the tec | Yes | Migrator | 0 |
| 160 | 34 | 30 | 37,8 | 128 | 26,25 | 62 | 220 | Balanced | Balanced – 'Once the tec | Yes | Migrator | 0 |
| 165 | 47 | 38 | 25,5 | 88 | 51 | 94 | 220 | Balanced | Balanced – 'Once the tec | Yes | Migrator | 0 |
| 296 | 14 | 12 | 42,3 | 138 | 34,5 | 70 | 220 | Balanced | Balanced – 'Once the tec | Yes | Migrator | 0 |
| 306 | 23 | 19 | 14,25 | 50 | 182,5 | 151 | 220 | Balanced | Balanced – 'Once the tec | Yes | Migrator | 0 |
| 411 | 14 | 12 | 48 | 144 | 28 | 64 | 220 | Balanced | Balanced – 'Once the tec | Yes | Migrator | 0 |
| 51 | 85 | 47 | 3,111111 | 11 | 330,5 | 161 | 219 | Balanced | An Early Adopter – 'I love | No | Pioneer | 0 |
| 109 | 17 | 15 | 60 | 158 | 17,5 | 46 | 219 | Balanced | Balanced – 'Once the tec | Yes | Migrator | 0 |
| 280 | 27 | 23 | 29,25 | 103 | 50,75 | 93 | 219 | Balanced | An Early Adopter – 'I love | No | Pioneer | 0 |
| 422 | 9 | 7 | 50,4 | 148 | 28 | 64 | 219 | Balanced | Balanced – 'Once the tec | Yes | Migrator | 0 |
| 28 | 23 | 20 | 37 | 126 | 36 | 72 | 218 | Balanced | Balanced – 'Once the tec | Yes | Migrator | 0 |
| 307 | 12 | 9 | 34,2 | 120 | 48 | 89 | 218 | Balanced | Balanced – 'Once the tec | Yes | Migrator | 0 |
| 81 | 93 | 49 | 15,9375 | 59 | 70 | 109 | 217 | Balanced | Balanced – 'Once the tec | Yes | Migrator | 0 |
| 130 | 29 | 26 | 14,25 | 50 | 143,188 | 141 | 217 | Balanced | Balanced – 'Once the tec | Yes | Migrator | 0 |
| 204 | 23 | 19 | 14,25 | 50 | 154 | 146 | 215 | Balanced | Balanced – 'Once the tec | Yes | Migrator | 0 |
| 143 | 38 | 32 | 28,4375 | 100 | 42,5 | 82 | 214 | Balanced | An Early Adopter – 'I love | No | Pioneer | 0 |
| 266 | 20 | 16 | 8,266667 | 28 | 481 | 167 | 211 | Balanced | Balanced – 'Once the tec | Yes | Migrator | 0 |
| 322 | 9 | 7 | 120 | 176 | 7,5 | 28 | 211 | Balanced | Balanced – 'Once the tec | Yes | Migrator | 0 |
| 382 | 8 | 6 | 54 | 152 | 18,75 | 50 | 208 | Balanced | Balanced – 'Once the tec | Yes | Migrator | 0 |
| 394 | 26 | 23 | 48 | 144 | 15 | 41 | 208 | Balanced | Balanced – 'Once the tec | Yes | Migrator | 0 |
| 427 | 10 | 7 | 38,4 | 129 | 36 | 72 | 208 | Balanced | Balanced – 'Once the tec | Yes | Migrator | 0 |
| 245 | 9 | 7 | 113,6 | 175 | 6,25 | 25 | 207 | Balanced | Balanced – 'Once the tec | Yes | Migrator | philipp.herz |
| 330 | 13 | 11 | 120 | 176 | 5 | 20 | 207 | Balanced | Balanced – 'Once the tec | Yes | Migrator | 0 |
| 440 | 35 | 30 | 48 | 144 | 10 | 33 | 207 | Balanced | Balanced – 'Once the tec | Yes | Migrator | 0 |
| 101 | 23 | 19 | 16,8 | 64 | 93,75 | 123 | 206 | Balanced | Balanced – 'Once the tec | Yes | Migrator | 0 |
| 195 | 18 | 16 | 27,73333 | 98 | 49,5 | 91 | 205 | Balanced | Balanced – 'Once the tec | Yes | Migrator | 0 |
| 388 | 7 | 5 | 48 | 144 | 22,5 | 56 | 205 | Balanced | An Early Adopter – 'I love | No | Pioneer | 0 |
| 372 | 11 | 9 | 84 | 171 | 6 | 24 | 204 | Balanced | Balanced – 'Once the tec | Yes | Migrator | 0 |
| 40 | 12 | 10 | 140,4 | 177 | 4 | 16 | 203 | Balanced | Balanced – 'Once the tec | Yes | Migrator | carsten@ot |
| 445 | 11 | 8 | 60 | 158 | 12,5 | 36 | 202 | Balanced | An Early Adopter – 'I love | No | Pioneer | 0 |
| 18 | 10 | 8 | 54 | 152 | 15 | 41 | 201 | Balanced | Balanced – 'Once the tec | Yes | Migrator | 0 |
| 389 | 25 | 21 | 48 | 144 | 12,5 | 36 | 201 | Balanced | Balanced – 'Once the tec | Yes | Migrator | 0 |
| 448 | 32 | 28 | 24 | 85 | 47,5 | 88 | 201 | Balanced | Balanced – 'Once the tec | Yes | Migrator | 0 |
| 95 | 73 | 43 | 11,82222 | 41 | 86,5833 | 115 | 199 | Balanced | Balanced – 'Once the tec | Yes | Migrator | 0 |
| 45 | 23 | 19 | 4 | 16 | 375 | 163 | 198 | Balanced | Balanced – 'Once the tec | Yes | Migrator | 0 |
| 402 | 22 | 19 | 43,2 | 138 | 15 | 41 | 198 | Balanced | Balanced – 'Once the tec | Yes | Migrator | 0 |
| 230 | 41 | 33 | 29,4 | 105 | 25 | 59 | 197 | Balanced | Balanced – 'Once the tec | Yes | Migrator | 0 |
| 368 | 11 | 8 | 48 | 144 | 17 | 45 | 197 | Balanced | Traditionalist – 'I don't ne | No | Settler | 0 |
| 362 | 9 | 7 | 50,4 | 148 | 15 | 41 | 196 | Balanced | Balanced – 'Once the tec | Yes | Migrator | 0 |
| 354 | 8 | 6 | 84 | 171 | 4,5 | 18 | 195 | Balanced | Balanced – 'Once the tec | Yes | Migrator | 0 |
| 356 | 7 | 5 | 120 | 176 | 3,5 | 14 | 195 | Balanced | An Early Adopter – 'I love | No | Pioneer | 0 |
| 224 | 20 | 17 | 18,9 | 69 | 68,25 | 107 | 193 | Balanced | Balanced – 'Once the tec | Yes | Migrator | 0 |
| 153 | 9 | 7 | 180 | 179 | 1,5 | 6 | 192 | Balanced | Balanced – 'Once the tec | Yes | Migrator | 0 |
| 193 | 21 | 18 | 97,2 | 173 | 0 | 1 | 192 | Balanced | Balanced – 'Once the tec | Yes | Migrator | 0 |
| 3 | 30 | 27 | 27,25 | 95 | 34 | 69 | 191 | Balanced | Balanced – 'Once the tec | Yes | Migrator | 0 |
| 133 | 18 | 16 | 52,8 | 151 | 6 | 24 | 191 | Balanced | Balanced – 'Once the tec | Yes | Migrator | 0 |
| 257 | 14 | 11 | 33,15 | 117 | 27 | 63 | 191 | Balanced | Balanced – 'Once the tec | Yes | Migrator | 0 |
| 308 | 55 | 40 | 14,45 | 53 | 58 | 98 | 191 | Balanced | Balanced – 'Once the tec | Yes | Migrator | 0 |
| 364 | 10 | 7 | 48 | 144 | 14,25 | 40 | 191 | Balanced | Balanced – 'Once the tec | Yes | Migrator | 0 |
| 420 | 8 | 6 | 36 | 125 | 26 | 60 | 191 | Balanced | Balanced – 'Once the tec | Yes | Migrator | 0 |
| 320 | 15 | 13 | 9,75 | 32 | 146,25 | 143 | 188 | Balanced | Balanced – 'Once the tec | Yes | Migrator | 0 |
| 326 | 18 | 16 | 54 | 152 | 4,75 | 19 | 187 | Balanced | Balanced – 'Once the tec | Yes | Migrator | 0 |
| 406 | 10 | 8 | 120 | 176 | 0 | 1 | 185 | Balanced | Balanced – 'Once the tec | Yes | Migrator | kopeke |
| 333 | 14 | 12 | 57,6 | 156 | 4 | 16 | 184 | Balanced | Traditionalist – 'I don't ne | No | Settler | 0 |
| 46 | 7 | 5 | 60 | 158 | 5 | 20 | 183 | Balanced | An Early Adopter – 'I low | No | Pioneer | 0 |
| 355 | 9 | 7 | 84 | 171 | 1 | 4 | 182 | Balanced | An Early Adopter – 'I low | No | Pioneer | 0 |
| 417 | 11 | 8 | 33,6 | 119 | 22 | 55 | 182 | Balanced | Balanced – 'Once the tec | Yes | Migrator | 0 |
| 238 | 18 | 16 | 37,8 | 128 | 13 | 37 | 181 | Balanced | Balanced – 'Once the tec | Yes | Migrator | 0 |
| 5 | 32 | 28 | 11,2125 | 40 | 79,25 | 112 | 180 | Balanced | Balanced – 'Once the tec | Yes | Migrator | 0 |
| 201 | 14 | 12 | 60 | 158 | 2,5 | 10 | 180 | Balanced | Balanced – 'Once the tec | Yes | Migrator | cd@access |
| 276 | 18 | 15 | 13,6 | 49 | 88,5 | 116 | 180 | Balanced | Balanced – 'Once the tec | Yes | Migrator | 0 |
| 331 | 17 | 15 | 60 | 158 | 1,75 | 7 | 180 | Balanced | Balanced – 'Once the tec | Yes | Migrator | 0 |
| 358 | 13 | 11 | 50,4 | 148 | 5,5 | 21 | 180 | Balanced | Balanced – 'Once the tec | Yes | Migrator | 0 |
| 194 | 18 | 16 | 25,2 | 88 | 37 | 73 | 177 | Balanced | Balanced – 'Once the tec | Yes | Migrator | 0 |
| 393 | 10 | 7 | 38,4 | 129 | 15 | 41 | 177 | Balanced | Balanced – 'Once the tec | Yes | Migrator | 0 |
| 20 | 11 | 9 | 72 | 166 | 0 | 1 | 176 | Balanced | Balanced – 'Once the tec | Yes | Migrator | sarah@ros |
| 53 | 42 | 34 | 44,45 | 141 | 0 | 1 | 176 | Balanced | An Early Adopter – 'I low | No | Pioneer | 0 |
| 207 | 14 | 12 | 60 | 158 | 1,5 | 6 | 176 | Balanced | Traditionalist – 'I don't ne | No | Settler | 0 |
| 223 | 30 | 27 | 27,375 | 96 | 20 | 53 | 176 | Balanced | Balanced – 'Once the tec | Yes | Migrator | 0 |
| 325 | 9 | 7 | 57,6 | 156 | 3,25 | 13 | 176 | Balanced | Traditionalist – 'I don't ne | No | Settler | 0 |
| 437 | 9 | 7 | 30 | 106 | 27 | 63 | 176 | Balanced | An Early Adopter – 'I low | No | Pioneer | 0 |
| 381 | 15 | 13 | 24,3 | 86 | 38,5 | 76 | 175 | Balanced | Balanced – 'Once the tec | Yes | Migrator | 0 |
| 345 | 12 | 9 | 48 | 144 | 5,5 | 21 | 174 | Balanced | Balanced – 'Once the tec | Yes | Migrator | 0 |
| 220 | 38 | 32 | 21,7 | 80 | 26,125 | 61 | 173 | Balanced | Balanced – 'Once the tec | Yes | Migrator | 0 |
| 177 | 12 | 9 | 48 | 144 | 4,5 | 18 | 171 | Balanced | An Early Adopter – 'I low | No | Pioneer | 0 |
| 365 | 7 | 5 | 38,4 | 129 | 10,5 | 34 | 168 | Balanced | Balanced – 'Once the tec | Yes | Migrator | 0 |
| 369 | 11 | 8 | 52,8 | 151 | 2 | 8 | 167 | Balanced | Balanced – 'Once the tec | Yes | Migrator | 0 |
| 391 | 7 | 5 | 38,4 | 129 | 10 | 33 | 167 | Balanced | Balanced – 'Once the tec | Yes | Migrator | 0 |
| 90 | 14 | 12 | 33,6 | 119 | 12 | 35 | 166 | Balanced | Balanced – 'Once the tec | Yes | Migrator | Marcus@G |
| 408 | 17 | 14 | 30 | 106 | 17,5 | 46 | 166 | Balanced | Balanced – 'Once the tec | Yes | Migrator | 0 |
| 469 | 9 | 7 | 60 | 158 | 0 | 1 | 166 | Balanced | Balanced – 'Once the tec | Yes | Migrator | 0 |
| 215 | 18 | 16 | 31,5 | 112 | 13 | 37 | 165 | Balanced | Balanced – 'Once the tec | Yes | Migrator | 0 |
| 424 | 13 | 11 | 33,6 | 119 | 12 | 35 | 165 | Balanced | Balanced – 'Once the tec | Yes | Migrator | 0 |
| 341 | 16 | 14 | 42 | 137 | 3,25 | 13 | 164 | Balanced | Balanced – 'Once the tec | Yes | Migrator | 0 |
| 379 | 33 | 29 | 33,6 | 119 | 4 | 16 | 164 | Balanced | Balanced – 'Once the tec | Yes | Migrator | 0 |
| 423 | 10 | 7 | 24 | 85 | 36 | 72 | 164 | Balanced | Traditionalist – 'I don't ne | No | Settler | 0 |
| 436 | 9 | 7 | 33,6 | 119 | 13,5 | 38 | 164 | Balanced | An Early Adopter – 'I low | No | Pioneer | 0 |
| 251 | 14 | 12 | 33,6 | 119 | 9,5 | 32 | 163 | Balanced | Balanced – 'Once the tec | Yes | Migrator | 0 |
| 253 | 17 | 14 | 30,6 | 108 | 15 | 41 | 163 | Balanced | Balanced – 'Once the tec | Yes | Migrator | 0 |
| 360 | 11 | 8 | 54 | 152 | 0,75 | 3 | 163 | Balanced | An Early Adopter – 'I low | No | Pioneer | 0 |
| 318 | 11 | 8 | 13,5 | 48 | 67,5 | 106 | 162 | Balanced | Balanced – 'Once the tec | Yes | Migrator | 0 |
| 350 | 11 | 8 | 46,2 | 143 | 2,5 | 10 | 161 | Balanced | Balanced – 'Once the tec | Yes | Migrator | cfaber1994 |
| 317 | 18 | 15 | 7,8 | 27 | 91,25 | 118 | 160 | Balanced | Balanced – 'Once the tec | Yes | Migrator | 0 |
| 367 | 10 | 7 | 38,4 | 129 | 6 | 24 | 160 | Balanced | Balanced – 'Once the tec | Yes | Migrator | 0 |
| 375 | 7 | 5 | 43,2 | 138 | 4 | 16 | 159 | Balanced | Balanced – 'Once the tec | Yes | Migrator | 0 |
| 128 | 22 | 19 | 42 | 137 | 0 | 1 | 157 | Balanced | Balanced – 'Once the tec | Yes | Migrator | 0 |
| 214 | 11 | 8 | 48 | 144 | 1,25 | 5 | 157 | Balanced | Balanced – 'Once the tec | Yes | Migrator | 0 |
| 335 | 22 | 19 | 38,4 | 129 | 2,25 | 9 | 157 | Balanced | Balanced – 'Once the tec | Yes | Migrator | 0 |
| 260 | 15 | 13 | 15,6 | 56 | 46,75 | 87 | 156 | Balanced | Balanced – 'Once the tec | Yes | Migrator | 0 |
| 227 | 20 | 17 | 15,75 | 58 | 42 | 80 | 155 | Balanced | Balanced – 'Once the tec | Yes | Migrator | 0 |
| 252 | 15 | 13 | 24 | 85 | 22,5 | 56 | 154 | Balanced | Balanced – 'Once the tec | Yes | Migrator | 0 |
| 399 | 7 | 5 | 24 | 85 | 27 | 63 | 153 | Balanced | Traditionalist – 'I don't ne | No | Settler | 0 |
| 426 | 21 | 17 | 21,6 | 78 | 24 | 58 | 153 | Balanced | Balanced – 'Once the tec | Yes | Migrator | 0 |
| 111 | 44 | 35 | 10,66667 | 36 | 42,125 | 81 | 152 | Balanced | Balanced – 'Once the tec | Yes | Migrator | 0 |
| 458 | 41 | 33 | 33 | 116 | 0 | 1 | 150 | Balanced | An Early Adopter – 'I low | No | Pioneer | 0 |
| 77 | 15 | 13 | 16,2 | 61 | 37,5 | 74 | 148 | Balanced | Balanced – 'Once the tec | Yes | Migrator | 0 |
| 162 | 47 | 38 | 14,7 | 54 | 22 | 55 | 147 | Balanced | Balanced – 'Once the tec | Yes | Migrator | 0 |
| 176 | 32 | 27 | 3,8 | 15 | 65 | 104 | 146 | Balanced | Balanced – 'Once the tec | Yes | Migrator | mail@mind |
| 327 | 7 | 5 | 43,2 | 138 | 0,75 | 3 | 146 | Balanced | Balanced – 'Once the tec | Yes | Migrator | 0 |
| 19 | 7 | 5 | 21,6 | 78 | 25 | 59 | 142 | Balanced | Balanced – 'Once the tec | Yes | Migrator | 0 |
| 124 | 29 | 26 | 12,6 | 44 | 35 | 72 | 142 | Balanced | Balanced – 'Once the tec | Yes | Migrator | 0 |
| 151 | 41 | 33 | 18 | 65 | 16,6667 | 44 | 142 | Balanced | Balanced – 'Once the tec | Yes | Migrator | 0 |
| 36 | 33 | 29 | 24 | 85 | 7 | 27 | 141 | Balanced | Balanced – 'Once the tec | Yes | Migrator | 0 |
| 107 | 32 | 27 | 10,8 | 37 | 39,375 | 77 | 141 | Balanced | An Early Adopter – 'I low | No | Pioneer | 0 |
| 344 | 17 | 14 | 28,8 | 101 | 6,5 | 26 | 141 | Balanced | Balanced – 'Once the tec | Yes | Migrator | 0 |
| 373 | 9 | 7 | 28,8 | 101 | 9 | 31 | 139 | Balanced | Balanced – 'Once the tec | Yes | Migrator | 0 |
| 181 | 17 | 14 | 18 | 65 | 25 | 59 | 138 | Balanced | Balanced – 'Once the tec | Yes | Migrator | Morgenster |
| 216 | 28 | 25 | 28,8 | 101 | 3 | 12 | 138 | Balanced | Balanced – 'Once the tec | Yes | Migrator | 0 |
| 359 | 20 | 17 | 30 | 106 | 3,75 | 15 | 138 | Balanced | Balanced – 'Once the tec | Yes | Migrator | 0 |
| 395 | 10 | 7 | 36 | 125 | 1,5 | 6 | 138 | Balanced | Balanced – 'Once the tec | Yes | Migrator | 0 |
| 161 | 6 | 4 | 1,8 | 6 | 100 | 127 | 137 | Balanced | Balanced – 'Once the tec | Yes | Migrator | 0 |
| 88 | 23 | 19 | 9,6 | 31 | 43,75 | 83 | 133 | Balanced | Balanced – 'Once the tec | Yes | Migrator | 0 |
| 255 | 15 | 13 | 13,5 | 48 | 36 | 72 | 133 | Balanced | Balanced – 'Once the tec | Yes | Migrator | 0 |
| 293 | 20 | 16 | 4,2 | 17 | 60 | 100 | 133 | Balanced | Balanced – 'Once the tec | Yes | Migrator | 0 |
| 400 | 9 | 6 | 24 | 85 | 15 | 41 | 132 | Balanced | Traditionalist – 'I don't ne | No | Settler | 0 |
| 376 | 13 | 11 | 27 | 93 | 7 | 27 | 131 | Balanced | Balanced – 'Once the tec | Yes | Migrator | 0 |
| 421 | 7 | 5 | 28,8 | 101 | 6 | 24 | 130 | Balanced | An Early Adopter – 'I low | No | Pioneer | 0 |
| 310 | 8 | 6 | 16,8 | 64 | 25 | 59 | 129 | Balanced | Balanced – 'Once the tec | Yes | Migrator | 0 |
| 451 | 26 | 22 | 12,6 | 44 | 27 | 63 | 129 | Balanced | Balanced – 'Once the tec | Yes | Migrator | 0 |
| 444 | 10 | 7 | 18 | 65 | 22,5 | 56 | 128 | Balanced | Balanced – 'Once the tec | Yes | Migrator | 0 |
| 2 | 15 | 13 | 7,2 | 26 | 46,875 | 88 | 127 | Balanced | Balanced – 'Once the tec | Yes | Migrator | 0 |
| 438 | 9 | 6 | 24 | 85 | 12,5 | 36 | 127 | Balanced | Balanced – 'Once the tec | Yes | Migrator | 0 |
| 425 | 26 | 22 | 12,6 | 44 | 26 | 60 | 126 | Balanced | Balanced – 'Once the tec | Yes | Migrator | 0 |
| 292 | 20 | 16 | 4,2 | 17 | 50 | 92 | 125 | Balanced | Balanced – 'Once the tec | Yes | Migrator | 0 |
| 392 | 10 | 7 | 18 | 65 | 20 | 53 | 125 | Balanced | Balanced – 'Once the tec | Yes | Migrator | 0 |
| 311 | 7 | 5 | 16,8 | 64 | 20 | 53 | 122 | Balanced | Balanced – 'Once the tec | Yes | Migrator | 0 |
| 349 | 35 | 30 | 25,2 | 88 | 1 | 4 | 122 | Balanced | Balanced – 'Once the tec | Yes | Migrator | 0 |
| 357 | 10 | 7 | 30 | 106 | 2 | 8 | 121 | Balanced | Balanced – 'Once the tec | Yes | Migrator | 0 |
| 190 | 14 | 11 | 10,8 | 37 | 36 | 72 | 120 | Balanced | Balanced – 'Once the tec | Yes | Migrator | 0 |
| 377 | 25 | 21 | 19,2 | 69 | 8 | 29 | 119 | Balanced | Balanced – 'Once the tec | Yes | Migrator | 0 |
| 12 | 15 | 12 | 1,4 | 6 | 60 | 100 | 118 | Balanced | Balanced – 'Once the tec | Yes | Migrator | Mkellie5083 |
| 336 | 8 | 6 | 28,8 | 101 | 2,75 | 11 | 118 | Balanced | Balanced – 'Once the tec | Yes | Migrator | 0 |
| 338 | 16 | 14 | 22,95 | 83 | 5,5 | 21 | 118 | Balanced | Balanced – 'Once the tec | Yes | Migrator | 0 |
| 361 | 7 | 5 | 24 | 85 | 7,5 | 28 | 118 | Balanced | Balanced – 'Once the tec | Yes | Migrator | 0 |
| 123 | 29 | 26 | 10,8 | 37 | 19,5 | 52 | 115 | Balanced | Balanced – 'Once the tec | Yes | Migrator | 0 |
| 337 | 10 | 7 | 24 | 85 | 5 | 20 | 112 | Balanced | Balanced – 'Once the tec | Yes | Migrator | 0 |
| 366 | 13 | 11 | 24 | 85 | 3,75 | 15 | 111 | Balanced | Balanced – 'Once the tec | Yes | Migrator | 0 |
| 166 | 15 | 13 | 7,8 | 27 | 34,5 | 70 | 110 | Balanced | Balanced – 'Once the tec | Yes | Migrator | 0 |
| 324 | 29 | 26 | 21 | 76 | 2 | 8 | 110 | Balanced | Balanced – 'Once the tec | Yes | Migrator | 0 |
| 348 | 9 | 6 | 21,6 | 78 | 6,5 | 26 | 110 | Balanced | Balanced – 'Once the tec | Yes | Migrator | 0 |
| 352 | 23 | 19 | 12,6 | 44 | 17,625 | 47 | 110 | Balanced | An Early Adopter – 'I low | No | Pioneer | 0 |
| 383 | 7 | 5 | 24 | 85 | 5 | 20 | 110 | Balanced | Balanced – 'Once the tec | Yes | Migrator | evelyn.borc |
| 89 | 9 | 7 | 28,8 | 101 | 0 | 1 | 109 | Balanced | Balanced – 'Once the tec | Yes | Migrator | 0 |
| 439 | 7 | 5 | 16,2 | 61 | 13,5 | 38 | 104 | Balanced | Balanced – 'Once the tec | Yes | Migrator | 0 |
| 446 | 8 | 6 | 11 | 39 | 25 | 59 | 104 | Balanced | Balanced – 'Once the tec | Yes | Migrator | 0 |
| 329 | 13 | 11 | 21,6 | 78 | 3,25 | 13 | 102 | Balanced | Balanced – 'Once the tec | Yes | Migrator | 0 |
| 363 | 10 | 7 | 16,2 | 61 | 10,5 | 34 | 102 | Balanced | Balanced – 'Once the tec | Yes | Migrator | 0 |
| 47 | 44 | 36 | 12,87778 | 47 | 4,5 | 18 | 101 | Balanced | Balanced – 'Once the tec | Yes | Migrator | 0 |
| 228 | 13 | 11 | 24 | 85 | 1 | 4 | 100 | Balanced | Balanced – 'Once the tec | Yes | Migrator | 0 |
| 159 | 45 | 37 | 6,3 | 23 | 13,5 | 38 | 98 | Balanced | Balanced – 'Once the tec | Yes | Migrator | 0 |
| 134 | 15 | 13 | 10 | 34 | 18,75 | 50 | 97 | Balanced | Balanced – 'Once the tec | Yes | Migrator | 0 |
| 262 | 10 | 7 | 19,2 | 69 | 5 | 20 | 96 | Balanced | Balanced – 'Once the tec | Yes | Migrator | 0 |
| 314 | 14 | 11 | 10,8 | 37 | 18 | 48 | 96 | Balanced | An Early Adopter – 'I low | No | Pioneer | 0 |
| 286 | 7 | 5 | 9,6 | 31 | 25 | 59 | 95 | Balanced | Balanced – 'Once the tec | Yes | Migrator | 0 |
| 404 | 12 | 9 | 16,2 | 61 | 6 | 24 | 94 | Balanced | Balanced – 'Once the tec | Yes | Migrator | 0 |
| 132 | 15 | 13 | 10,8 | 37 | 15,625 | 42 | 92 | Balanced | Balanced – 'Once the tec | Yes | Migrator | 0 |
| 199 | 7 | 5 | 24 | 85 | 0,5 | 2 | 92 | Balanced | Balanced – 'Once the tec | Yes | Migrator | 0 |
| 33 | 11 | 8 | 18 | 65 | 4,5 | 18 | 91 | Balanced | An Early Adopter – 'I low | No | Pioneer | 0 |
| 56 | 18 | 20 | 16 | 61 | 2,5 | 10 | 91 | Traditionalist | Traditionalist – 'I don't ne | Yes | Settler | 0 |
| 449 | 9 | 6 | 10 | 65 | 5 | 20 | 91 | Balanced | Balanced – 'Once the tec | Yes | Migrator | 0 |
| 386 | 32 | 27 | 4,2 | 17 | 17,5 | 46 | 90 | Balanced | Balanced – 'Once the tec | Yes | Migrator | 0 |
| 418 | 10 | 7 | 12 | 43 | 14 | 39 | 89 | Balanced | Balanced – 'Once the tec | Yes | Migrator | 0 |
| 180 | 18 | 15 | 10,26667 | 35 | 13,2 | 38 | 88 | Balanced | Balanced – 'Once the tec | Yes | Migrator | 0 |
| 374 | 12 | 9 | 9,6 | 31 | 18 | 48 | 88 | Balanced | Balanced – 'Once the tec | Yes | Migrator | 0 |
| 409 | 20 | 16 | 7,2 | 26 | 17,5 | 46 | 88 | Balanced | Balanced – 'Once the tec | Yes | Migrator | 0 |
| 55 | 10 | 7 | 21,6 | 78 | 0 | 1 | 86 | Balanced | Balanced – 'Once the tec | Yes | Migrator | 0 |
| 148 | 41 | 32 | 10,8 | 37 | 4,16667 | 17 | 86 | Balanced | Balanced – 'Once the tec | Yes | Migrator | 0 |
| 410 | 11 | 8 | 5,6 | 21 | 22,75 | 57 | 86 | Balanced | Balanced – 'Once the tec | Yes | Migrator | 0 |
| 456 | 34 | 29 | 15 | 55 | 0 | 1 | 85 | Balanced | Balanced – 'Once the tec | Yes | Migrator | 0 |
| 466 | 9 | 6 | 21,6 | 78 | 0 | 1 | 85 | Balanced | Balanced – 'Once the tec | Yes | Migrator | 0 |
| 49 | 20 | 16 | 5,366667 | 19 | 18,5 | 49 | 84 | Balanced | Balanced – 'Once the tec | Yes | Migrator | 0 |
| 396 | 29 | 25 | 1,2 | 5 | 21 | 54 | 84 | Balanced | Balanced – 'Once the tec | Yes | Migrator | 0 |
| 48 | 7 | 5 | 20 | 72 | 1,5 | 6 | 83 | Balanced | Traditionalist – 'I don't ne | No | Settler | 0 |
| 304 | 7 | 5 | 10,8 | 37 | 15 | 41 | 83 | Balanced | Balanced – 'Once the tec | Yes | Migrator | teodora.bak |
| 378 | 10 | 7 | 10,8 | 37 | 14 | 39 | 83 | Balanced | Balanced – 'Once the tec | Yes | Migrator | 0 |
| 432 | 12 | 9 | 4,8 | 18 | 22,5 | 56 | 83 | Balanced | Balanced – 'Once the tec | Yes | Migrator | 0 |
| 37 | 32 | 27 | 7,2 | 26 | 7,5 | 28 | 81 | Balanced | Balanced – 'Once the tec | Yes | Migrator | 0 |
| 384 | 32 | 27 | 2,4 | 9 | 16,25 | 43 | 79 | Balanced | Balanced – 'Once the tec | Yes | Migrator | 0 |
| 291 | 7 | 5 | 9,6 | 31 | 15 | 41 | 77 | Balanced | Balanced – 'Once the tec | Yes | Migrator | 0 |
| 340 | 16 | 13 | 11,875 | 42 | 5,75 | 22 | 77 | Balanced | Balanced – 'Once the tec | Yes | Migrator | 0 |
| 30 | 16 | 14 | 16,2 | 61 | 0 | 1 | 76 | Balanced | Balanced – 'Once the tec | Yes | Migrator | rainer.mieth |
| 211 | 7 | 5 | 19,2 | 69 | 0 | 1 | 75 | Balanced | Balanced – 'Once the tec | Yes | Migrator | rgrassick@ |
| 126 | 43 | 35 | 9 | 30 | 2,25 | 9 | 74 | Balanced | Balanced – 'Once the tec | Yes | Migrator | 0 |
| 202 | 25 | 21 | 9,6 | 31 | 5 | 20 | 72 | Balanced | Balanced – 'Once the tec | Yes | Migrator | 0 |
| 343 | 13 | 11 | 9,6 | 31 | 8,5 | 30 | 72 | Balanced | Balanced – 'Once the tec | Yes | Migrator | 0 |
| 24 | 20 | 17 | 12,8 | 46 | 2 | 8 | 71 | Balanced | An Early Adopter – 'I low | No | Pioneer | Timojroeme |
| 302 | 77 | 45 | 7 | 25 | 0 | 1 | 71 | Balanced | An Early Adopter – 'I low | No | Pioneer | 0 |
| 170 | 9 | 6 | 9,6 | 31 | 10 | 33 | 70 | Balanced | Balanced – 'Once the tec | Yes | Migrator | 0 |
| 347 | 18 | 15 | 6 | 21 | 10,5 | 34 | 70 | Balanced | Balanced – 'Once the tec | Yes | Migrator | 0 |
| 385 | 28 | 24 | 7,2 | 26 | 5 | 20 | 70 | Balanced | Balanced – 'Once the tec | Yes | Migrator | 0 |
| 287 | 7 | 5 | 9,6 | 31 | 10 | 33 | 69 | Balanced | Balanced – 'Once the tec | Yes | Migrator | 0 |
| 428 | 10 | 7 | 12,6 | 44 | 4,5 | 18 | 69 | Balanced | Balanced – 'Once the tec | Yes | Migrator | 0 |
| 415 | 13 | 11 | 8 | 28 | 8 | 29 | 68 | Balanced | Balanced – 'Once the tec | Yes | Migrator | 0 |
| 119 | 15 | 13 | 14,4 | 52 | 0 | 1 | 66 | Balanced | Balanced – 'Once the tec | Yes | Migrator | juliano.bras |
| 295 | 20 | 16 | 4,2 | 17 | 10 | 33 | 66 | Balanced | Balanced – 'Once the tec | Yes | Migrator | 0 |
| 197 | 25 | 21 | 4 | 16 | 7,5 | 28 | 65 | Balanced | Balanced – 'Once the tec | Yes | Migrator | 0 |
| 246 | 7 | 5 | 14,4 | 52 | 2 | 8 | 65 | Balanced | Balanced – 'Once the tec | Yes | Migrator | 0 |
| 457 | 42 | 33 | 9,6 | 31 | 0 | 1 | 65 | Balanced | An Early Adopter – 'I low | No | Pioneer | 0 |
| 87 | 20 | 16 | 1,8 | 6 | 15 | 41 | 63 | Balanced | Balanced – 'Once the tec | Yes | Migrator | 0 |
| 149 | 41 | 32 | 3,6 | 14 | 4,16667 | 17 | 63 | Balanced | Balanced – 'Once the tec | Yes | Migrator | 0 |
| 403 | 7 | 5 | 6 | 21 | 12 | 35 | 61 | Balanced | Balanced – 'Once the tec | Yes | Migrator | 0 |
| 401 | 13 | 11 | 10,8 | 37 | 3 | 12 | 60 | Balanced | Balanced – 'Once the tec | Yes | Migrator | 0 |
| 371 | 10 | 7 | 4,8 | 18 | 10 | 33 | 58 | Balanced | Balanced – 'Once the tec | Yes | Migrator | com.al.step |
| 464 | 38 | 31 | 7,2 | 26 | 0 | 1 | 58 | Balanced | Balanced – 'Once the tec | Yes | Migrator | 0 |
| 21 | 29 | 26 | 6 | 21 | 2,5 | 10 | 57 | Balanced | Balanced – 'Once the tec | Yes | Migrator | 0 |
| 39 | 77 | 45 | 2,5 | 10 | 0 | 1 | 56 | Balanced | Balanced – 'Once the tec | Yes | Migrator | 0 |
| 65 | 11 | 8 | 10 | 34 | 3 | 12 | 54 | Balanced | Balanced – 'Once the tec | Yes | Migrator | 0 |
| 397 | 12 | 9 | 10 | 34 | 2,5 | 10 | 53 | Balanced | Balanced – 'Once the tec | Yes | Migrator | 0 |
| 127 | 22 | 19 | 3,4 | 13 | 5 | 20 | 52 | Balanced | Balanced – 'Once the tec | Yes | Migrator | 0 |
| 206 | 7 | 5 | 12 | 43 | 1 | 4 | 52 | Balanced | Balanced – 'Once the tec | Yes | Migrator | 0 |
| 290 | 13 | 11 | 9,6 | 31 | 1 | 4 | 46 | Balanced | Balanced – 'Once the tec | Yes | Migrator | therese.jaco |
| 10 | 7 | 5 | 6,4 | 24 | 4 | 16 | 45 | Balanced | Traditionalist – 'I don't ne | No | Settler | 0 |
| 152 | 7 | 5 | 9,6 | 31 | 2 | 8 | 44 | Balanced | Traditionalist – 'I don't ne | No | Settler | 0 |
| 185 | 9 | 6 | 6,3 | 23 | 3,75 | 15 | 44 | Balanced | Balanced – 'Once the tec | Yes | Migrator | 0 |
| 258 | 9 | 6 | 9,6 | 31 | 1,25 | 5 | 42 | Balanced | An Early Adopter – 'I low | No | Pioneer | 0 |
| 50 | 8 | 6 | 6,3 | 23 | 3 | 12 | 41 | Balanced | Balanced – 'Once the tec | Yes | Migrator | 0 |
| 102 | 27 | 23 | 4 | 16 | 0,5 | 2 | 41 | Balanced | Balanced – 'Once the tec | Yes | Migrator | 0 |
| 85 | 46 | 38 | 0 | 1 | 0 | 1 | 40 | Balanced | Balanced – 'Once the tec | Yes | Migrator | 0 |
| 183 | 41 | 32 | 0,9 | 4 | 0 | 1 | 37 | Balanced | Balanced – 'Once the tec | Yes | Migrator | 0 |
| 465 | 38 | 31 | 0,5 | 2 | 0 | 1 | 34 | Balanced | Balanced – 'Once the tec | Yes | Migrator | 0 |
| 370 | 33 | 28 | 0,6 | 3 | 0 | 1 | 32 | Balanced | Balanced – 'Once the tec | Yes | Migrator | stefan.lange |
| 462 | 16 | 13 | 4,8 | 18 | 0 | 1 | 32 | Balanced | Traditionalist – 'I don't ne | No | Settler | 0 |
| 229 | 24 | 20 | 2,4 | 9 | 0 | 1 | 30 | Balanced | Balanced – 'Once the tec | Yes | Migrator | 0 |
| 187 | 30 | 26 | 0 | 1 | 0 | 1 | 28 | Balanced | Balanced – 'Once the tec | Yes | Migrator | 0 |
| 73 | 10 | 7 | 3,2 | 12 | 2 | 8 | 27 | Balanced | Balanced – 'Once the tec | Yes | Migrator | David.holge |
| 323 | 10 | 7 | 3,6 | 14 | 1,5 | 8 | 27 | Balanced | Balanced – 'Once the tec | Yes | Migrator | 0 |
| 203 | 23 | 19 | 0 | 1 | 0 | 1 | 21 | Traditionalist | Traditionalist – 'I don't ne | Yes | Settler | lukas.mechi |
| 481 | 23 | 19 | 0 | 1 | 0 | 1 | 21 | Traditionalist | Traditionalist – 'I don't ne | Yes | Settler | 0 |
| 498 | 23 | 19 | 0 | 1 | 0 | 1 | 21 | Traditionalist | Traditionalist – 'I don't ne | Yes | Settler | 0 |
| 513 | 23 | 19 | 0 | 1 | 0 | 1 | 21 | Traditionalist | Balanced – 'Once the tec | Yes | Migrator | 0 |
| 184 | 7 | 5 | 1,2 | 5 | 2,5 | 10 | 20 | Traditionalist | Traditionalist – 'I don't ne | Yes | Settler | jan.ennenba |
| 443 | 11 | 8 | 2,4 | 9 | 0,75 | 3 | 20 | Traditionalist | Traditionalist – 'I don't ne | Yes | Settler | skype |
| 455 | 22 | 18 | 0 | 1 | 0 | 1 | 20 | Traditionalist | Balanced – 'Once the tec | No | Migrator | bernd.zattle |
| 461 | 1 | 1 | 4,8 | 18 | 0 | 1 | 20 | Traditionalist | An Early Adopter – 'I low | No | Pioneer | 0 |
| 443 | 11 | 8 | 2,4 | 9 | 0,75 | 3 | 20 | Traditionalist | Traditionalist – 'I don't ne | Yes | Settler | annaroe@li |
| 455 | 22 | 18 | 0 | 1 | 0 | 1 | 20 | Traditionalist | Balanced – 'Once the tec | No | Migrator | bernd.zattle |
| 461 | 1 | 1 | 4,8 | 18 | 0 | 1 | 20 | Traditionalist | An Early Adopter – 'I lov | No | Pioneer | 0 |
| 487 | 22 | 18 | 0 | 1 | 0 | 1 | 20 | Traditionalist | Traditionalist – 'I don't ne | Yes | Settler | 0 |
| 504 | 22 | 18 | 0 | 1 | 0 | 1 | 20 | Traditionalist | Traditionalist – 'I don't ne | Yes | Settler | 0 |
| 518 | 22 | 18 | 0 | 1 | 0 | 1 | 20 | Traditionalist | Traditionalist – 'I don't ne | Yes | Settler | 0 |
| 459 | 12 | 9 | 2,1 | 8 | 0 | 1 | 18 | Traditionalist | Traditionalist – 'I don't ne | Yes | Settler | 0 |
| 198 | 18 | 15 | 0 | 1 | 0 | 1 | 17 | Traditionalist | Balanced – 'Once the tec | No | Migrator | 0 |
| 480 | 18 | 15 | 0 | 1 | 0 | 1 | 17 | Traditionalist | Traditionalist – 'I don't ne | Yes | Settler | 0 |
| 497 | 18 | 15 | 0 | 1 | 0 | 1 | 17 | Traditionalist | Balanced – 'Once the tec | No | Migrator | 0 |
| 512 | 18 | 15 | 0 | 1 | 0 | 1 | 17 | Traditionalist | Traditionalist – 'I don't ne | Yes | Settler | 0 |
| 189 | 17 | 14 | 0 | 1 | 0 | 1 | 16 | Traditionalist | Traditionalist – 'I don't ne | Yes | Settler | 0 |
| 79 | 16 | 13 | 0 | 1 | 0 | 1 | 15 | Traditionalist | Balanced – 'Once the tec | No | Migrator | 0 |
| 188 | 16 | 13 | 0 | 1 | 0 | 1 | 15 | Traditionalist | Traditionalist – 'I don't ne | Yes | Settler | m.seeger@ |
| 478 | 16 | 13 | 0 | 1 | 0 | 1 | 15 | Traditionalist | Traditionalist – 'I don't ne | Yes | Settler | 0 |
| 495 | 16 | 13 | 0 | 1 | 0 | 1 | 15 | Traditionalist | Traditionalist – 'I don't ne | Yes | Settler | 0 |
| 510 | 16 | 13 | 0 | 1 | 0 | 1 | 15 | Traditionalist | Traditionalist – 'I don't ne | Yes | Settler | 0 |
| 29 | 15 | 12 | 0 | 1 | 0 | 1 | 14 | Traditionalist | Traditionalist – 'I don't ne | Yes | Settler | 0 |
| 82 | 13 | 11 | 0 | 1 | 0 | 1 | 13 | Traditionalist | Balanced – 'Once the tec | No | Migrator | 0 |
| 218 | 14 | 11 | 0 | 1 | 0 | 1 | 13 | Traditionalist | Traditionalist – 'I don't ne | Yes | Settler | 0 |
| 83 | 12 | 9 | 0 | 1 | 0 | 1 | 11 | Traditionalist | Balanced – 'Once the tec | No | Migrator | 0 |
| 140 | 12 | 9 | 0 | 1 | 0 | 1 | 11 | Traditionalist | Traditionalist – 'I don't ne | Yes | Settler | 0 |
| 212 | 12 | 9 | 0 | 1 | 0 | 1 | 11 | Traditionalist | Traditionalist – 'I don't ne | Yes | Settler | 0 |
| 472 | 12 | 9 | 0 | 1 | 0 | 1 | 11 | Traditionalist | Traditionalist – 'I don't ne | Yes | Settler | 0 |
| 482 | 12 | 9 | 0 | 1 | 0 | 1 | 11 | Traditionalist | Traditionalist – 'I don't ne | Yes | Settler | 0 |
| 492 | 12 | 9 | 0 | 1 | 0 | 1 | 11 | Traditionalist | Balanced – 'Once the tec | No | Migrator | 0 |
| 499 | 12 | 9 | 0 | 1 | 0 | 1 | 11 | Traditionalist | Traditionalist – 'I don't ne | Yes | Settler | 0 |
| 507 | 12 | 9 | 0 | 1 | 0 | 1 | 11 | Traditionalist | Traditionalist – 'I don't ne | Yes | Settler | 0 |
| 514 | 12 | 9 | 0 | 1 | 0 | 1 | 11 | Traditionalist | Traditionalist – 'I don't ne | Yes | Settler | skype |
| 31 | 11 | 8 | 0 | 1 | 0 | 1 | 10 | Traditionalist | Traditionalist – 'I don't ne | Yes | Settler | 0 |
| 141 | 11 | 8 | 0 | 1 | 0 | 1 | 10 | Traditionalist | Balanced – 'Once the tec | No | Migrator | 0 |
| 192 | 11 | 8 | 0 | 1 | 0 | 1 | 10 | Traditionalist | Traditionalist – 'I don't ne | Yes | Settler | ahmadani7 |
| 470 | 11 | 8 | 0 | 1 | 0 | 1 | 10 | Traditionalist | Balanced – 'Once the tec | No | Migrator | 0 |
| 479 | 11 | 8 | 0 | 1 | 0 | 1 | 10 | Traditionalist | Traditionalist – 'I don't ne | Yes | Settler | 0 |
| 489 | 11 | 8 | 0 | 1 | 0 | 1 | 10 | Traditionalist | Traditionalist – 'I don't ne | Yes | Settler | 0 |
| 496 | 11 | 8 | 0 | 1 | 0 | 1 | 10 | Traditionalist | Traditionalist – 'I don't ne | Yes | Settler | 0 |
| 505 | 11 | 8 | 0 | 1 | 0 | 1 | 10 | Traditionalist | Traditionalist – 'I don't ne | Yes | Settler | 0 |
| 511 | 11 | 8 | 0 | 1 | 0 | 1 | 10 | Traditionalist | Traditionalist – 'I don't ne | Yes | Settler | 0 |
| 232 | 10 | 7 | 0 | 1 | 0 | 1 | 9 | Traditionalist | Traditionalist – 'I don't ne | Yes | Settler | andreas.po |
| 484 | 10 | 7 | 0 | 1 | 0 | 1 | 9 | Traditionalist | Balanced – 'Once the tec | No | Migrator | 0 |
| 500 | 10 | 7 | 0 | 1 | 0 | 1 | 9 | Traditionalist | Traditionalist – 'I don't ne | Yes | Settler | 0 |
| 516 | 10 | 7 | 0 | 1 | 0 | 1 | 9 | Traditionalist | Traditionalist – 'I don't ne | Yes | Settler | 0 |
| 398 | 5 | 3 | 0 | 1 | 0 | 1 | 5 | Traditionalist | Traditionalist – 'I don't ne | Yes | Settler | 0 |
| 485 | 5 | 3 | 0 | 1 | 0 | 1 | 5 | Traditionalist | Traditionalist – 'I don't ne | Yes | Settler | 0 |
| 501 | 5 | 3 | 0 | 1 | 0 | 1 | 5 | Traditionalist | Balanced – 'Once the tec | No | Migrator | 0 |
| 71 | 4 | 2 | 0 | 1 | 0 | 1 | 4 | Traditionalist | Traditionalist – 'I don't ne | Yes | Settler | 0 |
| 80 | 4 | 2 | 0 | 1 | 0 | 1 | 4 | Traditionalist | Traditionalist – 'I don't ne | Yes | Settler | 0 |
| 471 | 4 | 2 | 0 | 1 | 0 | 1 | 4 | Traditionalist | Traditionalist – 'I don't ne | Yes | Settler | 0 |
| 506 | 4 | 2 | 0 | 1 | 0 | 1 | 4 | Traditionalist | Balanced – 'Once the tec | No | Migrator | 0 |
| 173 | 1 | 1 | 0 | 1 | 0 | 1 | 3 | Traditionalist | Traditionalist – 'I don't ne | Yes | Settler | 0 |
| 476 | 1 | 1 | 0 | 1 | 0 | 1 | 3 | Traditionalist | Balanced – 'Once the tec | No | Migrator | 0 |
| 494 | 1 | 1 | 0 | 1 | 0 | 1 | 3 | Traditionalist | Traditionalist – 'I don't ne | Yes | Settler | 0 |
| 509 | 1 | 1 | 0 | 1 | 0 | 1 | 3 | Traditionalist | Traditionalist – 'I don't ne | Yes | Settler | 0 |